\documentclass[aps,floats]{revtex4-2}
\usepackage{amsmath,amssymb}
\usepackage{graphicx,epsfig}
\usepackage[greek,english]{babel}
\usepackage{bbold}

\usepackage{amsfonts}

\usepackage{mathtools}
\usepackage{makecell,tabularx}
\setcellgapes{1pt}
\makegapedcells

\begin{document}
\bibliographystyle {plain}

\pdfoutput=1
\def\oppropto{\mathop{\propto}} 
\def\opsimeq{\mathop{\simeq}}
\def\opoverderline{\mathop{\overline}}
\def\operarrow{\mathop{\longrightarrow}}
\def\opsim{\mathop{\sim}}

\def\opmin{\mathop{\min}} 
\def\opmax{\mathop{\max}} 
\def\oplim{\mathop{\lim}}

\title{ Large deviations for trajectory observables of diffusion processes in dimension $d>1$ \\
in the double limit of large time and small diffusion coefficient  } 


\author{C\'ecile Monthus}
\affiliation{Universit\'e Paris-Saclay, CNRS, CEA, Institut de Physique Th\'eorique, 91191 Gif-sur-Yvette, France}


\begin{abstract}

For diffusion processes in dimension $d>1$, the statistics of trajectory observables over the time-window $[0,T]$ can be studied via the Feynman-Kac deformations of the Fokker-Planck generator, that can be interpreted as euclidean non-hermitian electromagnetic quantum Hamiltonians. It is then interesting to compare the four regimes corresponding to the time $T$ either finite or large and to the diffusion coefficient $D$ either finite or small. (1) For finite $T$ and finite $D$, one needs to consider the full time-dependent quantum problem that involves the full spectrum of the Hamiltonian. (2) For large time $T \to + \infty$ and finite $D$, one only needs to consider the ground-state properties of the quantum Hamiltonian to obtain the generating function of rescaled cumulants and to construct the corresponding canonical conditioned processes. (3) For finite $T$  and $D \to 0$, one only needs to consider the dominant classical trajectory and its action satisfying the Hamilton-Jacobi equation, as in the semi-classical WKB approximation of quantum mechanics. (4) In the double limit $T \to + \infty$ and $D \to 0$, the simplifications in the large deviations in $\frac{T}{D}$ of trajectory observables can be analyzed via the two orders of limits, i.e. either from the limit $D \to 0$ of the ground-state properties of the quantum Hamiltonians of (2), or from the limit of long classical trajectories $T \to +\infty$ in the semi-classical WKB approximation of (3). This general framework is illustrated in dimension $d=2$ with rotational invariance.

\end{abstract}

\maketitle


\section{ Introduction  }

The theory of large deviations (see the reviews \cite{oono,ellis,review_touchette} and references therein)
has attracted a lot of interest recently, in particular in the field of nonequilibrium 
(see the reviews with different scopes \cite{derrida-lecture,harris_Schu,searles,harris,mft,sollich_review,lazarescu_companion,lazarescu_generic,jack_review}, 
the PhD Theses \cite{fortelle_thesis,vivien_thesis,chetrite_thesis,wynants_thesis,chabane_thesis,duBuisson_thesis} 
 and the Habilitation Thesis \cite{chetrite_HDR}).
 For Markov processes, the large deviations properties of 
 time-local trajectory observables over a large time-window $[0,T]$
can be analyzed via appropriate deformations of the Markov generators
 \cite{peliti,derrida-lecture,sollich_review,lazarescu_companion,lazarescu_generic,jack_review,vivien_thesis,lecomte_chaotic,lecomte_thermo,lecomte_formalism,lecomte_glass,kristina1,kristina2,jack_ensemble,simon1,simon2,tailleur,simon3,Gunter1,Engel_Seifert,Gunter2,Gunter3,Gunter4,chetrite_canonical,chetrite_conditioned,chetrite_optimal,chetrite_HDR,touchette_circle,touchette_langevin,touchette_occ,touchette_occupation,garrahan_lecture,Vivo,c_ring,c_detailed,chemical,derrida-conditioned,derrida-ring,bertin-conditioned,touchette-reflected,touchette-reflectedbis,c_lyapunov,previousquantum2.5doob,quantum2.5doob,quantum2.5dooblong,c_ruelle,lapolla,c_east,chabane,us_gyrator,duBuisson_gyrator,c_largedevpearson},
 and the construction of the corresponding canonical conditioned processes \cite{chetrite_conditioned,chetrite_optimal}.
 Another point of view involves the large deviations at the so-called level 2.5,
 where one can write explicit rates functions 
 for the joint distribution of the empirical density and of the empirical flows,
  either for discrete-time Markov chains 
\cite{fortelle_thesis,fortelle_chain,review_touchette,c_largedevdisorder,c_reset,c_inference,c_microcanoEnsembles,c_diffReg},
for continuous-time Markov jump processes with discrete configuration space
\cite{fortelle_thesis,fortelle_jump,maes_canonical,maes_onandbeyond,wynants_thesis,chetrite_formal,BFG1,BFG2,chetrite_HDR,c_ring,c_interactions,c_open,barato_periodic,chetrite_periodic,c_reset,c_inference,c_LargeDevAbsorbing,c_microcanoEnsembles,c_susyboundarydriven,c_diffReg,c_inverse},
for diffusion processes in continuous space
\cite{wynants_thesis,maes_diffusion,chetrite_formal,engel,chetrite_HDR,c_lyapunov,c_inference,c_susyboundarydriven,c_diffReg,c_inverse}, 
or for jump-diffusion or jump-drift processes \cite{c_reset,c_runandtumble,c_jumpdrift,c_SkewDB}.
Since any time-local trajectory observable can be rewritten in terms of the empirical density and of the empirical flows with appropriate coefficients, its rate function corresponds to
the optimization of the explicit rate function at level 2.5 over the empirical density and of the empirical flows
with appropriate constraints.

However in most physical applications, these large deviations of trajectory observables
 for large time $T \to + \infty$ are unfortunately not explicit, 
 since one needs to solve eigenvalue equations for deformed generators.
As discussed in detail in \cite{Engel_Seifert,derrida-ring,bertin-conditioned} for the particular case of diffusion processes
on a one-dimensional ring, it is then interesting to analyze what simplifications occur
when the diffusion coefficient becomes small $ D \to 0$. 
These large deviations of trajectory observables in the double limit $T \to + \infty$ and $D \to 0$ 
can be alternatively analyzed via the other order of limits \cite{Engel_Seifert,derrida-ring,bertin-conditioned},
where the first limit $D \to 0$ while the time $T$ remains finite
is somewhat analog to the semi-classical WKB approximation of quantum mechanics.
 The goal of the present paper is to discuss the same issue 
 of large deviations for trajectory observables in the double limit $T \to + \infty$ and $D \to 0$ 
 but for diffusion processes in dimension $d>1$.
 This problem was already studied in \cite{chabane,chabane_thesis}
 among many other cases involving a double limit 
in the large deviations for trajectory observables of various Markov processes,
as well as in the mathematical literature \cite{Bertini,renaud,DVmeetsFW}.
Here we will use instead the langage of euclidean non-hermitian electromagnetic quantum mechanics \cite{us_gyrator}
 in order to provide a self-contained elementary presentation for physicists
 of the various limits $T \to + \infty$ and/or $D \to 0$
   based only on basic knowledge of quantum mechanics
  (Schr\"odinger equation involving a scalar potential and a vector potential, Feynman path-integrals, perturbation theory for eigenvalues and eigenvectors, semi-classical WKB approximations, Hamilton-Jacobi equations of classical actions).

The paper is organized as follows.
In section \ref{seC_general}, we recall how the statistics of trajectory observables over a time-window $[0,T]$
 for diffusion processes with diffusion coefficient $D$ in dimension $d>1$ can be analyzed via the Feynman-Kac formula.
 We then describe how the large deviations in the double limit of large time $T \to + \infty$ and small diffusion coefficient $D \to 0$ can be studied via two different routes :
 
 (i) one can start from the large deviations properties for large time $T \to + \infty$ and finite diffusion coefficient
 $D$ described in section \ref{seC_largeT} 
 and then analyze the simplifications that occur in the limit $D \to 0$,
 as discussed in section \ref{seC_largeTsmallD}.
 
(ii) one can instead start from the large deviations properties for small diffusion coefficient $D \to 0$
 and finite time $T $ described in section \ref{seC_smallD}, 
 and then analyze the simplifications that occur 
 when the time-window becomes large $T \to +\infty$, as discussed in section \ref{seC_smallDlargeT}.
 
 This general framework is then illustrated in dimension $d=2$ with rotational invariance
 in section \ref{sec_2DPolar}.
  Our conclusions are summarized in \ref{seC_conclusions} with tables 
 in order to compare the different limits $ T \to + \infty$ and/or $D \to 0$
 from the point of view of various observables.
 Appendix \ref{app_2.5} describe how the large deviations at level 2.5 for large time $T \to + \infty$
and finite $D$ become simpler in the limit $D \to 0$
 in order to make the link with the statistics of trajectory observables
discussed in the main text.
 Appendix \ref{app_gauge} recalls the properties of gauge transformations 
for the euclidean non-hermitian electromagnetic quantum problems
that appear in the main text.


\section{ Trajectory observables for diffusions over a finite time-window $[0,T]$  }

\label{seC_general}

The Fokker-Planck dynamics corresponds to a conserved continuity equation
for the probability density $P_t(\vec x) $ to be at position $\vec x$ at time $t$
\begin{eqnarray}
 \partial_t P_t(\vec x)    =  -   \vec \nabla . \vec J_t(\vec x)  
\label{fokkerplanck}
\end{eqnarray}
where the current $\vec J_t(\vec x) $ involves the force $\vec F(\vec x)$ and the diffusion coefficient $D$
\begin{eqnarray}
\vec J_t(\vec x) \equiv P_t(\vec x )   \vec F(\vec x ) -D \vec \nabla   P_t(\vec x)  
\label{fokkerplanckj}
\end{eqnarray}
In order to analyse the statistics of trajectory observables over a finite time-window $[0,T]$,
it is useful to recall first the link between diffusion processes
and the euclidean non-Hermitian quantum mechanics in an electromagnetic potential \cite{us_gyrator}.


\subsection{ Reminder on the link with euclidean non-Hermitian quantum mechanics in an electromagnetic potential}

\subsubsection{ Fokker-Planck generator as an euclidean non-Hermitian electromagnetic quantum Hamiltonian }

The interpretation of the Fokker-Planck Eq. \ref{fokkerplanck}
 as an Euclidean Schr\"odinger equation
\begin{eqnarray}
 \partial_t P_t(\vec x)     = - H P_t(\vec x)
\label{fokkerplanckeuclidean}
\end{eqnarray}
involves the non-Hermitian quantum Hamiltonian $H \ne H^{\dagger}$
\begin{eqnarray}
H&&  =   \vec \nabla . \left(  - D  \vec \nabla +   \vec F(\vec x)\right)
=     - D  \vec \nabla^2
 +   \vec F(\vec x) . \vec \nabla
 +    [\vec \nabla . \vec F(\vec x) ]  
 \nonumber \\
 H^{\dagger} &&  = - \left( D  \vec \nabla +   \vec F(\vec x)\right). \vec \nabla
 =     - D  \vec \nabla^2 -   \vec F(\vec x) . \vec \nabla
\label{FPhamiltonian}
\end{eqnarray}
The identification with the Euclidean Hamiltonian
involving a real scalar potential $V(\vec x) $ and a 
purely imaginary vector potential $[-i \vec A(\vec x)] $ of real amplitude $\vec A(\vec x)$
(see more details in Appendix A of \cite{us_gyrator})
\begin{eqnarray}
H &&   =  -  D \left( \vec \nabla -   \vec A(\vec x) \right)^2    +  V(\vec x)
=  - D \vec \nabla^2 
+2 D \vec A ({\vec x}). \vec \nabla
+ D   \left( \vec \nabla.\vec A ({\vec x})\right)  - D  \vec A^2 (\vec x)  + V (\vec x)
\nonumber \\
H^{\dagger} &&   =
- D \left(  \vec \nabla  +  \vec A (\vec x) \right)^2  + V (\vec x) 
   =  - D  \vec \nabla^2 - 2 D \vec A ({\vec x}). \vec \nabla
-  D \left( \vec \nabla . \vec A ({\vec x})\right)  - D \vec A^2 (\vec x)  + V (\vec x)
\label{FPhamiltonianQuantum}
\end{eqnarray}
leads to the vector potential 
\begin{eqnarray}
  \vec A (\vec x ) \equiv  \frac{ \vec F( \vec x) }{2D} 
 \label{vectorpot}
\end{eqnarray}
and to the scalar potential $V(\vec x) $
\begin{eqnarray}
V(\vec x) \equiv  \frac{ \vec F ^2( \vec x) }{ 4D  }   + \frac{  [\vec \nabla . \vec F(\vec x) ]  }{2 }   
\label{scalarpot}
\end{eqnarray}

The magnetic 'field' corresponding to the 
the antisymmetric matrix $B_{\mu \nu}(\vec x )  =-B_{\nu \mu} (\vec x ) $
that can be computed from the vector potential $\vec A ( \vec x) $ of Eq. \ref{vectorpot}
 via the formula generalizing the three-dimensional curl
\begin{eqnarray}
 B_{\mu \nu}(\vec x )  =-B_{\nu \mu} (\vec x )  
 \equiv \partial_{\mu} A_{\nu} (\vec x ) -  \partial_{\nu} A_{\mu} (\vec x ) 
  = \frac{ \partial_{\mu} F_{\nu} (\vec x ) -  \partial_{\nu} F_{\mu} (\vec x ) }{2D}
 \label{magneticB}
\end{eqnarray}
characterizes the irreversibility of the diffusion \cite{us_gyrator}.

Since non-hermitian quantum Hamiltonians of the form of Eq. \ref{FPhamiltonianQuantum} 
are the only Hamiltonians that appear in the present paper,
please note that from now on, we will not always recall the "non-hermitian" character of this type of Hamiltonians,
and we will use the shorter name "the vector potential $\vec A ( \vec x) $ " and not 
"a purely imaginary vector potential $[-i \vec A(\vec x)] $ of real amplitude $\vec A(\vec x)$" anymore.


\subsubsection{ Propagator $P_T(\vec x \vert \vec x_0) $ via the path-integral involving the euclidean classical electromagnetic Lagrangian}

The path-integral for the propagator $P_T(\vec x \vert \vec x_0) $ from $\vec x_0$ towards $\vec x$ in time $T$
\begin{eqnarray}
P_T(\vec x \vert \vec x_0) \equiv \langle \vec x \vert e^{- T H} \vert \vec x_0 \rangle
= \int_{\vec x(t=0)=\vec x_0}^{\vec x(t=T)=\vec x} {\cal D}   \vec x(.)  
 e^{ - \displaystyle 
  \int_0^T dt {\cal L} (\vec x(t), \dot {\vec x}(t) )
 } 
\label{pathintegral}
\end{eqnarray}
involves the classical Lagrangian
\begin{eqnarray}
 {\cal L} (\vec x(t), \dot {\vec x}(t))
 && \equiv \frac{\left( \dot {\vec x } (t) - \vec F( \vec x(t)) \right)^2}{4 D} 
 + \frac{  [\vec \nabla . \vec F(\vec x (t) ) ]  }{2 }    
 \nonumber \\
 && = \frac{ \dot {\vec x}^2  (t)}{4 D} 
 - \dot {\vec x } (t) . \frac{ \vec F( \vec x (t)) }{2D} 
+\frac{ \vec F ^2( \vec x (t) ) }{ 4D  }   + \frac{  [\vec \nabla . \vec F(\vec x (t) ) ]  }{2 }   
 \nonumber \\
 && \equiv 
 \frac{ \dot {\vec x}^2  (t) }{4 D}   
 - \dot {\vec x } (t) . \vec A( \vec x(t))
 + V(\vec x(t))
\label{lagrangian}
\end{eqnarray}
The last rewriting as an Euclidean classical electromagnetic Lagrangian
involves as it should the vector potential $\vec A ( \vec x) $ of Eq. \ref{vectorpot}
and the scalar potential $V ( \vec x) $ of Eq. \ref{scalarpot}.


\subsection{ Statistics of trajectory observables ${\cal O} [\vec x(0 \leq t \leq T) ] $ via the Feynman-Kac formula}
 
An additive observable of the trajectory ${\cal O} [\vec x(0 \leq t \leq T) ]  $ can be parametrized by 
some scalar field $V^{[{\cal O}]}(\vec x)$ 
and some vector field $\vec A^{[{\cal O}]}(\vec x)$
in the Stratonovich interpretation.
\begin{eqnarray}
{\cal O} [\vec x(0 \leq t \leq T) ]  \equiv \int_0^T dt \left[   - V^{[{\cal O}]} (\vec x(t) ) 
+  \vec A^{[{\cal O}]}( \vec x(t)) . \dot {\vec x}(t)  \right]
\label{additive}
\end{eqnarray}
The standard method to study its statistics properties
is the famous Feynman-Kac formula \cite{feynman,kac,c_these,review_maj}
involving appropriate deformations of the Fokker-Planck generator.
Note that while the Feynman-Kac formula is often 
described only for the scalar field $V^{[{\cal O}]} (\vec x ) $, 
its application to the vector field case $\vec A^{[{\cal O}]}( \vec x) $
is essential to take into account topological constraints in the context of polymer physics \cite{edw67,wiegel}
and to analyze the winding properties of Brownian paths \cite{orsay_winding1,orsay_winding2,orsay_winding3,orsay_winding4,c_these,winding1,winding2}.
In the following, it is thus useful to recall the Feynman-Kac formula for the general case of Eq. \ref{additive}
where both $V^{[{\cal O}]} (\vec x ) $ and $\vec A^{[{\cal O}]}( \vec x) $ are present
and have a simple interpretation as deformed scalar and vector potentials
in the langage of non-hermitian quantum electromagnetic Hamiltonians \cite{us_gyrator}.


\subsubsection{ Generating functions of trajectory observables via path-integrals with deformed scalar and vector potentials}

Using the propagator of Eq. \ref{pathintegral}, one obtains that the joint distribution
${\cal P}_T({\cal O} ; \vec x \vert \vec x_0 ) $ to be at position $\vec x(T)= \vec x$ and to see the value $ {\cal O} [\vec x(0 \leq t \leq T) ]= {\cal O}$ when starting at position $\vec x_0$ at time 0 
corresponds to the constrained path-integral
\begin{eqnarray}
{\cal P}_T({\cal O} ; \vec x \vert \vec x_0 )
= \int_{\vec x(t=0)=\vec x_0}^{\vec x(t=T)=\vec x} {\cal D}   \vec x(.)  
 e^{ - \displaystyle 
  \int_0^T dt {\cal L} (\vec x(t), \dot {\vec x}(t) )
 } \delta \left( \int_0^T dt \left[   - V^{[{\cal O}]} (\vec x(t) ) 
+  \vec A^{[{\cal O}]}( \vec x(t)) . \dot {\vec x}(t)  \right] - {\cal O}\right)
\label{pathintegralPO}
\end{eqnarray}
where the delta function selects the trajectories having the correct value ${\cal O} $ for the additive observable.

It is thus useful to consider the generating function $Z^{[k]}_T(\vec x \vert \vec x_0)$
of the observable ${\cal O}[\vec x(0 \leq t \leq T) ] $ of Eq. \ref{additive}
over the stochastic trajectories $ \vec x(0 \leq t \leq T) $ starting at $\vec x(0)=\vec x_0$ and ending at $\vec x(T)=\vec x$,
where we will use the parameter $\frac{k}{D}$ as in \cite{bertin-conditioned} that will be more convenient
to analyze later the limit of small diffusion coefficient $D \to 0$
(instead of the standard choice $k$ when the diffusion coefficient $D$ remains finite)
\begin{eqnarray}
 Z^{[k]}_T(\vec x \vert \vec x_0) 
&& \equiv  \int d {\cal O}  \   {\cal P}_T({\cal O} ; \vec x \vert \vec x_0 ) \ e^{ \displaystyle \frac{k}{D}{\cal O}}
 \nonumber \\
 &&
= \int_{\vec x(t=0)=\vec x_0}^{\vec x(t=T)=\vec x} {\cal D}   \vec x(.)  
 e^{ - \displaystyle 
  \int_0^T dt {\cal L} (\vec x(t), \dot {\vec x}(t) )
+ \frac{k}{D} 
\int_{0}^T d t \left[   - V^{[{\cal O}]} (\vec x(t) )
+  \dot {\vec x} (t) . \vec A^{[{\cal O}]}( \vec x(t)) \right]
 }
  \nonumber \\
 &&
 \equiv
 \int_{\vec x(t=0)=\vec x_0}^{\vec x(t=T)=\vec x} {\cal D}   \vec x(.)  
 e^{ - \displaystyle \int_{0}^T d t
{\cal L}_k (\vec x(t), \dot {\vec x}(t) )
 }
 \label{gene}
\end{eqnarray}
where the $k$-deformed classical Lagrangian ${\cal L}_k (\vec x(t), \dot {\vec x}(t) )  $
with respect to the Lagrangian ${\cal L} (\vec x(t), \dot {\vec x}(t) ) $ of Eq. \ref{lagrangian}
\begin{eqnarray}
{\cal L}_k (\vec x(t), \dot {\vec x}(t) ) 
&& \equiv  {\cal L} (\vec x(t), \dot {\vec x}(t) ) + \frac{k}{D} V^{[{\cal O}]}( \vec x(t) - \frac{k}{D}  \dot {\vec x} (t) . \vec A^{[{\cal O}]}( \vec x(t))
\nonumber \\
&& = \frac{\left( \dot {\vec x } (t) - \vec F( \vec x(t)) \right)^2}{4 D} 
+ \frac{k}{D} V^{[{\cal O}]}( \vec x(t) - \frac{k}{D}  \dot {\vec x} (t) . \vec A^{[{\cal O}]}( \vec x(t))
 + \frac{  [\vec \nabla . \vec F(\vec x (t) ) ]  }{2 } 
 \nonumber \\
&& = \frac{\dot {\vec x}^2  (t) }{4D}  
 - \dot {\vec x } (t) . \vec A^{[k]}( \vec x(t))
 + V^{[k]}(\vec x(t))
\label{lagrangiank}
\end{eqnarray}
involves the
$k$-deformed vector potential  $\vec A^{[k]} (\vec x ) $ with respect to the initial vector potential  $ A( \vec x) $ of Eq. \ref{vectorpot}
\begin{eqnarray}
 \vec A^{[k]} (\vec x ) \equiv \vec A( \vec x) +\frac{k}{D} \vec A^{[{\cal O}]}( \vec x)
 =  \frac{ \vec F( \vec x) }{2D} +\frac{k}{D} \vec A^{[{\cal O}]}( \vec x)
 \label{vectorpotp}
\end{eqnarray}
and the $k$-deformed scalar potential $V^{[k]}(\vec x) $
with respect to the initial scalar potential $V ( \vec x)  $ of Eq. \ref{scalarpot}
\begin{eqnarray}
V^{[k]}(\vec x) \equiv V ( \vec x) + \frac{k}{D} V^{[{\cal O}]} ( \vec x)
=  \frac{ \vec F ^2( \vec x) }{ 4D  }   + \frac{  [\vec \nabla . \vec F(\vec x) ]  }{2 }   + \frac{k}{D} V^{[{\cal O}]} ( \vec x)
\label{scalarpotp}
\end{eqnarray}


\subsubsection{ Quantum Hamiltonian with deformed scalar and vector potentials}

The path-integral representation of Eq. \ref{gene} means that
 the generating function $Z^{[k]}_T(\vec x \vert \vec x_0) $ satisfies the Euclidean Schr\"odinger equation analogous to Eq. \ref{fokkerplanck}
\begin{eqnarray}
 \partial_T Z_T^{[k]}(\vec x \vert \vec x_0)  
 =  -  H_k Z^{[k]}_T(\vec x \vert \vec x_0)
\label{EuclideanZ}
\end{eqnarray}
where the $k$-deformed quantum Hamiltonian 
\begin{eqnarray}
 H_k   && =     -  D  \left( \vec \nabla -   \vec A^{[k]}(\vec x) \right)^2    +  V^{[k]}(\vec x)
 \label{FPhamiltonianp}
\end{eqnarray}
involves the deformed scalar and vector potentials 
of Eqs \ref{vectorpotp} and \ref{scalarpotp}. The expansion and refactorization 
of this Hamiltonian are useful
to see more clearly the $k$-deformation with respect to the initial Hamiltonian $ H= H_{[k=0]} $
of Eqs \ref{FPhamiltonian} involving the force $\vec F(\vec x)$
\begin{eqnarray}
 H_k   &&  =   - D \left( \vec \nabla  -\frac{k}{D} \vec A^{[{\cal O}]}( \vec x) -  \frac{ \vec F( \vec x) }{2D} \right)^2   
  +  \left[ \frac{ \vec F ^2( \vec x) }{ 4D  }   + \frac{  [\vec \nabla . \vec F(\vec x) ]  }{2 }   + \frac{k}{D} V^{[{\cal O}]} ( \vec x) \right] 
      \nonumber \\
 && = - \frac{1}{D}   \left( - D \vec \nabla  + k \vec A^{[{\cal O}]}( \vec x) \right)  
 \left[ - D  \vec \nabla  +k \vec A^{[{\cal O}]}( \vec x)  + \vec F( \vec x) \right]
  + \frac{k}{D} V^{[{\cal O}]} ( \vec x) 
 \label{FPhamiltonianpexpanded}
\end{eqnarray}
with the corresponding adjoint operator
\begin{eqnarray}
 H_k^{\dagger}  &&  =   - D \left( \vec \nabla  +\frac{k}{D} \vec A^{[{\cal O}]}( \vec x) +  \frac{ \vec F( \vec x) }{2D} \right)^2   
  +  \left[ \frac{ \vec F ^2( \vec x) }{ 4D  }   + \frac{  [\vec \nabla . \vec F(\vec x) ]  }{2 }   + \frac{k}{D} V^{[{\cal O}]} ( \vec x) \right]
    \nonumber \\
 && = -  \frac{1}{D} \left[  D   \vec \nabla  + k  \vec A^{[{\cal O}]}( \vec x)  + \vec F( \vec x) \right]
   \left( D \vec \nabla  + k \vec A^{[{\cal O}]}( \vec x) \right)  
  + \frac{k}{D} V^{[{\cal O}]} ( \vec x) 
 \label{FPhamiltonianpexpandedadjoint}
\end{eqnarray}


\subsection{ Simplifications that are expected for large time $T \to + \infty$ and/or small diffusion coefficient $D \to 0$} 

In summary, if one is interested into properties for finite time $T$ and finite diffusion coefficient $D$,
one needs to solve a full time-dependent quantum problem
governed by the Hamiltonian $H$ of Eq. \ref{FPhamiltonian}
for the propagator $P_T(\vec x \vert \vec x_0)$ 
and by the $k$-deformed Hamiltonians $H_k$ of Eq. \ref{FPhamiltonianpexpanded}
for the generating functions $Z^{[k]}(\vec x \vert \vec x_0) $ of trajectory observables.
In the four following sections, we will focus on the simplifications for large time $T \to + \infty$ and/or small diffusion coefficient $D \to 0$ as follows :

(i) In the limit of large time $T \to + \infty$ with finite diffusion coefficient $D$ described in section \ref{seC_largeT},
one only needs to consider the ground-state properties of the quantum Hamiltonians $H$ and $H_k$
(instead of their whole spectra that would be necessary to reconstruct the properties for finite $T$).

(ii) In the limit of small diffusion coefficient $D \to 0$ with finite time $T$ described in section \ref{seC_smallD},
one only needs to consider the dominant classical trajectory and its action (instead of the full path-integral over trajectories that is necessary for finite $D$), as in the semi-classical WKB approximation of quantum mechanics.

(iii) In the double limit of large time $T \to + \infty$ and small diffusion coefficient $D \to 0$,
there are even more simplifications that can be analyzed via the two orders of limit
as described in sections \ref{seC_largeTsmallD} and \ref{seC_smallDlargeT} respectively.


\section{ Large deviations for large time $T\to + \infty$ and finite diffusion coefficient $D$ }

\label{seC_largeT}

In this section, we focus on the simplifications for large time $T \to + \infty$
with respect to the properties described in the previous section.


\subsection{ Convergence of the Fokker-Planck dynamics 
 towards the steady state $P_*(\vec x) $ with its steady current $\vec J_*(\vec x) $  } 
 
We assume that the Fokker-Planck dynamics of Eq. \ref{fokkerplanckeuclidean}
converges towards some normalizable steady state $P_*(\vec x)$ satisfying
\begin{eqnarray}
0= \partial_t P_*(\vec x)     = - H P_*(\vec x)
\label{fokkerplanckeuclideanstar}
\end{eqnarray}
i.e. the steady state $P_*(\vec x) $ corresponds to the positive right eigenvector of the Hamiltonian $H$ of Eq. \ref{FPhamiltonian}
associated to zero energy.
The corresponding steady current $\vec J_*(\vec x) $ 
 satisfying the steady version of Eq. \ref{fokkerplanckj}
\begin{eqnarray}
\vec J_*(\vec x) \equiv P_*(\vec x)  \vec F(\vec x ) -D \vec \nabla   P_*(\vec x)  
\label{jsteady}
\end{eqnarray}
 should be divergenceless in order to to satisfy the steady version of the Fokker-Planck dynamics of Eq. \ref{fokkerplanck}
\begin{eqnarray}
0 = \partial_t P_*(\vec x)    =  -   \vec \nabla . \vec J_*(\vec x)  
\label{fokkerplanckst}
\end{eqnarray}


 \subsubsection{ Decompositions of the force $\vec F(\vec x)= \vec F^{rev} (\vec x)+\vec F^{irr} (\vec x)$ 
 into its reversible and irreversible contributions}

In the field of nonequilibrium, it is standard 
to decompose the force $\vec F(\vec x) = \vec F^{rev} (\vec x)+\vec F^{irr} (\vec x)$ 
into its reversible and irreversible contributions.
The reversible contribution $\vec F^{rev} ( \vec x) $ is the force 
that would produce a vanishing steady current in Eq. \ref{jsteady}
\begin{eqnarray}
\vec F^{rev} (\vec x ) \equiv    D \vec \nabla  \ln P_*(\vec x) 
\equiv - \vec \nabla U_*(\vec x)
\label{forceRev}
\end{eqnarray}
where we have introduced the steady potential $U_*(\vec x) $ parametrizing  
the steady state $P_*(\vec x) $ via 
\begin{eqnarray}
P_*(\vec x)  = \frac{ e^{ - \frac{ U_*(\vec x) }{  D }  } }{ \int d^d \vec y \ e^{ - \frac{ U_*(\vec y) }{  D }  }} 
\label{rhostarfromurev}
\end{eqnarray}

The remaining irreversible contribution of the force
\begin{eqnarray}
\vec F^{irr} (\vec x) \equiv \vec F(\vec x)  - \vec F^{rev} (\vec x)
\label{forceIrrev}
\end{eqnarray}
 is then directly responsible for the steady current of Eq. \ref{jsteady} via
\begin{eqnarray}
  \vec J_*(\vec x)  =  P_*(\vec x ) \vec F^{irr}( \vec x)  
 \label{jsteadyirrev}
\end{eqnarray}
The vanishing divergence for the steady current $\vec J_*(\vec x)$ of Eq. \ref{fokkerplanckst}
\begin{eqnarray}
0 =  \vec \nabla .  \vec J_*(\vec x) = \vec \nabla .   \left( P_*(\vec x ) \vec F^{irr}( \vec x)    \right)
= \vec F^{irr}( \vec x) . \vec \nabla P_*(\vec x ) +P_*(\vec x )    \left( \vec \nabla . \vec F^{irr}( \vec x)  \right)
 \label{divjsteadyirrev}
\end{eqnarray}
yields that the divergence of the irreversible force $\vec F^{irr}( \vec x) $ 
is directly related to the scalar product of the reversible and irreversible forces
\begin{eqnarray}
 \vec \nabla .  \vec F^{irr}( \vec x)  
  = -  \vec F^{irr}( \vec x) .   \vec \nabla \ln   P_*(\vec x )
  = -  \frac{ \vec F^{irr}( \vec x) . \vec F^{rev} (\vec x ) }{D}
  = \frac{ \vec F^{irr}( \vec x) . \vec \nabla U_*(\vec x) }{D}
 \label{divforceirrev}
\end{eqnarray}

In summary, the steady state $P_*(\vec x) $ of Eq. \ref{fokkerplanckeuclideanstar}
can be computed if one is able
to decompose the given force $\vec F (\vec x)$ into 
\begin{eqnarray}
  \vec F(\vec x)  = - \vec \nabla U_*(\vec x)  + \vec F^{irr} (\vec x)
\label{forcegradRevIrrev}
\end{eqnarray}
satisfying Eq. \ref{divforceirrev}
\begin{eqnarray}
0  && = D \vec \nabla .  \vec F^{irr}( \vec x)   -   \vec F^{irr}( \vec x) . \vec \nabla U_*(\vec x) 
\nonumber \\
&& =  D  \left( \vec \nabla . \vec F (\vec x) + \Delta U_*(\vec x) \right) 
  -  \left(\vec F (\vec x) + \vec \nabla U_*(\vec x) \right) . \vec \nabla U_*(\vec x) 
 \label{divforceirrevgrad}
\end{eqnarray}
For an arbitrary force $ \vec F (\vec x)$ in dimension $d$, this equation for the steady potential $U_*(\vec x) $ 
that parametrizes the steady state $P_*(\vec x) $ via Eq. \ref{rhostarfromurev}
is not easy to solve.
 As discussed in detail in \cite{Aurell},
it is interesting to compare the decomposition of Eq. \ref{forcegradRevIrrev}
with two other simpler decompositions of the force $\vec F (\vec x)  $.


  \subsubsection{ Comparison with the Helmholtz decomposition of the force $\vec F (\vec x)  $
  into a gradient and a divergence-free contribution }
 
 The Helmholtz decomposition of the force $\vec F (\vec x)  $
into a gradient and a divergence-free contribution 
\begin{eqnarray}
\vec F (\vec x) 
= - \vec \nabla U^H(\vec x)+ \vec F^H (\vec x) \ \ {\rm with \ \ zero \ divergence } \ \ \vec \nabla . \vec F^H (\vec x)  =0
\label{forcetothelmholtz}
\end{eqnarray}
requires to solve the Poisson equation involving the Laplacian of the Helmholtz potential $U^H(\vec x)$ 
\begin{eqnarray}
0  = \vec \nabla . \left( \vec F (\vec x) + \vec \nabla U^H(\vec x)\right)
= \vec \nabla .  \vec F (\vec x) + \Delta U^H(\vec x)
\label{poisson}
\end{eqnarray}
which is much simpler than Eq. \ref{divforceirrevgrad}.

Note that in dimension $d=3$, the divergence-free force $\vec F^{H} (\vec x) $  of the Helmholtz decomposition of
 Eq. \ref{forcetothelmholtz} is usually written as the curl of some field $\vec \Omega^H(\vec x)$
 \begin{eqnarray}
\vec F (\vec x) 
= - \vec \nabla U^H(\vec x)+ \vec \nabla \times \vec \Omega^H( \vec x)
\label{forcetothelmholtz3d}
\end{eqnarray}
Then the curl of the force $\vec F (\vec x)  $
\begin{eqnarray}
\vec \nabla \times \vec F(\vec x) && =\vec \nabla \times \left( \vec \nabla \times \vec \Omega^H( \vec x)\right)
 = \vec \nabla  \left( \vec \nabla. \vec \Omega^H( \vec x)\right) - \Delta  \vec \Omega^H( \vec x)
\nonumber \\
&& = - \Delta  \vec \Omega^H( \vec x) \ \ \ \text{ with the gauge choice } \ \  \vec \nabla. \vec \Omega^H( \vec x) =0
\label{forcetot3DrotR}
\end{eqnarray}
reduces to the Laplacian of $\Omega^H(\vec x) $ for the standard gauge choice of zero-divergence $\vec \nabla. \vec \Omega^H( \vec x) =0 $.

 
 \subsubsection{ Comparison with the orthogonal decomposition of the force $\vec F (\vec x)  $ into a gradient and a perpendicular contribution }

The orthogonal decomposition of the force $\vec F (\vec x)  $ into a gradient and a perpendicular contribution
\begin{eqnarray}
\vec F (\vec x) 
= - \vec \nabla U^{\perp}(\vec x)+  \vec F^{\perp}( \vec x) 
\ \ \ {\rm with \ orthogonality} \ \  \vec F^{\perp}( \vec x) . \vec \nabla U^{\perp}(\vec x)=0
\label{forceNormalDecomposition}
\end{eqnarray}
requires to solve the non-linear equation for the gradient of $U^{\perp}(\vec x) $
\begin{eqnarray}
0 =  \left( \vec F (\vec x) + \vec \nabla U^{\perp}(\vec x)\right) . \vec \nabla U^{\perp}(\vec x)
\label{forceNormalDecompositionEq}
\end{eqnarray}
which is simpler than Eq. \ref{divforceirrevgrad} and emerges as the limit of Eq. \ref{divforceirrevgrad}
for small diffusion coefficient $D \to 0$.

 Note that in dimension $d=3$, the perpendicular force $ \vec F^{\perp}( \vec x)$ 
 can be written as the cross product of the gradient with some field $\vec \Omega^{\perp}(\vec x)$
\begin{eqnarray}
\vec F (\vec x) 
= - \vec \nabla U^{\perp}(\vec x) - \left[ \vec \nabla U^{\perp}(\vec x)\right]    \times \vec \Omega^{\perp}(\vec x)
\label{forceperp3D}
\end{eqnarray}

 
 \subsubsection{ Comparison of the three decompositions }

The three decompositions of 
Eqs \ref{forcegradRevIrrev} \ref{forcetothelmholtz} \ref{forceNormalDecomposition} 
lead to the following expressions for the divergence of the force $ \vec F (\vec x) $ 
\begin{eqnarray}
\vec \nabla . \vec F(\vec x) 
&&= - \Delta U_*(\vec x)  + \vec \nabla . \vec F^{irr} (\vec x)
\nonumber \\
&& = - \Delta U^H(\vec x)
\nonumber \\
&& = - \Delta U^{\perp}(\vec x)  + \vec \nabla . \vec F^{\perp} (\vec x)
\label{forcediv3}
\end{eqnarray}
while the curl antisymmetric matrix of the force $ \vec F (\vec x) $
\begin{eqnarray}
  \partial_{\mu} F_{\nu} (\vec x ) -  \partial_{\nu} F_{\mu} (\vec x ) 
 && = \partial_{\mu} F^{irr}_{\nu} (\vec x ) -  \partial_{\nu} F^{irr}_{\mu} (\vec x ) 
  \nonumber \\
 && = \partial_{\mu} F^H_{\nu} (\vec x ) -  \partial_{\nu} F^H_{\mu} (\vec x ) 
 \nonumber \\
 && = \partial_{\mu} F_{\nu}^{\perp} (\vec x ) -  \partial_{\nu} F^{\perp}_{\mu} (\vec x ) 
  \nonumber \\
 \label{curlmatrix}
\end{eqnarray}
coincides with the curl antisymmetric matrices of the three complementary forces 
$\vec F^{irr} (\vec x) $, $\vec F^H (\vec x)$ and $\vec F^{\perp} (\vec x)$.

In practice, when one wishes to analyze a new given non-equilibrium diffusion with force $\vec F(\vec x)$,
it is useful to compute its divergence $\vec \nabla .\vec F(\vec x) $
to see if one is able to solve the Poisson Eq. \ref{poisson}
in order to obtain the Helmholtz potential $U^H(\vec x)$.
If $U^H(\vec x)$ is explicit, it is then easy to compute the scalar product 
\begin{eqnarray}
\vec F^H (\vec x) . \vec \nabla U^H(\vec x) 
= \left(\vec F (\vec x) + \vec \nabla U^H(\vec x) \right). \vec \nabla U^H(\vec x)
\label{scalarforcetothelmholtz}
\end{eqnarray}
with the following discussion :

(i) in general, the scalar product of Eq. \ref{scalarforcetothelmholtz} does not vanish,
and the three decompositions are different 
\begin{eqnarray}
\text{ General case } :  \ \ \vec F^H (\vec x) . \vec \nabla U^H(\vec x) \ne 0
\ \ \text{then the three decompositions are different } 
\label{3different}
\end{eqnarray}
Then the reversible-irreversible decomposition can be difficult to obtain,
even when the Helmholtz and the orthogonal decompositions have very simple explicit expressions
(see the first two-dimensional example with quadratic forces studied in \cite{Aurell}, 
where the Helmholtz potential $U^H(\vec x) $
and the orthogonal potential $U^{\perp}(\vec x) $ are cubic and quadratic potentials, 
while the steady potential $U_*(\vec x) $
could be studied only numerically).

(ii) for the particular forces $\vec F(\vec x)$ where the scalar product of Eq. \ref{scalarforcetothelmholtz} vanishes,
i.e. when the Helmholtz decomposition happens to be orthogonal,
then both the orthogonal decomposition
and the reversible-irreversible decomposition actually coincide with the Helmholtz decomposition,
i.e. the three decompositions coincide
\begin{eqnarray}
\text{ Special case } :  \ \ \vec F^H (\vec x) . \vec \nabla U^H(\vec x) =0
\ \ \text{then the three decompositions coincide } \ \ U_*(\vec x)=U^H(\vec x) = U^{\perp}(\vec x) 
\label{3coincide}
\end{eqnarray}
An example will be described in section \ref{sec_2DPolar}.


\subsection{ Generating function $Z^{[k]}_T(\vec x \vert \vec x_0)$ for large time $T$ : lowest energy $E(k)$ of the quantum Hamiltonian $H_k $ } 

For large time $T \to +\infty$, the generating function $Z^{[k]}_T(\vec x \vert \vec x_0)$ of Eq. \ref{gene}
rewritten as an Euclidean Schr\"odinger propagator associated to the k-deformed Hamiltonian $H_k $
of Eqs \ref{FPhamiltonianp} \ref{FPhamiltonianpexpanded}
will display the asymptotic behavior
\begin{eqnarray}
Z^{[k]}_T(\vec x \vert \vec x_0) = \langle \vec x \vert e^{- TH_k } \vert \vec x_0  \rangle
\opsimeq_{T \to +\infty}
 e^{- T E(k) } r_k(\vec x) l_k(\vec x_0)
\label{genelargeT}
\end{eqnarray}
where $E(k)$ is the ground-state energy of the Hamiltonian $H_k $ of Eq. \ref{FPhamiltonianpexpanded}.
The corresponding positive right eigenvector $r_k(\vec x) >0$  satisfying
\begin{eqnarray}
 E(k)  r_k( \vec x) && = H_k  r_k( \vec x) 
 \nonumber \\
 && = \frac{1}{D}  \bigg[  
  -   \left( - D \vec \nabla  + k \vec A^{[{\cal O}]}( \vec x) \right)  .
 \left[ - D  \vec \nabla  +k \vec A^{[{\cal O}]}( \vec x)  + \vec F( \vec x) \right]
  + k V^{[{\cal O}]} ( \vec x) 
     \bigg] r_k( \vec x) 
\label{eigenright}
\end{eqnarray}
 is the deformation of the steady state $r_{k=0}(\vec x) =P_*(\vec x)$
 satisfying Eq. \ref{fokkerplanckeuclideanstar}.
The corresponding positive left eigenvector $l_k(\vec x)>0$ satisfying
\begin{eqnarray}
E(k)  l_k( \vec x) && = H_k^{\dagger}  l_k( \vec x)
\nonumber \\
&& = \frac{1}{D}  \bigg(
-  \left[  D   \vec \nabla  + k  \vec A^{[{\cal O}]}( \vec x)  + \vec F( \vec x) \right] .
   \left( D \vec \nabla  + k \vec A^{[{\cal O}]}( \vec x) \right)  
  + kV^{[{\cal O}]} ( \vec x) 
\bigg) l_k( \vec x) 
\label{eigenleft}
\end{eqnarray}
is the deformation of the trivial left eigenvector $l_{k=0}(\vec x) =1$
associated to the conservation of probability of the Fokker-Planck dynamics.
The large-time behavior of Eq. \ref{genelargeT} requires the normalization
\begin{eqnarray}
\int d^d \vec x \   l_k( \vec x)  r_k( \vec x) =1
  \label{eigennorma}
\end{eqnarray}

For later purposes, it is useful to rewrite the eigenvalue Eq. \ref{eigenright} 
in terms of the logarithm of $ r_k( \vec x)$ 
\begin{eqnarray}
D E(k)   =
  -   \left( - D \vec \nabla  - D  \vec \nabla \ln r_k( \vec x) + k \vec A^{[{\cal O}]}( \vec x) \right)    .
 \left[ - D  \vec \nabla \ln r_k( \vec x) +k \vec A^{[{\cal O}]}( \vec x)  + \vec F( \vec x) \right]  
  + k V^{[{\cal O}]} ( \vec x)  
\label{eigenrightlog}
\end{eqnarray}
and to rewrite the eigenvalue Eq. \ref{eigenleft} in terms of the logarithm of $ l_k( \vec x)$
 \begin{eqnarray}
D E(k)  &&  =
-  \left[  D   \vec \nabla + D \vec \nabla \ln  l_k( \vec x) + k  \vec A^{[{\cal O}]}( \vec x)  + \vec F( \vec x) \right]   .
   \left( D \vec \nabla \ln  l_k( \vec x) + k \vec A^{[{\cal O}]}( \vec x) \right)  
  + kV^{[{\cal O}]} ( \vec x)  
\label{eigenleftlog}
\end{eqnarray}
 
 
 \subsection{ Canonical conditioned process : 
 conditioned steady state $P^{C[k]}_* (\vec x) $ and conditioned force $\vec  F^{C[k]}(\vec x) $ } 
 
 \label{subsec_canocond}
 
  \subsubsection{ Conditioned steady state $P^{C[k]}_* (\vec x) $ and conditioned force $\vec  F^{C[k]}(\vec x) $
  in terms of the eigenvectors $l_k( \vec x) $ and $r_k( \vec x) $ }

Since the generating function $Z^{[k]}_T(\vec x \vert \vec x_0)$ of Eq. \ref{genelargeT}
will decay exponentially in time if the ground-state energy is positive $ E(k)>0$
or grow exponentially in time if the ground-state energy is negative $ E(k)<0$,
it is useful to construct the canonical conditioned propagator
\begin{eqnarray}
P_T^{C[k]}(x \vert x_0)  \equiv e^{ T E(k) } \frac{ l_k( \vec x)}{ l_k( \vec x_0)} Z^{[k]}_T(\vec x \vert \vec x_0) 
  \opsimeq_{T \to + \infty} 
l_k( \vec x) r_k( \vec x) \equiv
P^{C[k]}_* (\vec x)
\label{conditionedpropagator}
\end{eqnarray}
that converges, independently of the initial condition $\vec x_0$,
 towards the conditioned steady state $P^{C[k]}_* (\vec x) $
given by the product of the left eigenvector $l_k( \vec x)$ of Eq.  \ref{eigenleft}
and the right eigenvector $r_k( \vec x)$ of Eq \ref{eigenright}, so that it is normalized as a consequence of Eq. \ref{eigennorma}
\begin{eqnarray}
1=   \int d^d \vec x \ P^{C[k]}_* (\vec x)
  \label{normarhocond}
\end{eqnarray}

The Fokker-Planck dynamics satisfied by the conditioned propagator
\begin{eqnarray}
 \partial_T P_T^{C[k]}(x \vert x_0)   
   = \vec \nabla . \left(   D  \vec \nabla  -   \vec F^{C[k]}(\vec x)\right) P_T^{C[k]}(x \vert x_0)
\label{fokkerplanckCond}
\end{eqnarray}
involves the same diffusion coefficient $D$ as the initial Fokker-Planck dynamics,
while the initial force $\vec F(\vec x)$ has been replaced by the conditioned force
\begin{eqnarray}
\vec  F^{C[k]}(\vec x)
  = \vec F(\vec x)+ 2 k \vec A^{[{\cal O}]}(\vec x) +2 D   \vec \nabla  \ln l_k ( \vec x)  
\label{forceDoobDiff}
\end{eqnarray}
that involves the gradient of the logarithm of the left eigenvector $l_k( \vec x)$ of Eq.  \ref{eigenleft}.


  \subsubsection{ Properties of the conditioned force $\vec  F^{C[k]}(\vec x) $ }
  
The curl antisymmetric matrix of the conditioned force $ \vec F^{C[k]} (\vec x) $ Eq. \ref{forceDoobDiff} 
\begin{eqnarray}
  \partial_{\mu} F^{C[k]}_{\nu} (\vec x ) -  \partial_{\nu} F^{C[k]}_{\mu} (\vec x ) 
 = \partial_{\mu} F_{\nu} (\vec x ) -  \partial_{\nu} F_{\mu} (\vec x ) 
  + 2k \left( \partial_{\mu} A^{[{\cal O}]}_{\nu} (\vec x ) -  \partial_{\nu} A^{[{\cal O}]}_{\mu} (\vec x ) \right)
 \label{curlmatrixcond}
\end{eqnarray}
is completely fixed by the curl antisymmetric matrix of the initial force $\vec F(\vec x)$
and by the curl antisymmetric matrix of $ \vec A^{[{\cal O}]} (\vec x )$ that parametrizes the observable of Eq. \ref{additive}
under study.

Using Eq. \ref{forceDoobDiff}
to replace
\begin{eqnarray}
D   \vec \nabla  \ln l_k ( \vec x) +k \vec A^{[{\cal O}]}(\vec x)    =\frac{ \vec F^{C[k]}(\vec x) - \vec F(\vec x)}{ 2 }  
\label{forceDoobDiffplug}
\end{eqnarray}
into Eq. \ref{eigenleftlog} leads to following equation for the conditioned force $F^{C[k]}(\vec x) $ of Eq. \ref{forceDoobDiff}
 \begin{eqnarray}
D E(k)  &&  =
-  \left[  D   \vec \nabla + \frac{ \vec F^{C[k]}(\vec x) + \vec F(\vec x)}{ 2 } \right]   .
   \left( \frac{ \vec F^{C[k]}(\vec x) - \vec F(\vec x)}{ 2 } \right)  
  + kV^{[{\cal O}]} ( \vec x)  
  \nonumber \\
  && = -   D   \vec \nabla    .  \left( \frac{ \vec F^{C[k]}(\vec x) - \vec F(\vec x)}{ 2 } \right)  
 -  \left[   \frac{ [\vec F^{C[k]}(\vec x)]^2 - \vec F^2(\vec x)}{ 4 } \right] 
    + kV^{[{\cal O}]} ( \vec x)  
\label{eigenleftlogcond}
\end{eqnarray}
The rewriting as
\begin{eqnarray}
E(k) +  \frac{ [\vec F^{C[k]} ( \vec x) ]^2 }{ 4D  }   + \frac{  [\vec \nabla . \vec F^{C[k]}(\vec x) ]  }{2 }   
= \frac{ \vec F ^2( \vec x) }{ 4D  }   + \frac{  [\vec \nabla . \vec F(\vec x) ]  }{2 }   
+ \frac{k}{D} V^{[{\cal O}]}(\vec x)
\label{eigenleftforcecond}
\end{eqnarray}
leads to a simple interpretation in terms of the scalar potential $V(\vec x)$ of Eq. \ref{scalarpot}
associated to the initial force $\vec F(\vec x)$ and in terms of the conditioned scalar potential 
\begin{eqnarray}
   V^{C[k]}(\vec x) \equiv  \frac{ [\vec F^{C[k]} ( \vec x) ]^2 }{ 4D  }   + \frac{  [\vec \nabla . \vec F^{C[k]}(\vec x) ]  }{2 }
\label{scalarpotcond}
\end{eqnarray}
associated to the conditioned force $\vec F^{C[k]}(\vec x)$, since Eq. \ref{eigenleftforcecond}
becomes
\begin{eqnarray}
E(k) +   V^{C[k]}(\vec x)
= V(\vec x)+ \frac{k}{D} V^{[{\cal O}]}(\vec x) \equiv V^{[k]}(\vec x)
\label{eigenleftforcecondscalarpot}
\end{eqnarray}
So the conditioned scalar potential $V^{C[k]}(\vec x) $ differs from the $k$-deformed scalar potential of Eq. \ref{scalarpotp}
only via the constant $E(k)$.


  \subsubsection{ Decomposition of the conditioned force $\vec  F^{C[k]}(\vec x) $ into its reversible and irreversible components}

As for the initial model, it is interesting to decompose the conditioned force $\vec F^{C[k]} ( \vec x)$
into its reversible and irreversible components.
The reversible contribution $\vec F^{C[k]rev} (\vec x ) $ is determined by the conditioned steady state $P^{C[k]}_* (\vec x) $ as in Eq. \ref{forceRev}
\begin{eqnarray}
\vec F^{C[k]rev} (\vec x ) \equiv    D \vec \nabla  \ln P^{C[k]}_* (\vec x)
\equiv - \vec \nabla U^{C[k]}_*(\vec x) 
\label{forceRevcond}
\end{eqnarray}
where we have introduced the conditioned steady potential $U^{C[k]}_*(\vec x) $ parametrizing  
the conditioned steady state  
\begin{eqnarray}
P^{C[k]}_*(\vec x)  = \frac{ e^{ - \frac{ U^{C[k]}_*(\vec x) }{  D }  } }{ \int d^d \vec y \ e^{ - \frac{ U^{C[k]}_*(\vec y) }{  D }  }} 
\label{rhostarfromurevCond}
\end{eqnarray}

The conditioned irreversible contribution
\begin{eqnarray}
\vec F^{C[k]irr} (\vec x) \equiv \vec F^{C[k]}(\vec x)  - \vec F^{C[k]rev} (\vec x) = \vec F^{C[k]}(\vec x)+ \vec \nabla U^{C[k]}_*(\vec x)
\label{forceIrrevCond}
\end{eqnarray}
is then directly related  to the conditioned steady current 
\begin{eqnarray}
  \vec J^{C[k]}_*(\vec x) =P^{C[k]}_*(\vec x ) \vec F^{C[k]}( \vec x)  - D \vec \nabla P^{C[k]}_*(\vec x )
   =  P^{C[k]}_*(\vec x ) \vec F^{C[k]irr}( \vec x)  
 \label{jsteadyirrevcond}
\end{eqnarray}
As in Eq. \ref{divforceirrev}, the vanishing divergence for the conditioned steady current $\vec J_*^{C[k]}(\vec x)$
yields that the divergence of the irreversible conditioned force $\vec F^{C[k]irr}( \vec x) $
is related to its scalar product with the reversible conditioned contribution $\vec F^{C[k]rev} (\vec x ) $ 
\begin{eqnarray}
 \vec \nabla .  \vec F^{C[k]irr}( \vec x)  
  = -  \vec F^{C[k]irr}( \vec x) .   \vec \nabla \ln   P^{C[k]}_*(\vec x )
  = -  \frac{ \vec F^{C[k]irr}( \vec x) . \vec F^{C[k]rev} (\vec x ) }{D}
  = \frac{ \vec F^{C[k]irr}( \vec x) . \vec \nabla U^{C[k]}_*(\vec x) }{D}
 \label{divforceirrevcond}
\end{eqnarray}

 
 \subsection{  Rate function $ I \left( O = \frac{{\cal O}}{T} \right) $ 
 for the intensive trajectory observable $O = \frac{{\cal O} [\vec x(0 \leq t \leq T) ]}{T} $ 
 }

For large time $T \to + \infty$, the probability distribution of Eq \ref{pathintegralPO}
for the additive trajectory observable ${\cal O}$
is expected to display the large deviation form
 \begin{eqnarray}
   {\cal P}_T({\cal O}  ) \oppropto_{ T \to + \infty} e^{- T I \left( O = \frac{{\cal O}}{T} \right) }
    \label{ratefunctionCalO}
\end{eqnarray}
 where the positive rate function $I(O) \geq 0 $ for the intensive value $O = \frac{{\cal O}}{T} $
 vanishes and is minimum
 \begin{eqnarray}
  0 =  I ( O_*) = I'(O_*)
    \label{ratefunctionvanish}
\end{eqnarray}
 for the steady value $O_*$ that can be evaluated using Eq. \ref{additive}
in terms of the steady state $P_*(\vec x) $ 
 and the steady current $\vec J_*(\vec x)$ 
 \begin{eqnarray}
\frac{ {\cal O} [\vec x(0 \leq t \leq T) ] }{T} && \equiv \frac{1}{T} \int_0^T dt \left[   - V^{[{\cal O}]} (\vec x(t) ) 
+  \vec A^{[{\cal O}]}( \vec x(t)) . \frac{ d \vec x(t)}{dt} \right]
\nonumber \\
&& \opsimeq_{T \to +\infty} \int d^d \vec x \left[   - V^{[{\cal O}]} (\vec x ) P_*(\vec x) 
+  \vec A^{[{\cal O}]}( \vec x) . \vec J_*(\vec x)\right] \equiv O_*
\label{additiveSteady}
\end{eqnarray}
Using the expression of Eq. \ref{jsteady} for the steady current $\vec J_*(\vec x) $,
one obtains that the steady value $ O_*$ 
reads in terms of the steady state $P_*(\vec x)  $ alone 
 \begin{eqnarray}
O_* && = \int d^d \vec x \left[   - V^{[{\cal O}]} (\vec x ) P_*(\vec x) 
+  \vec A^{[{\cal O}]}( \vec x) . \left(  P_*(\vec x)  \vec F(\vec x ) -D \vec \nabla   P_*(\vec x)\right) \right]  
\nonumber \\
&& =  \int d^d \vec x P_*(\vec x)
\left[   - V^{[{\cal O}]} (\vec x )  
+  \vec A^{[{\cal O}]}( \vec x) . \left(    \vec F(\vec x ) -D \vec \nabla  \ln P_*(\vec x)\right) \right] 
\nonumber \\
&& =  \int d^d \vec x P_*(\vec x)
\left[   - V^{[{\cal O}]} (\vec x )  
+  \vec A^{[{\cal O}]}( \vec x) .  \vec F^{irr}(\vec x )  \right]
\label{additiveSteadyrho}
\end{eqnarray}

For large time, the evaluation of the generating function of Eq. \ref{gene} using the large deviation form of Eq. \ref{ratefunctionCalO}
 \begin{eqnarray}
 Z^{[k]}_T 
 \equiv  \int d {\cal O}  \   {\cal P}_T({\cal O}  ) \ e^{ \displaystyle \frac{k}{D}{\cal O}}
\opsimeq_{T \to + \infty}  \int d O \ e^{ \displaystyle T \left[ \frac{k}{D}  O -I(O) \right]} \opsimeq_{T \to +\infty}
 e^{  \displaystyle - T E(k) }
 \label{genesaddle}
\end{eqnarray}
 via the saddle-point method leads to the Legendre transformation
  \begin{eqnarray}
 I(O)- \frac{k}{D}  O  && = E(k)
 \nonumber \\
  I'(O)- \frac{k}{D}   && = 0
 \label{legendre}
\end{eqnarray}
with the reciprocal Legendre transformation
   \begin{eqnarray}
 I(O)  && = E(k) +  \frac{k}{D}  O
 \nonumber \\
 0   && = E'(k) +  \frac{1}{D}  O
 \label{legendrereci}
\end{eqnarray}

In particular, the steady value $O_*$ of Eq. \ref{additiveSteady} where the rate function vanishes $I(O_*)=0$
is associated to $k=0$ where the energy vanishes $E(k=0)=0$, so that Eq. \ref{legendrereci}
determines the first derivative 
   \begin{eqnarray}
  E'(k=0) = -  \frac{O_*}{D}  
 \label{legendrerecistar}
\end{eqnarray}
in agreement with the perturbation theory for $E(k)$ that we now recall.

 
 \subsection{First cumulants of trajectory observables via the series expansion
  of the ground state energy $E(k)$ in $k$}

  \subsubsection{ Reminder on the first cumulants of trajectory observables 
   for large time $T \to + \infty$}

 The cumulants $C_n(T)$ of the intensive trajectory observable $O [x(0 \leq t \leq T) ] $
 are defined by the series expansion of the logarithm of the generating function $Z^{[k]}_T(\vec x \vert \vec x_0) $ of Eq. \ref{gene}
 \begin{eqnarray}
\ln Z^{[k]}_T(\vec x \vert \vec x_0)= \ln \left[  \overline{ e^{\frac{k}{D} T O  } }  \right]
&& =  \sum_{n=1}^{+\infty}  \frac{ \left(\frac{k}{D} T \right)^n}{n!} C_n(T)
\nonumber \\
&& = \frac{k}{D} T C_1 (T)+ \frac{  k^2 T^2}{2 D^2} C_2 (T)+ O(k^3)
\label{defcumulants}
\end{eqnarray}
The large-time behavior of Eq. \ref{genelargeT}
yields that $E(k)$ corresponds to the series
\begin{eqnarray}
 E(k)  && = \lim_{T \to + \infty} \left[ - \frac{\ln Z^{[k]}_T(\vec x \vert \vec x_0)}{T} \right]
 =   -  \sum_{n=1}^{+\infty}  \frac{ \left(\frac{k}{D}  \right)^n}{n!}  \lim_{T \to + \infty} \left[ T^{n-1}  C_n(T) \right]
\nonumber \\
&& = - k  \lim_{T \to + \infty} \left[  \frac{C_1 (T) }{D} \right]
-  k^2 \lim_{T \to + \infty} \left[ \frac{   T C_2 (T) }{2 D^2} \right] 
+ O(k^3)
\label{E0cumulants}
\end{eqnarray}
So the series expansion in $k$ 
of the ground state energy $E(k)$ that involves coefficients $E^{[n]}$ that do not depend on $T$
\begin{eqnarray}
  E(k)  = 0 + k E^{[1]} + k^2 E^{[2]} +O(k^3)
 \label{FPhamiltoniankseries}
\end{eqnarray} 
is useful to obtain the first cumulants via
the identification with Eq. \ref{E0cumulants} as follows :

(1) the first cumulant $C_1(T)$, i.e. the average of the intensive trajectory observable $O [x(0 \leq t \leq T) ] $,
 converges for $T \to +\infty$ towards a finite limit determined by $E^{[1]} $
\begin{eqnarray}
   C_1 (T) \equiv \overline{ O [x(0 \leq t \leq T) ] } \opsimeq_{T \to + \infty}  - D E^{[1]}
\label{cumulant1}
\end{eqnarray}

(2) the second cumulant $C_2(T)$, i.e. the variance of the intensive trajectory observable $O [x(0 \leq t \leq T) ] $, scales as $\frac{1}{T}$ with an amplitude determined by $E^{[2]} $
\begin{eqnarray}
   C_2(T)  \equiv \overline{ \left( O [x(0 \leq t \leq T) ]  -  \overline{ O [x(0 \leq t \leq T) ] }\right)^2 }  
    \opsimeq_{T \to + \infty} - \frac{2 D^2 E^{[2]}}{T}
    \label{cumulant2}
\end{eqnarray}

As a consequence, to obtain this variance, 
it is useful to consider the perturbation theory up to second order in $k$ as described in the next subsection.

 
  \subsubsection{ Perturbation theory for the ground state energy $E(k)$ up to second order in $k$} 
 
For the non-hermitian Hamiltonian $H_k$ of Eq. \ref{FPhamiltonianpexpanded} 
containing contributions of order $k$ and $k^2$ around the Hamiltonian $H_{k=0}=H$ of Eq. \ref{FPhamiltonian}
\begin{eqnarray}
  H_k   && = H + k h_1 + k^2 h_2 
    \nonumber \\
   H  && = \vec \nabla . \left(  - D  \vec \nabla +   \vec F(\vec x)\right)
  \nonumber \\
    h_1  &&  
    =  2 \vec A^{[{\cal O}]}( \vec x) .  \vec \nabla  
    + [ \vec \nabla  .   \vec A^{[{\cal O}]}( \vec x)   ]
    - \frac{1}{D}    \vec A^{[{\cal O}]}( \vec x) .  \vec F( \vec x)  
  + \frac{1}{D} V^{[{\cal O}]} ( \vec x) 
    \nonumber \\
     h_2  &&  = - \frac{1}{D}   \left[ \vec A^{[{\cal O}]}( \vec x)   \right]^2
 \label{Hkseries}
\end{eqnarray}
with the adjoint operators
\begin{eqnarray}
  H_k^{\dagger}   && = H^{\dagger} + k h_1^{\dagger} + k^2 h_2^{\dagger} 
    \nonumber \\
    H^{\dagger}  && =  - \left( D  \vec \nabla +   \vec F(\vec x)\right). \vec \nabla
  \nonumber \\
    h_1^{\dagger}  &&  
    =  - 2 \vec A^{[{\cal O}]}( \vec x) .  \vec \nabla  
    - [ \vec \nabla  .   \vec A^{[{\cal O}]}( \vec x)   ]
    - \frac{1}{D}    \vec A^{[{\cal O}]}( \vec x) .  \vec F( \vec x)  
  + \frac{1}{D} V^{[{\cal O}]} ( \vec x) 
    \nonumber \\
     h_2^{\dagger}  &&  = - \frac{1}{D}   \left[ \vec A^{[{\cal O}]}( \vec x)   \right]^2
 \label{Hkseriesdagger}
\end{eqnarray}
the perturbation theory for the ground state energy $E(k)$ of Eq. \ref{FPhamiltoniankseries}
is very similar to the standard perturbation theory of quantum mechanics with
the following adaptation to take into account the right and left eigenvectors of Eqs \ref{eigenright} and \ref{eigenleft}
with their perturbative expansions 
\begin{eqnarray}
r_k(\vec x) && =  P_*(\vec x) + k r^{[1]}(\vec x) + k^2 r^{[2]}(\vec x)+O(k^3)
\nonumber \\
l_k(\vec x) && = 1 + k  l^{[1]}(\vec x) + k^2 l^{[2]}(\vec x)+O(k^3)
  \label{rlkseries}
\end{eqnarray}
with the orthogonality relations coming from
the normalization of Eq. \ref{eigennorma} at first-order in $k$
\begin{eqnarray}
0 && = \langle l_0 \vert r^{[1]} \rangle = \int d^d \vec x \     r^{[1]}( \vec x) 
\nonumber \\
0 && =  \langle l^{[1]} \vert r_0 \rangle = \int d^d \vec x \     l^{[1]}( \vec x)  P_*(\vec x)
  \label{eigennormaseries}
\end{eqnarray}

Plugging the various series expansions into the eigenvalue Eq. \ref{eigenright}
for the right eigenvector $r_k(\vec x)$
\begin{eqnarray}
 0 && = \left[ H_k - E(k) \right] r_k(\vec x) 
 = \left[ H + k (h_1 -E^{[1]})+ k^2 (h_2 -E^{[2]}) +O(k^3) \right] 
 \left[ P_*(\vec x) + k r^{[1]}(\vec x) + k^2 r^{[2]}(\vec x)+O(k^3)\right]
 \nonumber \\
 && =  k \bigg( H r^{[1]}(\vec x) +  (h_1 -E^{[1]}) P_*(\vec x) \bigg)
  + k^2 \bigg( H r^{[2]}(\vec x)
 +  (h_1 -E^{[1]}) r^{[1]}(\vec x) 
+ (h_2 -E^{[2]}) P_*(\vec x) \bigg)
 +O(k^3)
\label{eigenrightseries}
\end{eqnarray}
yields after the integration over $\vec x$ that corresponds to the projection on $\langle l_0 \vert$ that is useful to simplify many terms
\begin{eqnarray}
 0  && = \int d^d \vec x \left[ H_k - E(k) \right] r_k(\vec x) 
\nonumber \\
&&   =  k \bigg(   \int d^d \vec x h_1 P_*(\vec x)  - E^{[1]} \bigg)
  + k^2 \bigg(   \int d^d \vec x h_1  r^{[1]}(\vec x) 
+ \int d^d \vec x h_2  P_*(\vec x)  - E^{[2]} \bigg)
 +O(k^3)
\label{eigenrightseriesinteg}
\end{eqnarray}
So the two first coefficients of the perturbative expansion of Eq. \ref{FPhamiltoniankseries} for the energy read
\begin{eqnarray}
 E^{[1]} && =   \int d^d \vec x h_1 P_*(\vec x)  
 \nonumber \\
 E^{[2]} && =  \int d^d \vec x h_1  r^{[1]}(\vec x) 
+ \int d^d \vec x h_2  P_*(\vec x) 
\label{e1e2r}
\end{eqnarray}
where the first correction $r^{[1]}(\vec x) $ for the right eigenvector 
should be computed from Eq. \ref{eigenrightseries} at order $k$
\begin{eqnarray}
0  = H r^{[1]}(\vec x) +  (h_1 -E^{[1]}) P_*(\vec x)
 \label{eigenrightseriesr1}
\end{eqnarray}

Similarly, plugging the various series expansions
into the eigenvalue Eq. \ref{eigenleft}
for the left eigenvector $l_k(\vec x)$
\begin{eqnarray}
 0 && = \left[ H_k^{\dagger} - E(k) \right] l_k( \vec x) 
 = \left[ H^{\dagger} + k (h_1^{\dagger}-E^{[1]}) + k^2 (h_2^{\dagger} - E^{[2]}) +O(k^3) \right] 
 \left[1 + k  l^{[1]}(\vec x) + k^2 l^{[2]}(\vec x)+O(k^3) \right]
 \nonumber \\
 && =   k  \bigg( H^{\dagger} l^{[1]}(\vec x) + (h_1^{\dagger} 1 -E^{[1]}) \bigg)
 + k^2 \bigg( H^{\dagger} l^{[2]}(\vec x)
+   (h_1^{\dagger}-E^{[1]}) l^{[1]}(\vec x) 
  +  (h_2^{\dagger} 1 - E^{[2]}) \bigg)
 +O(k^3)
\label{eigenleftseries}
\end{eqnarray}
yields after the multiplication by $P_*(\vec x) $ and
the integration over $\vec x$ that corresponds to the projection on $\langle r_0 \vert$ that is useful to simplify many terms
\begin{eqnarray}
 0 && = \int d^d \vec x P_*(\vec x)\left[ H_k^{\dagger} - E(k) \right] l_k( \vec x) 
\nonumber \\
&&  =   k  \bigg(  \int d^d \vec x P_*(\vec x) h_1^{\dagger} 1 -E^{[1]} \bigg)
 + k^2 \bigg(  \int d^d \vec x P_*(\vec x) h_1^{\dagger} l^{[1]}(\vec x) 
  +  \int d^d \vec x P_*(\vec x) h_2^{\dagger} 1 - E^{[2]} \bigg)
 +O(k^3)
\label{eigenleftseriesproj}
\end{eqnarray}
So the two first coefficients of the perturbative expansion of Eq. \ref{FPhamiltoniankseries} for the energy read
\begin{eqnarray}
 E^{[1]} && =   \int d^d \vec x P_*(\vec x) h_1^{\dagger} 1
 \nonumber \\
 E^{[2]} && = \int d^d \vec x P_*(\vec x) h_1^{\dagger} l^{[1]}(\vec x) 
  +  \int d^d \vec x P_*(\vec x) h_2^{\dagger} 1
  \label{e1e2l}
\end{eqnarray}
where the first correction $l^{[1]}(\vec x) $ for the left eigenvector 
should be computed from Eq. \ref{eigenleftseries} at order $k$
\begin{eqnarray}
 0  = H^{\dagger} l^{[1]}(\vec x) + (h_1^{\dagger} 1 -E^{[1]}) 
 \label{eigenleftseriesl1}
\end{eqnarray}


  \subsubsection{ Conclusion on the first-order correction $E^{[1]}$ and on the first cumulant $C_1(T)$}

The two expressions of Eqs \ref{e1e2r} and \ref{e1e2l}
for the first-order correction $E^{[1]} $ are of course consistent with each other and read 
using the explicit expressions of Eqs \ref{Hkseries} \ref{Hkseriesdagger}
for $h_1$ and $h_1^{\dagger}$
\begin{eqnarray}
E^{[1]} && =  \int d^d x h_1 P_*(\vec x) = \int d^d \vec x P_*(\vec x) h_1^{\dagger} 1
\nonumber \\
&& = \int d^d \vec x P_*(\vec x) \left[  
    - [ \vec \nabla  .   \vec A^{[{\cal O}]}( \vec x)   ]
    - \frac{1}{D}    \vec A^{[{\cal O}]}( \vec x) .  \vec F( \vec x)  
  + \frac{1}{D} V^{[{\cal O}]} ( \vec x) 
 \right] 
  \label{energy1}
\end{eqnarray}
It is useful to make an integration by parts for the first contribution
\begin{eqnarray}
E^{[1]} &&  = \frac{1}{D} \int d^d \vec x \left[  
    \vec A^{[{\cal O}]}( \vec x) . D \vec \nabla  P_*(\vec x)  
    -     \vec A^{[{\cal O}]}( \vec x) .  \vec F( \vec x)  P_*(\vec x) 
  + V^{[{\cal O}]} ( \vec x)  P_*(\vec x)
 \right] 
 \nonumber \\
 && =\frac{1}{D} \int d^d \vec x  P_*(\vec x) \left[  
    \vec A^{[{\cal O}]}( \vec x) . \left( D \vec \nabla  \ln P_*(\vec x)  
    -     \vec F( \vec x)  \right)
  + V^{[{\cal O}]} ( \vec x) 
 \right] 
 \equiv  -  \frac{O_*}{D}  
  \label{energy1final}
\end{eqnarray}
in order to recognize the steady value $O_*$ of Eq. \ref{additiveSteadyrho}
as it should for consistency with Eq. \ref{legendrerecistar}
   \begin{eqnarray}
E^{[1]} =  E'(k=0) = -  \frac{O_*}{D}  
 \label{legendrerecistare1}
\end{eqnarray}
and to make the link with the first cumulant of Eq. \ref{cumulant1}
\begin{eqnarray}
   C_1 (T) \equiv \overline{ O [x(0 \leq t \leq T) ] } \opsimeq_{T \to + \infty}  - D E^{[1]} = O_*
\label{cumulant1Ostar}
\end{eqnarray}


  \subsubsection{ Computing the second-order correction $E^{[2]}$ in terms of the 
  first-order left eigenvector $ l^{[1]}(\vec x) $ }

To compute the second-order correction $E^{[2]} $, one can use either 
Eq. \ref{e1e2r} based on the right eigenvector,
or Eq. \ref{e1e2l} based on the left eigenvector.
We will now focus on this second expression that has two advantages : it is technically simpler 
and it will have an interesting interpretation in terms of the conditioned force
of Eq. \ref{forceDoobDiff}.
 
Using the explicit expressions of Eq. \ref{Hkseriesdagger} for $h_1^{\dagger}$ and $h_2^{\dagger}$,
Eq. \ref{e1e2l} reads
\begin{eqnarray}
&& E^{[2]}   =  \int d^d \vec x P_*(\vec x) h_2^{\dagger} 1
+ \int d^d \vec x P_*(\vec x) h_1^{\dagger} l^{[1]}(\vec x) 
\nonumber \\
&& = - \frac{1}{D} \int d^d \vec x P_*(\vec x)    \left[ \vec A^{[{\cal O}]}( \vec x)   \right]^2
+ \int d^d \vec x P_*(\vec x) \left[ - 2 \vec A^{[{\cal O}]}( \vec x) .  \vec \nabla  
    - [ \vec \nabla  .   \vec A^{[{\cal O}]}( \vec x)   ]
    - \frac{1}{D}    \vec A^{[{\cal O}]}( \vec x) .  \vec F( \vec x)  
  + \frac{1}{D} V^{[{\cal O}]} ( \vec x) \right]  l^{[1]}(\vec x) 
  \label{energy2froml}
\end{eqnarray}
where the first correction $l^{[1]}(\vec x) $ satisfies Eq. \ref{eigenleftseriesl1} 
\begin{eqnarray}
 0 && = H^{\dagger} l^{[1]}(\vec x) + (h_1^{\dagger} 1 -E^{[1]}) 
 \nonumber \\
&& =  - \left( D  \vec \nabla +   \vec F(\vec x)\right). \vec \nabla l^{[1]}(\vec x) 
+ \left[ 
    - [ \vec \nabla  .   \vec A^{[{\cal O}]}( \vec x)   ]
    - \frac{1}{D}    \vec A^{[{\cal O}]}( \vec x) .  \vec F( \vec x)  
  + \frac{1}{D} V^{[{\cal O}]} ( \vec x)   -E^{[1]} \right] 
 \label{eigenleftseriesl1expli}
\end{eqnarray}

The multiplication of Eq. \ref{eigenleftseriesl1expli} by $P_*(\vec x) l^{[1]}(\vec x)$
and the integration over $\vec x$ is useful to eliminate the contribution in $E^{[1]}  $
using Eq. \ref{eigennormaseries} 
and leads to the following equality that can be further simplified using integration by parts
\begin{eqnarray}
&& \int d^d \vec x P_*(\vec x) l^{[1]}(\vec x) 
\left[ 
    - [ \vec \nabla  .   \vec A^{[{\cal O}]}( \vec x)   ]
    - \frac{1}{D}    \vec A^{[{\cal O}]}( \vec x) .  \vec F( \vec x)  
  + \frac{1}{D} V^{[{\cal O}]} ( \vec x)   \right]
  \nonumber \\
&& = \int d^d \vec x P_*(\vec x) l^{[1]}(\vec x) \left( D  \vec \nabla +   \vec F(\vec x)\right). \vec \nabla l^{[1]}(\vec x) 
  \nonumber \\
&& = \int d^d \vec x P_*(\vec x) l^{[1]}(\vec x)    \vec F(\vec x). \vec \nabla l^{[1]}(\vec x) 
- \int d^d \vec x \vec \nabla l^{[1]}(\vec x) . \left[  l^{[1]}(\vec x) D \vec \nabla P_*(\vec x)
+ P_*(\vec x) D \vec \nabla l^{[1]}(\vec x) 
\right] 
  \nonumber \\
&& =  \frac{1}{2} \int d^d \vec x   \left[ P_*(\vec x) \vec F(\vec x) - D \vec \nabla P_*(\vec x) \right] . \vec \nabla [ l^{[1]}(\vec x) ]^2
- D \int d^d \vec x P_*(\vec x)  \left[  \vec \nabla l^{[1]}(\vec x) \right]^2
  \nonumber \\
&& = -  \frac{1}{2} \int d^d \vec x  l^{[1]}(\vec x) \vec \nabla . J_*(\vec x)
- D \int d^d \vec x P_*(\vec x)  \left[  \vec \nabla l^{[1]}(\vec x) \right]^2
= 0 - D \int d^d \vec x P_*(\vec x)  \left[  \vec \nabla l^{[1]}(\vec x) \right]^2
 \label{eigenleftseriesl1multiply}
\end{eqnarray}
where the vanishing of the first contribution comes from the vanishing divergence $\vec \nabla . J_*(\vec x) $ of the steady current.

Plugging Eq. \ref{eigenleftseriesl1multiply} into Eq. \ref{energy2froml}
is useful to obtain the following simpler factorized expression for the second-order correction
\begin{eqnarray}
 E^{[2]}  &&  =   - \frac{1}{D} \int d^d \vec x P_*(\vec x)    \left[ \vec A^{[{\cal O}]}( \vec x)   \right]^2
-2 \int d^d \vec x P_*(\vec x)  \vec A^{[{\cal O}]}( \vec x) .  \vec \nabla  l^{[1]}(\vec x) 
  - D \int d^d \vec x P_*(\vec x)  \left[  \vec \nabla l^{[1]}(\vec x) \right]^2
  \nonumber \\
  && =  - \frac{1}{D} \int d^d \vec x P_*(\vec x)    \left[ \vec A^{[{\cal O}]}( \vec x) + D  \vec \nabla l^{[1]}(\vec x) \right]^2
  \label{energy2fromlfactorized}
\end{eqnarray}
 
 Let us now turn to the interpretation in terms of the conditioned force.
  The perturbation expansion of Eq. \ref{rlkseries} for the left eigenvector $l_k(\vec x)$
  can be translated for its logarithm
\begin{eqnarray}
\ln l_k(\vec x)  && =  k  l^{[1]}(\vec x) +O(k^2)
  \label{loglkseries}
\end{eqnarray}
to obtain 
that the conditioned force of Eq. \ref{forceDoobDiff}
\begin{eqnarray}
\vec  F^{C[k]}(\vec x) = 
   \vec F(\vec x) + k \vec  F^{C_1}(\vec x) +O(k^2)
\label{forceDoobseries}
\end{eqnarray}
involves the first-order correction
\begin{eqnarray}
\vec  F^{C_1}(\vec x) \equiv  2  \vec A^{[{\cal O}]}(\vec x) +2 D   \vec \nabla    l^{[1]}(\vec x) 
\label{forceDoob1}
\end{eqnarray}
So the second-order correction of Eq. \ref{energy2fromlfactorized}
is directly related to the square of first-order conditioned force $\vec  F^{C_1}(\vec x) $
\begin{eqnarray}
 E^{[2]}   =  -  \int d^d \vec x P_*(\vec x)   \frac{ [\vec  F^{C_1}(\vec x) ]^2}{4 D}
  \label{energy2fromlfactorizedcond}
\end{eqnarray}
while Eq. \ref{eigenleftseriesl1} satisfied by $l^{[1]}(\vec x) $ 
with $E^{[1]}=- \frac{O_*}{D} $ of Eq. \ref{energy1final}
translates for $\vec  F^{C_1}(\vec x) $ into
\begin{eqnarray}
 V^{[{\cal O}]} ( \vec x)   +O_*  &&
  =    \left(  D \vec \nabla +  \vec F(\vec x) \right). \frac{\vec  F^{C_1}(\vec x)}{2}
\nonumber \\
&&   =    \frac{ D \vec \nabla . \vec  F^{C_1}(\vec x)}{2}
  +   \frac{ \vec F(\vec x) . \vec  F^{C_1}(\vec x)}{2}
 \label{eigenleftseriesl1F1}
\end{eqnarray}
that corresponds indeed to the first-order in $k$ of Eq. \ref{eigenleftforcecond}.


\section{ Large deviations for large time $T\to + \infty$ and then small diffusion $D\to 0$ }

\label{seC_largeTsmallD}

In this section, we describe how the properties for large time $T\to + \infty$ and finite $D$
described in the previous section \ref{seC_largeT}
can be further simplified when the diffusion coefficient is small $D \to 0$.
From the point of view of notations, we will use small letters for the various observables 
in the limit $D \to 0$ in order to stress the differences with the big letters used in the previous sections
for finite $D$.


\subsection{Simplifications for the steady state $p_*(\vec x)$ with the corresponding reversible and irreversible forces}

In the limit of small diffusion coefficient $D \to 0$, the steady state potential
$U_*(\vec x) $ introduced in Eq. \ref{forceRev} 
is expected to have a finite limit $u_*(\vec x)$, so that 
the steady state of Eq. \ref{rhostarfromurev}
is expected to display the following leading behavior  
\begin{eqnarray}
 p_*(\vec x)   =
 \frac{ e^{ - \frac{ u_*(\vec x) }{  D }  } }{ \int d^d \vec y \ e^{ - \frac{ u_*(\vec y) }{  D }  }} 
 \ \ \ \text { with } u_*(\vec x) \equiv \lim_{D \to 0}U_*(\vec x)
\label{rhosteadyD}
\end{eqnarray}
The gradient of $u_*(\vec x) $ determines
the finite limit $f^{rev} (\vec x ) $ when $D \to 0$
of the reversible force $\vec F^{rev} (\vec x ) $  of Eq. \ref{forceRevD}
\begin{eqnarray}
\vec f^{rev} (\vec x ) =  -  \vec \nabla u_* (\vec x) 
\label{forceRevD}
\end{eqnarray}
So the finite limit $\vec f^{irr} (\vec x ) $
of the irreversible force $\vec F^{irr} (\vec x ) $ of Eq. \ref{forceIrrev} 
reads
\begin{eqnarray}
\vec f^{irr} (\vec x) = \vec F(\vec x)  - \vec f^{rev} (\vec x)
= \vec F(\vec x)+  \vec \nabla u_* (\vec x)  
\label{forceIrrevD}
\end{eqnarray}
Then the leading order $\frac{1}{D}$ of Eq. \ref{divforceirrev}
reduces to the orthogonality condition
between the finite reversible and irreversible forces given in Eqs \ref{forceRevD} and \ref{forceIrrevD}
\begin{eqnarray}
0  = - \vec f^{irr}( \vec x) . \vec f^{rev}( \vec x) 
 =    \left[ \vec F(\vec x)+  \vec \nabla u_* (\vec x) \right] .   \vec \nabla u_* (\vec x) 
  \label{divforceirrevD1}
\end{eqnarray}
So one recognizes the orthogonal decomposition of the force $F(\vec x)$ discussed
 in Eqs \ref{forceNormalDecomposition} \ref{forceNormalDecompositionEq},
where $u_* (\vec x) $ coincides with $U^{\perp}(\vec x) $
and where $ \vec f^{irr} (\vec x ) $ coincides with $F^{\perp}( \vec x) $
\begin{eqnarray}
 u_* (\vec x) && = U^{\perp}(\vec x)
 \nonumber \\
 \vec f^{irr} (\vec x ) && = F^{\perp}( \vec x)
  \label{Ustar0isUperp}
\end{eqnarray}


\subsection{Simplifications for the lowest eigenvalue $E(k)$ of the Hamiltonian $H_k$
with its right and left eigenvectors } 

Since the right eigenvector $r_k(\vec x)$ is a deformation of the steady state $r_{k=0}(\vec x) =P_*(\vec x)$
that display the behavior of Eq. \ref{rhosteadyD} for small $D \to 0$,
it is natural to expect the same scaling 
\begin{eqnarray}
 r_k(\vec x)    \oppropto_{ D \to 0}  e^{ \displaystyle - \frac{ u_k (\vec x)}{D}  }
\label{rkD}
\end{eqnarray}
where $u_k (\vec x) $ will be a deformation of $u_{k=0} (\vec x)  =u_* (\vec x) $ discussed in the previous subsection.
The left eigenvector is also expected to display the analogous scaling
\begin{eqnarray}
 l_k(\vec x)    \oppropto_{ D \to 0}  e^{ \displaystyle - \frac{ {\tilde u}_k (\vec x)}{D}  }
\label{lkD}
\end{eqnarray}
where the function ${\tilde u}_k(\vec x)$ satisfies ${\tilde u}_{k=0}(\vec x) =0 $.

Finally the eigenvalue $E(k)$ of Eqs \ref{eigenright}
\ref{eigenleft}
is expected to display the scaling
\begin{eqnarray}
E(k) \opsimeq_{ D \to 0} \frac{ e(k)}{D}
\label{ekD}
\end{eqnarray}

Plugging Eqs \ref{rkD} and \ref{ekD}
into Eq. \ref{eigenrightlog} for the logarithm of $ r_k( \vec x)$
yields at leading order for $D \to 0$
\begin{eqnarray}
e(k)  =
  -   \left(   \vec \nabla u_k( \vec x) + k \vec A^{[{\cal O}]}( \vec x) \right)    .
 \left[   \vec \nabla u_k( \vec x) +k \vec A^{[{\cal O}]}( \vec x)  + \vec F( \vec x) \right]  
  + k V^{[{\cal O}]} ( \vec x)  
\label{eigenrightlogD}
\end{eqnarray}
that generalizes Eq. \ref{divforceirrevD1} corresponding to the special case $k=0$.

Similarly, plugging Eqs \ref{lkD} and \ref{ekD}
into Eq. \ref{eigenleftlog} for the logarithm of $ l_k( \vec x)$
yields at leading order for $D \to 0$
 \begin{eqnarray}
e(k)  =
-  \left[   - \vec \nabla  {\tilde u}_k( \vec x) + k  \vec A^{[{\cal O}]}( \vec x)  + \vec F( \vec x) \right]   .
   \left( - \vec \nabla  {\tilde u}_k( \vec x) + k \vec A^{[{\cal O}]}( \vec x) \right)  
  + kV^{[{\cal O}]} ( \vec x)  
\label{eigenleftlogD}
\end{eqnarray}

 Finally, Eqs \ref{rkD} \ref{lkD} and \ref{ekD} can be plugged into the large time behavior of Eq. \ref{genelargeT}
 for the generating function 
 in order to obtain the following further simplifications for small diffusion coefficient $D \to 0$
 \begin{eqnarray}
Z^{[k]}_T(\vec x \vert \vec x_0) 
\opsimeq_{T \to +\infty} e^{- T E(k) } r_k(\vec x) l_k(\vec x_0)
\opsimeq_{D \to 0} e^{ \displaystyle - \frac{T}{D}  e(k) - \frac{u_k(\vec x)}{D} - \frac{{\tilde u}_k(\vec x_0)}{D}} 
\label{genelargeTD}
\end{eqnarray}


\subsection{Simplifications for the conditioned force with its reversible and irreversible contributions  }

\label{subsec_canocondD}

For small $D \to 0$, the conditioned force $\vec F^{C[k]}(\vec x) $ of Eq. \ref{forceDoobDiff} becomes
using Eqs \ref{rkD} and \ref{lkD}
 \begin{eqnarray}
 \vec f^{C[k]}(\vec x)
 =   \vec F(\vec x)+ 2 k \vec A^{[{\cal O}]}(\vec x) - 2    \vec \nabla {\tilde u}_k  ( \vec x)  
\label{forceDoobDiffsimply}
\end{eqnarray}
while
Eq. \ref{eigenleftforcecond} becomes  at leading order $\frac{1}{D}$
\begin{eqnarray}
 e(k) +  \frac{ [\vec f^{C[k]} ( \vec x) ]^2 }{ 4  }     
= \frac{ \vec F ^2( \vec x) }{ 4  }     
+ k V^{[{\cal O}]}(\vec x)
\label{eigenleftforcecondsimply}
\end{eqnarray}
so that the dependence on $\vec x$ of the square of the conditioned force is determined by
\begin{eqnarray}
  [\vec f^{C[k]} ( \vec x) ]^2      =  \vec F ^2( \vec x)      + 4 k V^{[{\cal O}]}(\vec x) - 4 e(k)
\label{eigenleftforcecondsimply2}
\end{eqnarray}

The corresponding reversible conditioned force $\vec F^{C[k]rev} (\vec x ) $  of Eq. \ref{forceRevcond}
has for finite limit when $D \to 0$
\begin{eqnarray}
\vec f^{C[k]rev} (\vec x ) 
=  -  \vec \nabla u_k (\vec x) -  \vec \nabla {\tilde u}_k (\vec x) 
\label{forceRevDcond}
\end{eqnarray}
so that the finite limit $\vec f^{C[k]irr} (\vec x ) $
of the irreversible force $\vec F^{C[k]irr} (\vec x ) $ of Eq. \ref{forceIrrev} 
reads using Eqs \ref{forceDoobDiffsimply} and \ref{forceRevDcond}
\begin{eqnarray}
\vec f^{C[k]irr} (\vec x) \equiv \vec f^{C[k]}(\vec x)  - \vec f^{C[k]rev} (\vec x)
= \vec F(\vec x)+ 2 k \vec A^{[{\cal O}]}(\vec x) 
  +  \vec \nabla u_k (\vec x) -  \vec \nabla {\tilde u}_k (\vec x) 
\label{forceIrrevDcond}
\end{eqnarray}
Then the leading order $\frac{1}{D}$ of Eq. \ref{divforceirrevcond}
reduces to the orthogonality condition
between the finite reversible and irreversible conditioned forces given in Eqs \ref{forceRevDcond} and \ref{forceIrrevDcond}
\begin{eqnarray}
0  =  \vec f^{C[k]irr}( \vec x) . \vec f^{C[k]rev}( \vec x) 
  \label{divforceirrevD1cond}
\end{eqnarray}
i.e. the reversible-irreversible decomposition of the conditioned force $\vec f^{C[k]}(\vec x) $ coincides with its orthogonal decomposition.

 
 \subsection{ Rescaling of the rate function $ I (O) $ for small diffusion coefficient $ D \to 0$  }

 Plugging the rescaling of the ground state energy of Eq. \ref{ekD} into the Legendre transformations of Eq. \ref{legendre} and \ref{legendrereci}
 yields that the rate function $I(O)$ governing Eq. \ref{ratefunctionCalO}
 displays the analog rescaling
 \begin{eqnarray}
I(O) \opsimeq_{ D \to 0} \frac{ i(O)}{D}
\label{ioD}
\end{eqnarray}
 where the rescaled rate function $i(O)$ is related to the rescaled eigenvalue $e(k)$
 via the Legendre transformation obtained from of Eq. \ref{legendre}
  \begin{eqnarray}
 i(O)- k  O  && = e(k)
 \nonumber \\
  i'(O)- k  && = 0
 \label{legendrerescaled}
\end{eqnarray}
and the reciprocal Legendre transformation
   \begin{eqnarray}
 i(O)  && = e(k) +  k  O
 \nonumber \\
 0   && = e'(k) +   O
 \label{legendrerecirescaped}
\end{eqnarray}

For $ D \to 0$, the steady value $O_*$ of Eq. \ref{additiveSteadyrho}
becomes using Eq. \ref{rhosteadyD}
 \begin{eqnarray}
o_*  && = \int d^d \vec x  p_*(\vec x) \left[   - V^{[{\cal O}]} (\vec x ) 
+  \vec A^{[{\cal O}]}( \vec x) . \left(   \vec F(\vec x ) + \vec \nabla  u_* (\vec x) \right) \right]  
\nonumber \\
&& = \int d^d \vec x  p_*(\vec x) \left[   - V^{[{\cal O}]} (\vec x ) 
+  \vec A^{[{\cal O}]}( \vec x) .  \vec f^{irr}(\vec x )  \right]  
\label{additiveSteadyrhoD}
\end{eqnarray}
where $p_*(\vec x) $ is the steady state of Eq. \ref{rhosteadyD} parametrized by $u_* (\vec x) $.
This value $o_*$ where rate function vanishes $i(o_*)=0$
is associated to $k=0$ where the rescaled energy vanishes $e(k=0)=0$, so that Eq. \ref{legendrerecirescaped}
determines the first derivative 
   \begin{eqnarray}
  e'(k=0) = -  o_* 
 \label{legendrerecistarderie}
\end{eqnarray}

In conclusion, the large deviation form of Eq. \ref{ratefunctionCalO} for large $T$
for the probability distribution of Eq \ref{ratefunctionCalO}
of the additive trajectory observable ${\cal O}$
becomes for small diffusion coefficient
 \begin{eqnarray}
   {\cal P}_T({\cal O} = T O  ) \oppropto_{ T \to + \infty} e^{- T I (O) }   
   \oppropto_{ D \to 0} e^{- \frac{T}{D} i (O) }   
    \label{ratefunctionCalOrecaled}
\end{eqnarray}


  \subsection{ Simplifications for the first cumulants of trajectory observables in the double limit $T \to + \infty$ and $D \to 0$}
  
  \label{subsec_conclusione1e2}
  
   The series expansion in $k$ of the rescaled energy $e(k)$ of Eq. \ref{ekD}
  \begin{eqnarray}
  e(k)  = 0 + k e^{[1]} + k^2 e^{[2]} +O(k^3)
 \label{FPhamiltoniankseriesrescaled}
\end{eqnarray} 
  can be derived from the series expansion of Eq. \ref{FPhamiltoniankseries} for $E(k)$ 
 
 The result of Eq. \ref{energy1final} for $E^{[1]}  $ yields at leading order $\frac{1}{D}$ 
 that $e_1$ reads using Eq. \ref{rhosteadyD}
 \begin{eqnarray}
e^{[1]} &&  = \int d^d \vec x  p_*(\vec x) \left[  V^{[{\cal O}]} ( \vec x) 
   - \vec A^{[{\cal O}]}( \vec x) . \left(  \vec F( \vec x) +  \vec \nabla  u_* (\vec x) \right)
 \right] \equiv  -  o_*   
  \label{energy1finalrescaled}
\end{eqnarray}
 in agreement with Eqs \ref{additiveSteadyrhoD} \ref{legendrerecistarderie}.
 
 The result of Eq. \ref{energy2fromlfactorizedcond} for $E^{[2]}  $ yields at leading order $\frac{1}{D}$ 
 that $e_2$ is given by
 \begin{eqnarray}
 e^{[2]}   =  -  \int d^d \vec x p_*(\vec x)   \frac{ [\vec  f^{C_1}(\vec x) ]^2}{4 }
  \label{energy2rescaled}
\end{eqnarray}
where $\vec  f^{C_1}(\vec x)$
corresponds to the first-order conditioned force of Eq. \ref{forceDoob1}
in the limit $D \to 0$ that reads using Eq \ref{lkD}
\begin{eqnarray}
\vec  f^{C_1}(\vec x) 
=2  \vec A^{[{\cal O}]}(\vec x) - 2    \vec \nabla    {\tilde u}^{[1]}(\vec x)
\label{C1assumdiffD}
\end{eqnarray}
while Eq. \ref{eigenleftseriesl1F1} simplifies into
 \begin{eqnarray}
 V^{[{\cal O}]} ( \vec x)   +o_*    = \vec F(\vec x) .   \frac{ \vec  f^{C_1}(\vec x)}{2}
 \label{eqsimplifiedCondf}
\end{eqnarray}
 
 So the second cumulant $C_2(T)$ of Eq. \ref{cumulant2}, i.e. the variance of the intensive trajectory observable $O [x(0 \leq t \leq T) ] $, 
becomes
in the double limit $T \to + \infty$ and $D \to 0$
\begin{eqnarray}
   C_2(T) &&  \equiv \overline{ \left( O [x(0 \leq t \leq T) ]  -  \overline{ O [x(0 \leq t \leq T) ] }\right)^2 }  
   \opsimeq_{T \to + \infty} - \frac{2 D^2 E^{[2]}}{T}
 \opsimeq_{D \to 0} - \frac{2 D e^{[2]}}{T}
  = \frac{ D }{T}  \int d^d \vec x p_*(\vec x)   \frac{ [\vec  f^{C_1}(\vec x) ]^2}{2 }
    \label{cumulant2conclusionD}
\end{eqnarray}


\section{ Large deviations for small diffusion coefficient $D\to 0$ and finite time $T$  }

\label{seC_smallD}

In this section, we restart from the section \ref{seC_general}
concerning finite time $T$ and finite diffusion coefficient $D$
 in order to describe the simplifications for $D \to 0$.

\subsection{Propagator $ P_T(\vec x \vert \vec x_0) $ for small diffusion coefficient $D \to 0$ and finite time $T$} 

The analysis of diffusion processes in the weak-noise limit of small diffusion coefficients
has a very long history both in physics \cite{graham1983,graham1984,graham1985,graham1986,graham1991}
with the analogy with the semiclassical WKB approximation of quantum mechanics,
and in the mathematical Freidlin-Wentzell theory \cite{FW} of large deviations 
(see the recent review \cite{FWreview} and references therein).

\subsubsection{ Path-integral for the propagator $P_T(\vec x \vert \vec x_0) $
dominated by the appropriate classical action }

The classical Lagrangian of Eq. \ref{lagrangian} contains the following leading contribution of order $\frac{1}{D}$
\begin{eqnarray}
 {\cal L} (\vec x(t), \dot {\vec x}(t)) && =
 \frac{ {\hat {\cal L}} (\vec x(t), \dot {\vec x}(t)) }{ D} 
 + O(D^0)  
 \label{lagrangianDexp}
\end{eqnarray}
where the rescaled Lagrangian ${\hat {\cal L}} (\vec x(t), \dot {\vec x}(t)) $ reduces to
\begin{eqnarray}
{\hat {\cal L}} (\vec x(t), \dot {\vec x}(t)) && =
 \frac{\left( \dot {\vec x } (t) - \vec F( \vec x(t)) \right)^2}{4 }  
  \label{lagrangianD}
\end{eqnarray}

For small diffusion coefficient $D \to 0$,
the path-integral of Eq. \ref{pathintegral}  
\begin{eqnarray}
P_T(\vec x \vert \vec x_0) 
\opsimeq_{ D \to 0} \int_{\vec x(t=0)=\vec x_0}^{\vec x(t=T)=\vec x} {\cal D}   \vec x(.)  
 e^{ - \displaystyle   \frac{ {\hat {\cal S}}(\vec x (0 \leq t \leq T)) }{ D}
 } 
\label{pathintegralD}
\end{eqnarray}
is thus 
governed by the rescaled action ${\hat S}(\vec x (0 \leq t \leq T) ) $ for trajectories $\vec x (0 \leq t \leq T) $
obtained from 
the rescaled Lagrangian of Eq. \ref{lagrangian}
\begin{eqnarray}
{\hat {\cal S}}(\vec x (0 \leq t \leq T) )
 \equiv  \int_0^T dt {\hat {\cal L}} (\vec x(t), \dot {\vec x}(t))
=  \int_0^T dt \frac{\left( \dot {\vec x } (t) - \vec F( \vec x(t)) \right)^2}{4 } 
\label{actionD}
\end{eqnarray}
For small diffusion coefficient $D \to 0$, 
 the sum ${\cal D}   \vec x(0 \leq t \leq T)  $ over trajectories  in the path-integral of Eq. \ref{pathintegralD}
will be dominated by the single
classical trajectory $\vec x^{Class} (0 \leq t \leq T) $
that minimizes the action ${\hat {\cal S}}(\vec x (0 \leq t \leq T) ) $ of Eq. \ref{actionD}
while satisfying
the two boundary conditions 
\begin{eqnarray}
\vec x^{Class}(0) && = \vec x_0
\nonumber \\
\vec x^{Class}(T) && = \vec x
\label{BCclassical}
\end{eqnarray}
so that Eq. \ref{pathintegralD} reduces to 
\begin{eqnarray}
P_T(\vec x \vert \vec x_0) 
\oppropto_{ D \to 0}  
 e^{ - \displaystyle   \frac{ {\hat S}_T(\vec x  \vert \vec x_0) }{ D}
 } 
\label{pathintegralDclassical}
\end{eqnarray}
where the action ${\hat S}_T(\vec x \vert \vec x_0)  $ associated to the classical trajectory satisfying the 
boundary conditions of Eq. \ref{BCclassical}
\begin{eqnarray}
 {\hat S}_T(\vec x \vert \vec x_0) 
 \equiv  \int_0^T dt \frac{\left( \dot {\vec x }^{Class} (t) - \vec F( \vec x^{Class}(t)) \right)^2}{4 } \geq 0
\label{Actionclassical}
\end{eqnarray}
can be considered as a positive rate function for the final position $\vec x$ at time $T$ once the initial position $\vec x_0$ 
is given.
This classical action $ {\hat S}_T(\vec x \vert \vec x_0) $ 
vanishes only if the zero-noise deterministic trajectory ${\vec x }^{deter} (t) $
satisfying
\begin{eqnarray}
\dot {\vec x }^{deter} (t) && = \vec F( \vec x^{deter}(t)) 
\nonumber \\
{\vec x }^{deter} (t=0) && =\vec x_0
\label{detertraj}
\end{eqnarray}
is at position ${\vec x }^{deter} (t=T)  =\vec x $ at time $T$
\begin{eqnarray}
 {\hat S}_T(\vec x \vert \vec x_0)  =0 \text{ if } \ \ \vec x ={\vec x }^{deter} (t=T)
\label{ActionclassicalVanish}
\end{eqnarray}
For all the other position $\vec x \ne {\vec x }^{deter} (t=T)$, the positive classical action ${\hat S}_T(\vec x  \vert \vec x_0)  >0 $ characterizes how rare it is for small $D$ to see the position $\vec x$ at time $T$ via Eq. \ref{pathintegralDclassical}.

If one plugs the asymptotic behavior of Eq. \ref{pathintegralDclassical} for small $D$ 
into the Fokker-Planck Eq \ref{fokkerplanckeuclidean}
satisfied by the propagator $P_T(\vec x \vert \vec x_0)$
\begin{eqnarray}
 \partial_T P_T(\vec x \vert \vec x_0)    = - H P_T(\vec x \vert \vec x_0)  
 =  \vec \nabla . \left(   D  \vec \nabla -   \vec F(\vec x)\right) P_T(\vec x \vert \vec x_0)
\label{fokkerplanckeuclideanpropagator}
\end{eqnarray}
one obtains at leading order $ \frac{1}{D}$ that 
the classical action $ {\hat S}_T(\vec x \vert \vec x_0) $ is governed by the dynamics
\begin{eqnarray}
\partial_T {\hat S}_T(\vec x  \vert \vec x_0) 
 = - \left[ \vec F( \vec x)  + \vec \nabla {\hat S}_T(\vec x  \vert \vec x_0)\right] .  \vec \nabla {\hat S}_T(\vec x  \vert \vec x_0)
\label{HamiltonJacobi}
\end{eqnarray}
that corresponds to the Hamilton-Jacobi equation as recalled in the next subsection.

Similarly, the backward Fokker-Planck equation
for the propagator $P_T(\vec y \vert \vec x)$ when $\vec x$ is the initial position 
that involves the adjoint operator $H^{\dagger}$ of Eq. \ref{FPhamiltonian} 
\begin{eqnarray}
 \partial_T P_T(\vec y \vert \vec x)   = - H^{\dagger} P_T(\vec y \vert \vec x) 
 = \left[   D  \vec \nabla +   \vec F(\vec x)\right] . \vec \nabla  P_T(\vec y \vert \vec x)
\label{fokkerplanckeuclideanpropagatorbackward}
\end{eqnarray}
yields at leading order $\frac{1}{D} $ the backward equation satisfied by the
classical action $ {\hat S}_T(\vec y \vert \vec x) $
\begin{eqnarray}
\partial_T {\hat S}_T(\vec y  \vert \vec x) 
 =  \left[ \vec F( \vec x) - \vec \nabla {\hat S}_T(\vec y  \vert \vec x) \right] .    \vec \nabla  {\hat S}_T(\vec y  \vert \vec x)
\label{HamiltonJacobiBackward}
\end{eqnarray}

Besides the forward and backward dynamics of Eqs \ref{HamiltonJacobi} and \ref{HamiltonJacobiBackward}
satisfied by the classical action ${\hat S}_T(\vec x  \vert \vec x_0) $, it can be useful 
to analyze the properties of the corresponding classical trajectories as discussed in the next subsection.


\subsubsection{ Properties of the classical trajectories }

The classical moment $\vec p(t)$ can be computed from the Lagrangian of Eq. \ref{lagrangianD}
\begin{eqnarray}
 \vec p(t) \equiv \frac{\partial {\hat {\cal L}} (\vec x(t), \dot {\vec x}(t) ) } {\partial  \dot {\vec x}(t)} 
=\frac{ \dot {\vec x } (t) - \vec F( \vec x(t))}{2} 
\label{lagrangianderidot}
\end{eqnarray}
while the classical Hamiltonian ${\hat {\cal H}}(\vec x, \vec p)  $ 
is obtained from the Lagrangian ${\hat {\cal L}} (\vec x, \dot {\vec x} ) $ of Eq. \ref{lagrangianD}
via the Legendre transform 
\begin{eqnarray}
{\hat {\cal H}}(\vec x, \vec p) && = \vec p . \dot {\vec x} -  {\hat {\cal L}} (\vec x, \dot {\vec x})
=  \vec p . \left[ 2 \vec p + \vec F( \vec x) \right] -   {\vec p \ }^2
 =    {\vec p \ }^2+  \vec p . \vec F( \vec x)
\label{HamiltonienClassique}
\end{eqnarray}
The Hamiltonian classical equations of motion 
correspond to the following first-order differential equations
for the $2d$ canonical coordinates $[x_{\mu}(t); p_{\mu}(t)]$ in phase-space 
\begin{eqnarray}
\dot x_{\mu}(t) && = \frac{\partial {\hat {\cal H}} ({\vec x}(t),  {\vec p}  (t)) } {\partial  p_{\mu} (t)} 
= 2  p_{\mu}(t) + F_{\mu}(\vec x (t)) 
\nonumber \\
      \dot p_{\mu}(t)  && = - \frac{\partial {\hat {\cal H}} ({\vec x}(t),  {\vec p}  (t)) } {\partial  x_{\mu} (t)} 
 = - \sum_{\nu=1}^d   p_{\nu}(t)  \frac{ \partial F_{\nu}( \vec x(t)) } {\partial   x_{\mu}(t)} 
\label{Hamilton12}
\end{eqnarray}
that ensure the conservation of energy along the classical trajectory
\begin{eqnarray}
{\hat {\cal E}} =  {\hat {\cal H}} ({\vec x}(t),  {\vec p}  (t)) && =  {\vec p \ }^2(t) +  \vec p (t) . \vec F( \vec x (t) )
 = \frac{ \dot {\vec x }^2 (t) - \vec F^2( \vec x(t))}{4} 
\label{HamiltonEnergyConserv}
\end{eqnarray}

In this Hamiltonian language, the zero-noise deterministic trajectory of Eq. \ref{detertraj}
corresponds to zero-momentum and zero energy
\begin{eqnarray}
\dot {\vec x }^{deter} (t) && = \vec F( \vec x^{deter}(t)) 
\nonumber \\
{\vec p }^{deter} (t) && =0
\nonumber \\
{\hat {\cal E}} && =0
\label{detertrajHamilton}
\end{eqnarray}

Instead of the Hamiltonian classical equations of motion of Eq \ref{Hamilton12},
one can write
the Lagrangian classical equations of motion 
that correspond to following second-order differential equations for the $d$ coordinates $[x_{\mu}(t)]$ in real space
\begin{eqnarray}
0  = \ddot x_{\mu}(t) +\sum_{\nu=1}^d   \dot x_{\nu}(t) \left[  \frac{ \partial F_{\nu}( \vec x(t)) } {\partial   x_{\mu}(t)} 
 -  \frac{ \partial F_{\mu}(\vec x (t)) }{\partial  x_{\nu}(t)} \right]
   -  \frac{ \partial  } {\partial   x_{\mu}(t)} \left[  \sum_{\nu=1}^d   \frac{  F_{\nu}^2( \vec x(t)) } {2} \right]
\label{Lagrange12}
\end{eqnarray}

If the classical trajectory satisfying the boundary conditions of Eq. \ref{BCclassical}
can be computed from the equations of motion of Eq. \ref{Hamilton12} or Eq. \ref{Lagrange12},
one can evaluate the corresponding classical action of Eq. \ref{Actionclassical}
that can be rewritten using the momentum of Eq. \ref{lagrangianderidot}
\begin{eqnarray}
 {\hat S}_T(\vec x, \vert \vec x_0) 
 =  \int_0^T dt \frac{\left( \dot {\vec x } (t) - \vec F( \vec x(t)) \right)^2}{4 } 
 = \int_0^T dt   {\vec p \ }^2(t)
\label{ClassicalActionp2}
\end{eqnarray}
or using the Legendre transform of Eq. \ref{HamiltonienClassique}
and the conservation of the energy of Eq. \ref{HamiltonEnergyConserv}
\begin{eqnarray}
 {\hat S}_T(\vec x  \vert \vec x_0) 
 =  \int_0^T dt {\hat {\cal L}} (\vec x(t), \dot {\vec x}(t))
 = \int_0^T dt  \left[   \vec p (t) . \dot {\vec x} (t)- {\hat {\cal H}}(\vec x (t) , \vec p (t) ) \right]
 = \int_0^T dt   \vec p(t) . \dot {\vec x}(t) - T  {\hat {\cal E}}
\label{ClassicalActionp2H}
\end{eqnarray}

Another possibility is to consider the Hamilton-Jacobi equation satisfied by the classical action
${\hat S}_T(\vec x, \vert \vec x_0)$ as recalled in the next subsection.


\subsubsection{ Hamilton-Jacobi equations for the classical action ${\hat S}_T(\vec x  \vert \vec x_0) $ 
and for its time-energy Legendre transform ${\hat W}_{\hat {\cal E}}(\vec x  \vert \vec x_0) $ }

The Hamilton-Jacobi equation for the classical action ${\hat S}_T(\vec x  \vert \vec x_0) $
involves the classical Hamiltonian ${\hat {\cal H}} ({\vec x},  {\vec p} )$ of Eq. \ref{HamiltonienClassique}
where the momentum $ {\vec p} $ is replaced by the gradient of the action $  {\hat S}_T(\vec x  \vert \vec x_0) $ with respect to the final position $\vec x$
\begin{eqnarray}
\partial_T {\hat S}_T(\vec x  \vert \vec x_0) 
&& = - {\hat {\cal H}} \left({\vec x},  {\vec p} = \vec \nabla {\hat S}_T(\vec x  \vert \vec x_0) \right)
 \nonumber \\
 && = - \left[ \vec F( \vec x)  + \vec \nabla {\hat S}_T(\vec x  \vert \vec x_0)\right] .  \vec \nabla {\hat S}_T(\vec x  \vert \vec x_0)
\label{HamiltonJacobibis}
\end{eqnarray}
that coincides with Eq. \ref{HamiltonJacobi} as it should for consistency.
Similarly, Eq. \ref{HamiltonJacobiBackward} corresponds to the backward 
Hamilton-Jacobi equation for the classical action ${\hat S}_T(\vec y  \vert \vec x) $
when $\vec x$ is the initial position
\begin{eqnarray}
\partial_T {\hat S}_T(\vec y  \vert \vec x) 
&& = - {\hat {\cal H}} \left({\vec x},  {\vec p} = - \vec \nabla {\hat S}_T(\vec y  \vert \vec x) \right)
 \nonumber \\
 && =  \left[ \vec F( \vec x) - \vec \nabla {\hat S}_T(\vec y  \vert \vec x) \right] .    \vec \nabla  {\hat S}_T(\vec y  \vert \vec x)
\label{HamiltonJacobiBackwardbis}
\end{eqnarray}

The time-energy Legendre transform between the classical action ${\hat S}_T(\vec x  \vert \vec x_0) $
and Hamilton's characteristic function ${\hat W}_{\hat {\cal E}}(\vec x  \vert \vec x_0) $
\begin{eqnarray}
{\hat W}_{\hat {\cal E}}(\vec x  \vert \vec x_0) && = {\hat S}_T(\vec x  \vert \vec x_0) + {\hat {\cal E}} T
 \nonumber \\
 0 && = \partial_T  {\hat S}_T(\vec x  \vert \vec x_0) + {\hat {\cal E}} 
\label{HJLegendre}
\end{eqnarray}
with its reciprocal Legendre transformation
\begin{eqnarray}
 {\hat S}_T(\vec x  \vert \vec x_0) && = {\hat W}_{\hat {\cal E}}(\vec x  \vert \vec x_0) - {\hat {\cal E}} T
  \nonumber \\
 0 && = \partial_{{\hat {\cal E}}} {\hat W}_{\hat {\cal E}}(\vec x  \vert \vec x_0) - T
\label{HJLegendrereci}
\end{eqnarray}
is useful to transform the time-dependent Hamilton-Jacobi Eq. \ref{HamiltonJacobibis} for ${\hat S}_T(\vec x  \vert \vec x_0) $
into the energy-dependent Hamilton-Jacobi equation for ${\hat W}_{\hat {\cal E}}(\vec x  \vert \vec x_0) $
\begin{eqnarray}
 {\hat {\cal E}} 
&& =  {\hat {\cal H}} \left({\vec x},  {\vec p} = \vec \nabla {\hat W}_{\hat {\cal E}}(\vec x  \vert \vec x_0) \right)
 \nonumber \\
 && =  \left[ \vec F( \vec x)  + \vec \nabla {\hat W}_{\hat {\cal E}}(\vec x  \vert \vec x_0)\right] .  \vec \nabla {\hat W}_{\hat {\cal E}}(\vec x  \vert \vec x_0)
\label{HamiltonJacobiW}
\end{eqnarray}

Similarly, the backward Eq. \ref{HamiltonJacobiBackwardbis} satisfied by the $ {\hat S}_T(\vec y \vert \vec x) $
can be translated for ${\hat W}_{\hat {\cal E}}(\vec y  \vert \vec x) $ into
\begin{eqnarray}
 {\hat {\cal E}} 
 && =  {\hat {\cal H}} \left({\vec x},  {\vec p} = - \vec \nabla {\hat W}_{\hat {\cal E}}(\vec y  \vert \vec x) \right)
 \nonumber \\
&& =    - \left[ \vec F( \vec x) - \vec \nabla {\hat W}_{\hat {\cal E}}(\vec y  \vert \vec x) \right] .    \vec \nabla  {\hat W}_{\hat {\cal E}}(\vec y  \vert \vec x)
\label{HamiltonJacobiWbackward}
\end{eqnarray}


\subsection{Generating functions $Z^{[k]}_T(\vec x \vert \vec x_0) $ of trajectory observables for small diffusion coefficient $D \to 0$ and finite $T$}

As discussed in \cite{Engel_Seifert,derrida-ring,bertin-conditioned} for the particular case of diffusion processes
on a one-dimensional ring, the statistics of trajectory observables for $D \to 0$ and finite time $T$
can be analyzed via the analog of the
semi-classical WKB approximation for the appropriate deformed generator.
Let us now describe how it works with our present notations in dimension $d>1$.

 \subsubsection{ Path-integral for the generating functions $Z^{[k]}_T(\vec x \vert \vec x_0) $
dominated by the appropriate classical action }

Using the deformed electromagnetic potentials of Eqs \ref{vectorpotp} and \ref{scalarpotp},
the $k$-deformed Lagrangian of Eq. \ref{lagrangiank} contains the following leading contribution of order $\frac{1}{D}$
\begin{eqnarray}
 {\cal L}_k (\vec x(t), \dot {\vec x}(t)) && =
 \frac{  {\hat {\cal L}}_k (\vec x(t), \dot {\vec x}(t)) }{ D} 
 + O(D^0)  
 \label{lagrangianDexpk}
\end{eqnarray}
where the rescaled Lagrangian
\begin{eqnarray}
{\hat {\cal L}}_k (\vec x(t), \dot {\vec x}(t)) && =
\frac{\dot {\vec x}^2  (t) }{4}  
 - \dot {\vec x } (t) . \left[ \frac{ \vec F( \vec x) }{2} +k \vec A^{[{\cal O}]}( \vec x) \right]
 + \left[\frac{ \vec F ^2( \vec x) }{ 4  }      + k V^{[{\cal O}]} ( \vec x) \right]
 \nonumber \\
&& =
\frac{\dot {\vec x}^2  (t) }{4}  
 - \dot {\vec x } (t) .  \vec a^{[k]} (\vec x )
+v^{[k]}(\vec x)
 \label{lagrangianDk}
\end{eqnarray}
involves the rescaled vector and scalar potentials
\begin{eqnarray}
  \vec a^{[k]} (\vec x ) \equiv  \frac{ \vec F( \vec x) }{2} +k \vec A^{[{\cal O}]}( \vec x)
\nonumber \\
v^{[k]}(\vec x) \equiv  \frac{ \vec F ^2( \vec x) }{ 4  }  + k V^{[{\cal O}]} ( \vec x)
\label{vectorscalarpotD}
\end{eqnarray}
with the corresponding rescaled magnetic matrix of Eq. \ref{magneticB}
\begin{eqnarray}
 b_{\mu \nu}^{[k]}(\vec x )  =-b_{\nu \mu}^{[k]} (\vec x )  
 \equiv \partial_{\mu} a_{\nu}^{[k]}(\vec x ) -  \partial_{\nu} a_{\mu}^{[k]} (\vec x ) 
 \label{magneticBD}
\end{eqnarray}
So the ideas are the same as in the previous subsection concerning the propagator around Eqs \ref{pathintegralD} \ref{pathintegralDclassical} :
the path-integral of Eq. \ref{gene} is dominated for small diffusion coefficient $D \to 0$ by
\begin{eqnarray}
 Z^{[k]}_T(\vec x \vert \vec x_0) 
\opsimeq_{D \to 0}
 \int_{\vec x(t=0)=\vec x_0}^{\vec x(t=T)=\vec x} {\cal D}   \vec x(.)  
 e^{ - \displaystyle \frac{ {\hat {\cal S}}_k(\vec x (0 \leq t \leq T) )}{D} }
 \opsimeq_{ D \to 0}  
 e^{ - \displaystyle   \frac{ {\hat S}_T^{[k]}(\vec x  \vert \vec x_0) }{ D}
 } 
 \label{geneD}
\end{eqnarray}
where the rescaled action ${\hat {\cal S}}_k(\vec x (0 \leq t \leq T) ) $ of the trajectory $\vec x (0 \leq t \leq T) $ is associated to 
the rescaled Lagrangian of Eq. \ref{lagrangian}
\begin{eqnarray}
{\hat {\cal S}}_k(\vec x (0 \leq t \leq T) )
 \equiv  \int_0^T dt {\hat {\cal L}}_k (\vec x(t), \dot {\vec x}(t))
\label{actionDk}
\end{eqnarray}
while ${\hat S}_T^{[k]}(\vec x  \vert \vec x_0)$ is the action of the classical trajectory
satisfying the 
boundary conditions of Eq. \ref{BCclassical}.

If one plugs the asymptotic behavior of Eq. \ref{geneD} for small $D$ 
into the dynamics of Eq. \ref{EuclideanZ} governed by the Hamiltonian of Eq. \ref{FPhamiltonianpexpanded}
\begin{eqnarray}
 \partial_T Z_T^{[k]}(\vec x \vert \vec x_0)  
&& =  -  H_k Z^{[k]}_T(\vec x \vert \vec x_0) 
\nonumber \\
&&  =  \frac{1}{D}   \left( - D \vec \nabla  + k \vec A^{[{\cal O}]}( \vec x) \right)  
 \left[ - D  \vec \nabla  +k \vec A^{[{\cal O}]}( \vec x)  + \vec F( \vec x) \right] Z^{[k]}_T(\vec x \vert \vec x_0)
  - \frac{k}{D} V^{[{\cal O}]} ( \vec x)  Z^{[k]}_T(\vec x \vert \vec x_0)
\label{geneforward}
\end{eqnarray}
one obtains at leading order $ \frac{1}{D}$ that the action $ {\hat S}^{[k]}_T(\vec x \vert \vec x_0) $ is governed by the dynamics
\begin{eqnarray}
\partial_T {\hat S}^{[k]}_T(\vec x  \vert \vec x_0) 
 = - \left[ \vec F( \vec x)  +k \vec A^{[{\cal O}]}( \vec x) + \vec \nabla {\hat S}^{[k]}_T(\vec x  \vert \vec x_0))\right] . 
 \left[ k \vec A^{[{\cal O}]}( \vec x)  + \vec \nabla {\hat S}^{[k]}_T(\vec x  \vert \vec x_0))\right]
 + k V^{[{\cal O}]} ( \vec x)
\label{HamiltonJacobik}
\end{eqnarray}
that corresponds to the $k$-deformation of the Hamilton-Jacobi Eq. \ref{HamiltonJacobi}.

Similarly, the backward dynamics
for the generating function $Z^{[k]}_T(\vec y \vert \vec x)  $ when $\vec x$ is the initial position
that involves the adjoint operator $H^{\dagger}_k$ of Eq. \ref{FPhamiltonianpexpandedadjoint} 
\begin{eqnarray}
 \partial_T Z^{[k]}_T(\vec y \vert \vec x) && = - H^{\dagger}_k Z^{[k]}_T(\vec y \vert \vec x)
 \nonumber \\
&& = \frac{1}{D} \left[  D   \vec \nabla  + k  \vec A^{[{\cal O}]}( \vec x)  + \vec F( \vec x) \right]
   \left( D \vec \nabla  + k \vec A^{[{\cal O}]}( \vec x) \right)  Z^{[k]}_T(\vec y \vert \vec x)
  - \frac{k}{D} V^{[{\cal O}]} ( \vec x) Z^{[k]}_T(\vec y \vert \vec x)
 \label{fokkerplanckeuclideanpropagatorbackwardk}
\end{eqnarray}
yields at leading order $\frac{1}{D} $ the backward equation satisfied by the $ {\hat S}^{k]}_T(\vec y \vert \vec x) $
\begin{eqnarray}
 \partial_T {\hat S}_T^{k]}(\vec y  \vert \vec x) 
&& = - \left[   \vec F( \vec x)  + k  \vec A^{[{\cal O}]}( \vec x)     - \vec \nabla {\hat S}^{k]}_T(\vec y \vert \vec x)\right]
   \left(  k \vec A^{[{\cal O}]}( \vec x) -   \vec \nabla {\hat S}^{k]}_T(\vec y \vert \vec x)   \right)  
  + k V^{[{\cal O}]} ( \vec x) 
\label{HamiltonJacobiBackwardk}
\end{eqnarray}


\subsubsection{ Properties of the classical trajectories } 

The classical moment $\vec p(t)$ obtained from the Lagrangian of Eq. \ref{lagrangianDk} with components
\begin{eqnarray}
 \vec p(t) \equiv \frac{\partial {\hat {\cal L}}_k (\vec x(t), \dot {\vec x}(t) ) } {\partial  \dot {\vec x}(t)} 
=\frac{ \dot {\vec x } (t)}{2}  - \vec a^{[k]}( \vec x(t)) 
=  \frac{ \dot  {\vec x } (t) - \vec F( \vec x) }{2} -k \vec A^{[{\cal O}]}( \vec x)
\label{lagrangianderidotk}
\end{eqnarray}
is useful to construct the classical electromagnetic Hamiltonian ${\hat {\cal H}}_k(\vec x, \vec p)  $ associated
 to the Lagrangian ${\hat {\cal L}}_k (\vec x, \dot {\vec x} ) $ of Eq. \ref{lagrangianDk}
via the Legendre transform 
\begin{eqnarray}
{\hat {\cal H}}_k(\vec x, \vec p) && = \vec p . \dot {\vec x} -  {\hat {\cal L}}_k (\vec x, \dot {\vec x})
 =  \left[ \vec p + \vec a^{[k]}(\vec x) \right]^2 - v^{[k]}(\vec x)
 =  {\vec p \ }^2+  2 \vec p . \vec a^{[k]}( \vec x) +  \left[ \vec a^{[k]}(\vec x) \right]^2 - v^{[k]}(\vec x)
\label{HamiltonienClassiquek}
\end{eqnarray}
with the following explicit expression in $k$ for the electromagnetic potentials of Eq. \ref{vectorscalarpotD} 
\begin{eqnarray}
{\hat {\cal H}}_k(\vec x, \vec p) 
&& =  \left[ \vec p + \frac{ \vec F( \vec x) }{2} +k \vec A^{[{\cal O}]}( \vec x) \right]^2 - \frac{ \vec F ^2( \vec x) }{ 4  }  - k V^{[{\cal O}]} ( \vec x)
\nonumber \\
&& =  \left[ \vec p  +k \vec A^{[{\cal O}]}( \vec x) \right]
 \left[ \vec p +  \vec F( \vec x)  +k \vec A^{[{\cal O}]}( \vec x) \right]   - k V^{[{\cal O}]} ( \vec x)
\label{HamiltonienClassiquekk}
\end{eqnarray}

The Hamiltonian classical equations of motion
for the $2d$ canonical coordinales $[x_{\mu}(t); p_{\mu}(t)]$ in phase-space 
\begin{eqnarray}
\dot x_{\mu}(t) && = \frac{\partial {\cal H}_k ({\vec x}(t),  {\vec p}  (t)) } {\partial  p_{\mu} (t)} 
\nonumber \\
      \dot p_{\mu}(t) &&  =  - \frac{\partial {\cal H}_k ({\vec x}(t),  {\vec p}  (t)) } {\partial  x_{\mu} (t)} 
\label{Hamilton12k}
\end{eqnarray}
ensure the conservation of energy along the classical trajectory
\begin{eqnarray}
{\hat {\cal E}}_k =  {\hat {\cal H}}_k ({\vec x}(t),  {\vec p}  (t)) 
&& =    \left[ \vec p (t) + \vec a^{[k]}(\vec x (t)) \right]^2 - v^{[k]}(\vec x (t))
=\frac{ \dot {\vec x }^2 (t) }{4}  - v^{[k]}(\vec x (t))
\nonumber \\
&& 
 = \frac{ \dot {\vec x }^2 (t) - \vec F^2( \vec x(t))}{4}  - k V^{[{\cal O}]} ( \vec x(t))
\label{HamiltonEnergyConservk}
\end{eqnarray}

The Lagrangian classical equations of motion 
\begin{eqnarray}
0    = \frac{ \ddot  x_{\mu}(t) }{2} 
+ \sum_{\nu=1}^d \dot  x_{\nu} (t)  b_{\mu \nu}^{[k]}( \vec x(t)) 
 - \frac{\partial v^{[k]}(\vec x(t)) } {\partial   x_{\mu}(t)} 
\label{lagrangeFPmotionDk}
\end{eqnarray}
involve the magnetic matrix $b_{\mu \nu}^{[k]}( \vec x) $ of Eq. \ref{magneticBD}.


\subsubsection{ Hamilton-Jacobi equations for the classical action ${\hat S}^{[k]}_T(\vec x  \vert \vec x_0) $ 
and for its time-energy Legendre transform ${\hat W}^{[k]}_{\hat {\cal E}}(\vec x  \vert \vec x_0) $ } 

The
Hamilton-Jacobi equation satisfied by the classical action
${\hat S}^{[k]}_T(\vec x, \vert \vec x_0)$
involves the classical Hamiltonian ${\hat {\cal H}}_k ({\vec x},  {\vec p} ) $ of Eq. \ref{HamiltonienClassiquekk}
where the momentum $ {\vec p} $ is replaced by the gradient of the action $  {\hat S}^{[k]}_T(\vec x  \vert \vec x_0) $ with respect to the final position $\vec x$
\begin{eqnarray}
\partial_T {\hat S}^{[k]}_T(\vec x  \vert \vec x_0) 
&& = - {\hat {\cal H}}_k ({\vec x},  {\vec p} = \vec \nabla {\hat S}^{[k]}_T(\vec x  \vert \vec x_0))
 \nonumber \\
 && =  - \left[  \vec \nabla {\hat S}^{[k]}_T(\vec x  \vert \vec x_0)  +k \vec A^{[{\cal O}]}( \vec x) \right]
 \left[  \vec \nabla {\hat S}^{[k]}_T(\vec x  \vert \vec x_0) +  \vec F( \vec x)  +k \vec A^{[{\cal O}]}( \vec x) \right]   + k V^{[{\cal O}]} ( \vec x)
 \label{HamiltonJacobikbis}
\end{eqnarray}
that coincides with Eq. \ref{HamiltonJacobik}
as its should for consistency.
Similarly, Eq. \ref{HamiltonJacobiBackwardk} corresponds to the backward 
Hamilton-Jacobi equation for the classical action ${\hat S}_T(\vec y  \vert \vec x) $
when $\vec x$ is the initial position
\begin{eqnarray}
\partial_T {\hat S}^{[k]}_T(\vec y  \vert \vec x) 
&& = - {\hat {\cal H}}_k \left({\vec x},  {\vec p} = - \vec \nabla {\hat S}_T^{[k]}(\vec y  \vert \vec x) \right)
 \nonumber \\
 && =  - \left[   \vec F( \vec x)  + k  \vec A^{[{\cal O}]}( \vec x)     - \vec \nabla {\hat S}^{k]}_T(\vec y \vert \vec x)\right]
   \left(  k \vec A^{[{\cal O}]}( \vec x) -   \vec \nabla {\hat S}^{k]}_T(\vec y \vert \vec x)   \right)  
  + k V^{[{\cal O}]} ( \vec x) 
\label{HamiltonJacobiBackkwardbis}
\end{eqnarray}

The time-energy Legendre transform between the classical action ${\hat S}_T^{[k]}(\vec x  \vert \vec x_0) $
and Hamilton's characteristic function ${\hat W}^{[k]}_{\hat {\cal E}}(\vec x  \vert \vec x_0) $
\begin{eqnarray}
{\hat W}^{[k]}_{\hat {\cal E}}(\vec x  \vert \vec x_0) && = {\hat S}^{[k]}_T(\vec x  \vert \vec x_0) + {\hat {\cal E}} T
 \nonumber \\
 0 && = \partial_T  {\hat S}^{[k]}_T(\vec x  \vert \vec x_0) + {\hat {\cal E}} 
\label{HJLegendrek}
\end{eqnarray}
with its reciprocal Legendre transformation
\begin{eqnarray}
 {\hat S}_T^{[k]}(\vec x  \vert \vec x_0) && = {\hat W}^{[k]}_{\hat {\cal E}}(\vec x  \vert \vec x_0) - {\hat {\cal E}} T
  \nonumber \\
 0 && = \partial_{{\hat {\cal E}}} {\hat W}^{[k]}_{\hat {\cal E}}(\vec x  \vert \vec x_0) - T 
\label{HJLegendrerecik}
\end{eqnarray}
is useful to transform the time-dependent Hamilton-Jacobi Eq. \ref{HamiltonJacobikbis} for ${\hat S}^{[k]}_T(\vec x  \vert \vec x_0) $
into the energy-dependent Hamilton-Jacobi equation for ${\hat W}^{[k]}_{\hat {\cal E}}(\vec x  \vert \vec x_0) $
\begin{eqnarray}
 {\hat {\cal E}} 
&& =  {\hat {\cal H}}^{[k]} \left({\vec x},  {\vec p} = \vec \nabla {\hat W}^{[k]}_{\hat {\cal E}}(\vec x  \vert \vec x_0) \right)
 \nonumber \\
 && =    \left[  \vec \nabla {\hat W}^{[k]}_{\hat {\cal E}}(\vec x  \vert \vec x_0)  +k \vec A^{[{\cal O}]}( \vec x) \right]
 \left[  \vec \nabla {\hat W}^{[k]}_{\hat {\cal E}}(\vec x  \vert \vec x_0) +  \vec F( \vec x)  +k \vec A^{[{\cal O}]}( \vec x) \right]  
  - k V^{[{\cal O}]} ( \vec x)
\label{HamiltonJacobiWk}
\end{eqnarray}

Similarly, the backward Eq. \ref{HamiltonJacobiBackkwardbis} satisfied by the $ {\hat S}^{[k]}_T(\vec y \vert \vec x) $
can be translated for ${\hat W}^{[k]}_{\hat {\cal E}}(\vec y  \vert \vec x) $ into
\begin{eqnarray}
 {\hat {\cal E}} && =  {\hat {\cal H}}^{[k]} \left({\vec x},  {\vec p} = - \vec \nabla {\hat W}^{[k]}_{\hat {\cal E}}(\vec y  \vert \vec x) \right)
 \nonumber \\
&& =    \left(  - \vec \nabla  {\hat W}^{[k]}_{\hat {\cal E}}(\vec y  \vert \vec x)  + k \vec A^{[{\cal O}]}( \vec x) \right)  
 \left[  - \vec \nabla  {\hat W}^{[k]}_{\hat {\cal E}}(\vec y  \vert \vec x)    + k  \vec A^{[{\cal O}]}( \vec x)  + \vec F( \vec x) \right]
  - k V^{[{\cal O}]} ( \vec x) 
\label{HamiltonJacobiWkbackward}
\end{eqnarray}


\subsubsection{Rate function ${\hat I}_T({\cal O} ;\vec x  \vert \vec x_0) $ for the joint probability distribution ${\cal P}_T({\cal O} ; \vec x \vert \vec x_0 ) $ for small $D \to 0$ }

For small diffusion coefficient $D \to 0$, the joint probability distribution ${\cal P}_T({\cal O} ; \vec x \vert \vec x_0 ) $
of Eq. \ref{pathintegralPO}
is expected to display the large deviation form
 \begin{eqnarray}
 {\cal P}_T({\cal O} ; \vec x \vert \vec x_0 ) \oppropto_{ D \to 0} e^{ \displaystyle 
 - \frac{{\hat I}_T({\cal O} ;\vec x  \vert \vec x_0)}{D}  }
    \label{ratefunctionD}
\end{eqnarray}
 where the positive rate function ${\hat I}_T({\cal O} ;\vec x  \vert \vec x_0) \geq 0 $
vanishes only if the zero-noise deterministic trajectory ${\vec x }^{deter} (t) $ of Eq. \ref{detertraj}
is at position ${\vec x }^{deter} (t=T)  =\vec x $ at time $T$,
 and if ${\cal O} $
corresponds to the value of Eq. \ref{additive}
associated to this deterministic trajectory
\begin{eqnarray}
{\cal O} [\vec x^{deter}(0 \leq t \leq T) ]  && = \int_0^T dt \left[   - V^{[{\cal O}]} (\vec x^{deter}(t) ) 
+  \vec A^{[{\cal O}]}( \vec x^{deter}(t)) . \dot {\vec x}^{deter}(t)  \right]
\nonumber \\
&& = \int_0^T dt \left[   - V^{[{\cal O}]} (\vec x^{deter}(t) ) 
+  \vec A^{[{\cal O}]}( \vec x^{deter}(t)) . \vec F( \vec x^{deter}(t))  \right]
\label{additivedeter}
\end{eqnarray}
For all the other cases $\big({\cal O} ; \vec x \big) \ne \big( {\cal O} [\vec x^{deter}(0 \leq t \leq T) ] ; {\vec x }^{deter} (T)\big)$, the rate 
function ${\hat I}_T({\cal O} ;\vec x  \vert \vec x_0) >0  $ characterizes how rare it is 
for small $D$ to see the joint values $\big({\cal O} ; \vec x \big)$ at time $T$ via Eq. \ref{ratefunctionD}.

For small $D$, the evaluation via the saddle-point method
of the generating function of Eq. \ref{gene} using the large deviation form of Eq. \ref{ratefunctionD}
 \begin{eqnarray}
 Z^{[k]}_T (\vec x \vert \vec x_0 )
 \equiv  \int d {\cal O}  \   {\cal P}_T({\cal O} ; \vec x \vert \vec x_0 ) \ e^{ \displaystyle \frac{k}{D}{\cal O}}
 \oppropto_{D \to 0}  \int d {\cal O} \ e^{ \displaystyle \frac{1}{D} \left[ k {\cal O} - {\hat I}_T({\cal O} ;\vec x  \vert \vec x_0) \right] }
 \oppropto_{D \to 0} e^{ \displaystyle -    \frac{ {\hat S}_T^{[k]}(\vec x  \vert \vec x_0) }{ D} }
 \label{genesaddleD}
\end{eqnarray}
  leads to the Legendre transformation
  \begin{eqnarray}
{\hat I}_T({\cal O} ;\vec x  \vert \vec x_0)- k {\cal O}  && = {\hat S}_T^{[k]}(\vec x  \vert \vec x_0)
 \nonumber \\
 \partial_{\cal O} {\hat I}_T({\cal O} ;\vec x  \vert \vec x_0) - k   && = 0
 \label{legendreD}
\end{eqnarray}
with the reciprocal Legendre transformation
   \begin{eqnarray}
{\hat I}_T({\cal O} ;\vec x  \vert \vec x_0)&& = k {\cal O}  + {\hat S}_T^{[k]}(\vec x  \vert \vec x_0)
 \nonumber \\
 0   && = {\cal O}  + \partial_k {\hat S}_T^{[k]}(\vec x  \vert \vec x_0)
 \label{legendrereciD}
\end{eqnarray}


\section{ Large deviations for small diffusion $D\to 0$ and then large time $T\to + \infty$  }

\label{seC_smallDlargeT}

In this section, we describe how the properties for small diffusion coefficient $D\to 0$ 
and finite time $T$
described in the previous section \ref{seC_largeT}
can be further simplified when the time-window becomes large $T\to + \infty$.

\subsection{Propagator $ P_T(\vec x \vert \vec x_0)\propto  
 e^{ -    \frac{ {\hat S}_T(\vec x  \vert \vec x_0) }{ D} }  $ 
 governed by the action ${\hat S}_T(\vec x  \vert \vec x_0) $
of a long classical trajectory $T \to + \infty$   }

 For finite time $T$, the initial position $\vec x_0$ and the final position $\vec x$ 
 are strongly correlated in the action ${\hat S}_T(\vec x  \vert \vec x_0) $ 
 as a consequence of the finite-time classical trajectory $x^{Class}( 0 \leq t \leq T)$ connecting them, 
 and the corresponding conserved energy of this classical trajectory will depend on $(\vec x, \vec x_0,T)$.
 However for very long classical trajectories $T \to + \infty$,
 the best strategy to minimize the action is 
 to remain, for the most part of the time-window $[0,T]$, as close as possible 
 to the deterministic classical trajectory of Eq. \ref{detertraj}
 that would make the action vanish (Eq \ref{ActionclassicalVanish}) and that corresponds to zero energy 
 $ {\hat {\cal E}}=0 $ in the Hamiltonian language of Eq. \ref{detertrajHamilton}.
 As a consequence for large time $ T \to + \infty$, the Legendre transformation of Eq. \ref{HJLegendre}
 yields that 
 the classical action ${\hat S}_T(\vec x  \vert \vec x_0) $
does not depend on the time $T$
\begin{eqnarray}
{\hat S}_T(\vec x  \vert \vec x_0) \opsimeq_{ T \to +\infty} {\hat W}_{\hat {\cal E}=0}(\vec x  \vert \vec x_0) 
\label{HJLegendreE0}
\end{eqnarray}
and reduces to the zero-energy Hamilton's characteristic function ${\hat W}_{\hat {\cal E}=0}(\vec x  \vert \vec x_0) $
satisfying the 
forward Hamilton-Jacobi equation of Eq. \ref{HamiltonJacobiW}
\begin{eqnarray}
0 =  \left[ \vec F( \vec x)  + \vec \nabla {\hat W}_0(\vec x  \vert \vec x_0)\right] .  \vec \nabla {\hat W}_0(\vec x  \vert \vec x_0)
\label{HamiltonJacobiWZeroE}
\end{eqnarray}
and the backward Hamilton-Jacobi equation of Eq. \ref{HamiltonJacobiWbackward}
\begin{eqnarray}
0 =    - 
 \left[  \vec F( \vec x_0) - \frac{ \partial {\hat W}^{[k]}_0(\vec x  \vert \vec x_0)}{ \partial \vec x_0}       \right]
 \frac{ \partial {\hat W}_0(\vec x  \vert \vec x_0)}{ \partial \vec x_0}    
\label{HamiltonJacobiWbackwardZeroE}
\end{eqnarray}

The small-$D$ propagator of Eq. \ref{pathintegralDclassical} 
is then governed by ${\hat W}_0(\vec x  \vert \vec x_0) $ for large time $T \to +\infty$
\begin{eqnarray}
P_T(\vec x \vert \vec x_0) 
\oppropto_{ D \to 0}  
 e^{ - \displaystyle   \frac{ {\hat S}_T(\vec x  \vert \vec x_0) }{ D} } 
 \oppropto_{ T \to + \infty }  
 e^{ - \displaystyle   \frac{ {\hat W}_0(\vec x  \vert \vec x_0) }{ D} } 
\label{pathintegralDclassicalTlarge}
\end{eqnarray}

The comparison with the steady state $ p_*(\vec x) $ of Eq. \ref{rhosteadyD}
yields that ${\hat W}_0(\vec x  \vert \vec x_0) $ should coincide with $u_* (\vec x) $
\begin{eqnarray}
{\hat W}_0(\vec x  \vert \vec x_0) = u_* (\vec x)
\label{identifWU}
\end{eqnarray}
So ${\hat W}_0(\vec x  \vert \vec x_0) $ is actually independent of the initial position $\vec x_0 $,
which is indeed the simplest way to solve the backward of Hamilton-Jacobi Eq. \ref{HamiltonJacobiWbackwardZeroE},
while the forward Hamilton-Jacobi equation of Eq. \ref{HamiltonJacobiWZeroE}
then coincides with Eq. \ref{divforceirrevD1} satisfied by $u_*(\vec x)$.


\subsection{Generating function $ Z_T^{[k]}(\vec x \vert \vec x_0)\propto  
 e^{ -    \frac{ {\hat S}_T^{[k]}(\vec x  \vert \vec x_0) }{ D}
 }  $  
 governed by the action ${\hat S}_T^{[k]}(\vec x  \vert \vec x_0) $
of a long classical trajectory   }

 Similarly for large time $ T \to + \infty$, one expects that the boundary conditions
 of Eq. \ref{BCclassical} will only affect the very beginning and the very final part of the classical trajectory,
 while the main part of the time-window $[0,T]$ will optimize the classical action ${\hat S}_T^{[k]}(\vec x  \vert \vec x_0) $
 with a well-defined classical energy $\hat {\cal E}_k$ in the transformation of Eq. \ref{HJLegendrek}
 \begin{eqnarray}
{\hat S}^{[k]}_T(\vec x  \vert \vec x_0) \opsimeq_{T \to + \infty} 
{\hat W}^{[k]}_{\hat {\cal E}_k}(\vec x  \vert \vec x_0) - {\hat {\cal E}}_k T
\label{HJLegendrekTlarge}
\end{eqnarray}
where  ${\hat W}_{\hat {\cal E}_k}(\vec x  \vert \vec x_0) $ satisfies Eqs \ref{HamiltonJacobiWk} and \ref{HamiltonJacobiWkbackward}.

The small-$D$ generating function of of Eq. \ref{geneD}
 then becomes for large time $T \to +\infty$
 \begin{eqnarray}
 Z^{[k]}_T(\vec x \vert \vec x_0) 
 \opsimeq_{ D \to 0}  
 e^{ - \displaystyle   \frac{ {\hat S}_T^{[k]}(\vec x  \vert \vec x_0) }{ D}
 } \opsimeq_{ T \to +\infty}  
 e^{  \displaystyle \frac{   T }{ D}{\hat {\cal E}}_k
  - \frac{ {\hat W}^{[k]}_{\hat {\cal E}_k}(\vec x  \vert \vec x_0)  }{ D} } 
 \label{geneDTlarge}
\end{eqnarray}

The comparison with Eq. \ref{genelargeTD}
leads to the identifications
 \begin{eqnarray}
 {\hat {\cal E}}_k && = - e(k)
 \nonumber \\
 {\hat W}^{[k]}_{\hat {\cal E}_k}(\vec x  \vert \vec x_0) && = u_k(\vec x) + {\tilde u}_k(\vec x_0)
\label{identificationk}
\end{eqnarray}
that can be plugged into the Hamilton-Jacobi Eqs \ref{HamiltonJacobiWk} 
\begin{eqnarray}
- e(k)
&& =  {\hat {\cal H}}^{[k]} \left({\vec x},  {\vec p} = \vec \nabla u_k(\vec x) \right)
 \nonumber \\
 && =    \left[   \vec \nabla u_k(\vec x) +k \vec A^{[{\cal O}]}( \vec x) \right]
 \left[   \vec \nabla u_k(\vec x) +  \vec F( \vec x)  +k \vec A^{[{\cal O}]}( \vec x) \right]  
  - k V^{[{\cal O}]} ( \vec x)
\label{HamiltonJacobiWkbis}
\end{eqnarray}
and into \ref{HamiltonJacobiWkbackward}
\begin{eqnarray}
 - e(k) && =  {\hat {\cal H}}^{[k]} \left({\vec x},  {\vec p} = - \vec \nabla {\tilde u}_k ( \vec x) \right)
 \nonumber \\
&& =    \left(  - \vec \nabla  {\tilde u}_k ( \vec x)  + k \vec A^{[{\cal O}]}( \vec x) \right)  
 \left[  - \vec \nabla  {\tilde u}_k ( \vec x)   + k  \vec A^{[{\cal O}]}( \vec x)  + \vec F( \vec x) \right]
  - k V^{[{\cal O}]} ( \vec x) 
\label{HamiltonJacobiWkbackwardbis}
\end{eqnarray}
to recover Eqs \ref{eigenrightlogD} and \ref{eigenleftlogD} satisfied by $u_k(\vec x) $ and by ${\tilde u}_k(\vec x) $ respectively.


\subsection{Joint distribution $ {\cal P}_T({\cal O} ; \vec x \vert \vec x_0 ) \propto e^{  
 - \frac{{\hat I}_T({\cal O} ;\vec x  \vert \vec x_0)}{D}  }  $ 
 governed by the rate function ${\hat I}_T({\cal O} ;\vec x  \vert \vec x_0) $ for $T \to + \infty$  }

Plugging the large-time bahaviors of Eqs \ref{HJLegendrekTlarge} and \ref{identificationk}
for the classical action ${\hat S}_T^{[k]}(\vec x  \vert \vec x_0) $
  \begin{eqnarray}
{\hat S}^{[k]}_T(\vec x  \vert \vec x_0) \opsimeq_{T \to + \infty} 
 u_k(\vec x) + {\tilde u}_k(\vec x_0) + T e(k)
\label{ActionSkTlarge}
\end{eqnarray}
 into the Legendre transformation of Eq. \ref{legendrereciD}
 yields that the rate function ${\hat I}_T({\cal O} ;\vec x  \vert \vec x_0) $ displays the large-time behavior
   \begin{eqnarray}
{\hat I}_T({\cal O} ;\vec x  \vert \vec x_0) \opsimeq_{T \to + \infty} T i \left( O = \frac{{\cal O}}{T} \right)
 \label{rateDpuisT}
\end{eqnarray}
where the rescaled rate function $i(O) $ of the intensive trajectory observable $O = \frac{{\cal O}}{T} $ satisfies 
    \begin{eqnarray}
 i(O) && = k  O  + e(k)
 \nonumber \\
 0   && = O  +  e'(k)
 \label{legendrereciDpuisT}
\end{eqnarray}
 in agreement with Eq. \ref{legendrerecirescaped} found via the other order of limits.


\section{ Application to dimension $d=2$ with rotational invariance  }

\label{sec_2DPolar}

In this section, the general formalism described in the previous sections is applied
to the simplest case in dimension $d>1$, namely
the dimension $d=2$ in the presence of rotational invariance.

\subsection{ Two-dimensional diffusion with rotational invariance in polar coordinates  }

\subsubsection{ The three decompositions of the force $\vec F(\vec x)$ coincide with the radial-orthoradial decomposition}

In polar coordinates $(r,\theta)$ with the corresponding orthonormal basis $(\vec e_r,\vec e_{\theta})$,
the rotational invariance means that both
the radial component $F_r (r,\theta )= F_r (r )$ and the orthoradial component $F_{\theta} (r,\theta)=F_{\theta} (r) $ of the force $\vec F(\vec x) $ depend only on the radial coordinate $r$ and not on the angle $\theta$
\begin{eqnarray}
\vec F(\vec x) && = \vec F^{Radial}(\vec x) + \vec F^{Ortho}(\vec x)
\nonumber \\
\vec F^{Radial}(\vec x) && \equiv F_r (r ) \vec e_r 
\nonumber \\
\vec F^{Ortho}(\vec x) && \equiv F_{\theta} (r) \vec e_{\theta} 
\label{forcePolar}
\end{eqnarray}

It is then useful to introduce the radial potential 
\begin{eqnarray}
U(r) && \equiv - \int_0^r d F_r (r )  
\nonumber \\
 U'(r) && = - F_r(r)
\label{forcePolarderi}
\end{eqnarray}
in order to rewrite the radial contribution of the force in terms of the gradient of $U(r)$
\begin{eqnarray}
\vec F^{Radial}(\vec x) \equiv F_r (r ) \vec e_r = - U'(r) \vec e_r = - \vec \nabla U(r)
\label{forceradialgradient}
\end{eqnarray}
The orthoradial component $ \vec F^{Ortho}(\vec x) $
is of course orthogonal to the radial component of Eq. \ref{forceradialgradient} 
and its divergence vanishes since $F_{\theta} (r) $ does not depend on $\theta$
\begin{eqnarray}
\vec F^{Ortho}(\vec x) . \vec \nabla U(r) && =0
\nonumber \\
\nabla . \left( \vec F^{Ortho}(\vec x) \right) &&=0
\label{forcePolarprop}
\end{eqnarray}
As a consequence for the force of Eq. \ref{forcePolar}, the orthogonal decomposition of Eq. \ref{forceNormalDecomposition},
the Hemlholtz decomposition of Eq. \ref{forcetothelmholtz3d}, and the reversible-irreversible decomposition of Eq. \ref{divforceirrevgrad}
  coincide
\begin{eqnarray}
   U^{\perp} (\vec x) = U^H(\vec x) = U_*(\vec x) && = U(r)
   \nonumber \\
   \vec F^{\perp} (\vec x) = \vec F^H(\vec x) = \vec F^{irr}(\vec x) && =F_{\theta} (r) \vec e_{\theta}
   \label{Upolarcoincide}
\end{eqnarray}

So the steady state of Eq. \ref{rhostarfromurev} depends only on the radial coordinate $r$
\begin{eqnarray}
P_*(\vec x)  = \frac{ e^{ - \frac{ U(r) }{  D }  } }{  \int_0^{+\infty} dr' 2 \pi r' e^{- \frac{U(r')}{D} }} \equiv P_*(r)
\label{Polarsteady}
\end{eqnarray}
while the steady current of Eq. \ref{jsteadyirrev} is orthoradial 
\begin{eqnarray}
  \vec J^*(r,\theta)  =  P_*(r ) F_{\theta} (r) \vec e_{\theta} \equiv J^*_{\theta}(r) \vec e_{\theta}
 \label{Polarjsteadyirrev}
\end{eqnarray}
with the amplitude $J^*_{\theta}(r)= P_*(r ) F_{\theta} (r)$ depending only on $r$.


\subsubsection{ Fokker-Planck generator : Scalar and vector potentials of the euclidean non-Hermitian quantum Hamiltonian $H$   }

The scalar potential $V(\vec x) $ of Eq. \ref{scalarpot}
associated to the force of Eq. \ref{forcePolar} 
\begin{eqnarray}
V(\vec x) \equiv  \frac{ \vec F ^2( \vec x) }{ 4D  }   + \frac{  [\vec \nabla . \vec F(\vec x) ]  }{2 }   
= \frac{ [U'(r)]^2 + [F_{\theta} (r)]^2}{ 4D  }  - \frac{1}{2 r} \frac{d [ r U'(r) ]}{dr}   \equiv V(r)
\label{scalarpotpolar}
\end{eqnarray}
depends only on the radial coordinate $r$ as expected.

The vector potential of Eq. \ref{vectorpot} associated to the force of Eq. \ref{forcePolar}
\begin{eqnarray}
  \vec A ( \vec x ) \equiv  \frac{ \vec F(\vec x ) }{2D} =  - \frac{U'(r) }{2D}\vec e_r + \frac{F_{\theta} (r) }{2D}\vec e_{\theta}
 \label{vectorpotpolar}
\end{eqnarray}
corresponds to the following magnetic field
in the direction orthogonal to the plane 
\begin{eqnarray}
 B(\vec x)  = [\vec \nabla \times \vec A(\vec x) ]. \vec e_3
  =  \frac{1}{2D r} \frac{d [  r  F_{\theta} (r) ]}{dr}  \equiv B(r)
 \label{magneticBpolar}
\end{eqnarray}
that depends only on the radial coordinate $r$ as expected.


\subsection{ Trajectory observables respecting the rotational invariance  }

We will focus on trajectory observables respecting the rotational invariance,
so the parametrization of Eq. \ref{additive}
\begin{eqnarray}
{\cal O} [ r (0 \leq t \leq T) ; \theta (0 \leq t \leq T)]  \equiv \int_0^T dt \left[   - V^{[{\cal O}]} (r(t) ) 
+  A^{[{\cal O}]}_r( r)  \dot r(t) 
+   A^{[{\cal O}]}_{\theta}( r)  r \dot \theta (t) \right]
\label{additivepolar}
\end{eqnarray}
involves a scalar potential $V^{[{\cal O}]} (r ) $ depending only $r$, 
and a vector potential whose radial component $A^{[{\cal O}]}_r( r) $ and 
orthoradial component $A^{[{\cal O}]}_{\theta}( r) $ depend only on $r$.


\subsubsection{ Scalar and vector potentials of the $k$-deformed euclidean non-Hermitian quantum Hamiltonian $H_k$  }

The generating function $  Z^{[k]}_T(\vec x \vert \vec x_0) $ of Eq. \ref{gene}
is associated to the following quantum problem.
The $k$-deformed scalar potential of Eq. \ref{scalarpotp} with respect to Eq. \ref{scalarpotpolar}
\begin{eqnarray}
V^{[k]}(r) \equiv V ( r) + \frac{k}{D} V^{[{\cal O}]} ( r) 
=  \frac{ [U'(r)]^2 + [F_{\theta} (r)]^2}{ 4D  }  - \frac{1}{2 r} \frac{d [ r U'(r)]}{dr} 
+ \frac{k}{D} V^{[{\cal O}]} ( r) 
\label{scalarpotppolar}
\end{eqnarray}
depends only on $r$,
while the $k$-deformed vector potential of Eq. \ref{vectorpotp} with respect to Eq. \ref{vectorpotpolar}
has radial and orthoradial components that depend only on $r$
\begin{eqnarray}
A^{[k]}_r (r ) && \equiv A_r( r) +\frac{k}{D} A^{[{\cal O}]}_r(r)
=- \frac{U'(r) }{2D} +\frac{k}{D} A^{[{\cal O}]}_r(r)
 \nonumber \\
A^{[k]}_{\theta} (r ) && \equiv A_{\theta}( r) +\frac{k}{D}  A^{[{\cal O}]}_{\theta}(r) 
=  \frac{F_{\theta}( r) }{2D} +\frac{k}{D}  A^{[{\cal O}]}_{\theta}(r) 
 \label{vectorpotppolark}
\end{eqnarray}
with the $k$-deformed magnetic field with respect to Eq. \ref{magneticBpolar}
\begin{eqnarray}
 B^{[k]}(r)  =  \frac{1 }{  r} \frac{d [ r A^{[k]}_{\theta} (r )]}{dr} 
  =  \frac{1}{2D r} \frac{d [ r  F_{\theta} (r) ]}{dr} 
  +   \frac{k }{ D r} \frac{d [ r A^{[{\cal O}]}_{\theta} (r ) ]}{dr} 
  \label{magneticBpolark}
\end{eqnarray}


\subsubsection{ Gauge transformation towards the Coulomb gauge where the new vector potential $ \vec {\mathring A}^{[k]}(\vec x ) $ is orthoradial }

As explained in Appendix \ref{app_gauge}, it is useful in practice to make
a gauge transformation of the form of Eq. \ref{vectorpotgaugeirr}
in order to obtain a simpler vector potential $\vec {\mathring A}^{[k]}(\vec x ) $.
In the present case where the radial component of the vector potential of Eq. \ref{vectorpotppolark}
can be rewritten the gradient of some function $ {\mathring \Phi}_k(r)$
\begin{eqnarray}
A^{[k]}_r (r ) \vec e_r && \equiv - \vec \nabla  {\mathring \Phi}_k(r) = -  {\mathring \Phi}_k'(r) \vec e_r
 \nonumber \\
 {\mathring \Phi}_k'(r) && \equiv - A_r( r) - \frac{k}{D} A^{[{\cal O}]}_r(r)
= \frac{U'(r) }{2D} - \frac{k}{D} A^{[{\cal O}]}_r(r)
 \label{vectorpotppolar}
\end{eqnarray}
it is convenient to consider the gauge transformation of Eq. \ref{vectorpotgaugeirr}
\begin{eqnarray}
  \vec A^{[k]} (\vec x )   = - \vec \nabla  {\mathring \Phi}_k(r) + \vec {\mathring A}^{[k]}(\vec x )
 \label{vectorpotgaugeirrpolar}
\end{eqnarray}
where the new vector potential $\vec {\mathring A}^{[k]}(\vec x ) $ 
producing the same magnetic field $B^{[k]}(r)  $ of Eq. \ref{magneticBpolark}
 reduces to the orthoradial component of Eq. \ref{vectorpotppolark}
\begin{eqnarray}
\vec {\mathring A}^{[k]}(\vec x ) \equiv  A^{[k]}_{\theta} (r ) \vec e_{\theta}
    \label{vectorpotirr}
\end{eqnarray}
and is divergenceless
\begin{eqnarray}
\vec \nabla . \vec {\mathring A}^{[k]}(\vec x ) =0
    \label{divvectorpotirr}
\end{eqnarray}
so that it corresponds to the standard choice of the Coulomb gauge.

The corresponding Hamiltonian of Eq. \ref{FPhamiltonianring} 
\begin{eqnarray}
{\mathring H}_k &&   =  -  D \left( \vec \nabla -   \vec {\mathring A}^{[k]}(\vec x) \right)^2    +  V^{[k]}(\vec x)
=  - D \vec \nabla^2 
+2 D \vec {\mathring A}^{[k]} ({\vec x}). \vec \nabla
+ D   \left( \vec \nabla.\vec {\mathring A}^{[k]} ({\vec x})\right) 
 - D  [\vec {\mathring A}^{[k]}]^2 (\vec x)  + V^{[k]} (\vec x)
 \label{FPhamiltonianringexpanded}
\end{eqnarray}
reads in polar coordinates using Eqs \ref{vectorpotirr} and \ref{divvectorpotirr}
\begin{eqnarray}
 {\mathring H}_k    
 =  - D \left[ \frac{1}{r} \frac{\partial}{\partial r} r \frac{\partial}{\partial r} + \frac{1}{r^2} \frac{\partial^2}{\partial \theta^2 } \right] 
  + 2 D   \frac{ A^{[k]}_{\theta} (r )  }{r} \frac{\partial}{\partial \theta }  
 - D [  A^{[k]}_{\theta} (r ) ]^2
    +  V^{[k]}(r)  
 \label{FPhamiltonianringexpandedpolar}
\end{eqnarray}
with its adjoint operator
\begin{eqnarray}
   {\mathring H}^{\dagger}_k  =  - D \left[ \frac{1}{r} \frac{\partial}{\partial r} r \frac{\partial}{\partial r} + \frac{1}{r^2} \frac{\partial^2}{\partial \theta^2 } \right] 
  - 2 D   \frac{ A^{[k]}_{\theta} (r )  }{r} \frac{\partial}{\partial \theta }  
 - D [  A^{[k]}_{\theta} (r ) ]^2
    +  V^{[k]}(r)  
    \label{FPhamiltonianringexpandedpolaradjoint}
\end{eqnarray}


\subsection{ Large deviations  for large time $T \to + \infty$ and finite diffusion coefficient $D$  }

The steady value $O_*$ of the intensive observable of Eqs \ref{additiveSteady} and \ref{additiveSteadyrho}
reads in terms of the steady state $P_*(r) $ of Eq. \ref{Polarsteady}
 \begin{eqnarray}
O_* &&  =  \int d^2 \vec x P_*(\vec x) \left[   - V^{[{\cal O}]} (\vec x )  +  \vec A^{[{\cal O}]}( \vec x) .  \vec F^{irr}(\vec x )  \right]
\nonumber \\
&& = \int_0^{+\infty} dr 2 \pi r P_*(r) \left[   - V^{[{\cal O}]} (r )  +  A^{[{\cal O}]}_{\theta}( r)  F_{\theta} (r)  \right]
\label{additiveSteadyrhoPolar}
\end{eqnarray}
In this subsection, the goal is to analyze the fluctuations around this steady value for large time $T \to + \infty$.

\subsubsection{ Ground state energy $E(k)$ of the Hamiltonian ${\mathring H}_k $ via eigenvalues equations}

As discussed in Eq. \ref{eigenrightleftring} of the Appendix, 
the ground state energy $E(k)$ of the initial Hamiltonian $H_k$
can be equivalently studied for the gauge-transformed Hamiltonian ${\mathring H}_k $
with the change of eigenvectors of Eq. \ref{changetowardshatrl}
that reads in our present case where ${\mathring \Phi}_k(\vec x) $ depends only on $r$
\begin{eqnarray}
 r_k(r,\theta)  && = e^{ - {\mathring \Phi}_k(r)}  {\mathring r}_k (r,\theta ) 
 \nonumber \\
  l_k(r,\theta)  && = e^{  {\mathring \Phi}_k(r)}  {\mathring l}_k (r,\theta ) 
\label{changetowardshatrlPolar}
\end{eqnarray}
The commutation of the Hamiltonian $ {\mathring H}_k $ of Eq. \ref{FPhamiltonianringexpandedpolar}
with the operator $\frac{\partial}{\partial \theta }   $ 
\begin{eqnarray}
[ {\mathring H}_k   , \frac{\partial}{\partial \theta }  ]=0
 \label{commutation}
\end{eqnarray}
is associated to the rotational invariance.
As a consequence, these two operators can be diagonalized simultaneously,
so that their common eigenvectors depend on the angle $\theta$ via factors $e^{i m \theta}$  with integers $m$
in order to ensure the periodicity $\theta \to \theta+2 \pi$.
In particular, the positive right and left eigenvectors 
associated to the ground state energy $E(k)$
are expected to depend only on $r$ (case $m=0$).
Since the Hamiltonian ${\mathring H}_k $ of Eq. \ref{eigenrightleftringpolar}
and its adjoint of Eq. \ref{FPhamiltonianringexpandedpolaradjoint}
coincide in this sector $m=0$ where they reduce to the radial hermitian Hamiltonian
\begin{eqnarray}
\text{ Sector } m=0 : \ \ \  {\mathring H}^{Radial}_k 
= \left( {\mathring H}^{Radial}_k \right)^{\dagger}
 =  - D  \frac{1}{r} \frac{d}{d r} r \frac{d}{d r} 
 - D [  A^{[k]}_{\theta} (r ) ]^2    +  V^{[k]}(r)  
 \label{FPhamiltonianringexpandedpolarm0}
\end{eqnarray}
the right and left eigenvectors coincide
\begin{eqnarray}
 {\mathring r}_k (r ) && =   {\mathring \psi}_k( r)
 \nonumber \\
   {\mathring l}_k( r) && =   {\mathring \psi}_k( r)
\label{eigenrightleftringpolar}
\end{eqnarray}
where ${\mathring \psi}_k( r) $ is the positive eigenvector of the radial Hamiltonian of Eq. \ref{FPhamiltonianringexpandedpolarm0} associated to the lowest eigenvalue $E(k)$
\begin{eqnarray}
E(k)  {\mathring \psi}_k( r)  && =  {\mathring H}^{Radial}_k {\mathring \psi}_k( r)
\nonumber \\
&& = \left[ - D  \frac{1}{r} \frac{d}{d r} r \frac{d}{d r}  
     +  V^{[k]}(r) - D [  A^{[k]}_{\theta} (r ) ]^2 \right ] {\mathring \psi}_k( r)
\label{eigenpsipolar}
\end{eqnarray}
that reads using the explicit expressions of 
the $k$-deformed scalar  potential $V^{[k]}(r) $ of Eq. \ref{scalarpotppolar}
and of the $k$-deformed scalar  potential$  A^{[k]}_{\theta} (r )$
of Eq. \ref{vectorpotppolark}
\begin{eqnarray}
E(k)  {\mathring \psi}_k( r)  && =  - D  \frac{1}{r} \frac{d}{d r} \left( r \frac{d {\mathring \psi}_k( r)}{d r} \right)
\nonumber \\
&& + \left[  \frac{ [U'(r)]^2 + [F_{\theta} (r)]^2 - \left(  F_{\theta}( r) +2 k A^{[{\cal O}]}_{\theta}(r)  \right)^2}{ 4D  }  
+ \frac{k}{D} V^{[{\cal O}]} ( r)  
- \frac{1}{2 r} \frac{d [ r U'(r)]}{dr}       \right ] {\mathring \psi}_k( r)
\label{eigenrightleftringpolarexplicit}
\end{eqnarray}
while the normalization of Eq. \ref{eigennormaring} becomes
\begin{eqnarray}
1 = \int d^d \vec x \   l_k( \vec x)  r_k( \vec x) =\int d^d \vec x \ {\mathring l}_k (\vec x )  {\mathring r}_k (\vec x ) 
= \int d^d \vec x  {\mathring \psi}^2_k( \vec x)
= \int_0^{+\infty} dr 2 \pi r {\mathring \psi}^2_k( \vec x)
  \label{eigennormaringpolar}
\end{eqnarray}


\subsubsection{ Ground state energy $E(k)$ via the conditioned force 
$\vec  F^{C[k]}(\vec x) $ and the conditioned steady potential $U^{C[k]}_*(r) $   }

As explained in subsection \ref{subsec_canocond}, one can alternatively analyze 
the ground state energy $E(k)$ via the properties of the conditioned conditioned force 
$\vec  F^{C[k]}(\vec x) $.
In the present example
the radial and orthoradial components of the conditioned force 
$\vec  F^{C[k]}(\vec x) $ of Eq. \ref{forceDoobDiff} 
\begin{eqnarray}
  F^{C[k]}_r(r)
&&  = F_r(r)+ 2 k  A^{[{\cal O}]}_r(r) +2 D  \frac{ d  \ln l_k (r)  }{dr}
  \nonumber \\
  F^{C[k]}_{\theta}(r)
&&  = F_{\theta}(r)+ 2 k  A^{[{\cal O}]}_{\theta}(r)   
\label{forceDoobDiffpolar}
\end{eqnarray}
depend only on $r$ as a consequence of rotational invariance.
So the three decompositions of the conditioned force coincide as in Eq. \ref{Upolarcoincide},
where the reversible conditioned force corresponds to the radial part 
\begin{eqnarray}
\vec F^{C[k]rev} (\vec x ) =  F^{C[k]}_r(r) \vec e_r
\left( F_r(r)+ 2 k  A^{[{\cal O}]}_r(r) +2 D  \frac{ d  \ln l_k (r)  }{dr} \right) \vec e_r
\equiv - \frac{ d U^{C[k]}_*(r)  }{dr} \vec e_r
\label{forcerevCondpolar}
\end{eqnarray}
and only involves the derivative of the conditioned steady potential $U^{C[k]}_*(r) $,
while the irreversible conditioned force $\vec F^{C[k]irr} (\vec x ) $ is given by the orthoradial part
\begin{eqnarray}
\vec F^{C[k]irr} (\vec x ) =  F^{C[k]}_{\theta}(r)\vec e_{\theta}
  = \left( F_{\theta}(r)+ 2 k  A^{[{\cal O}]}_{\theta}(r)   \right)\vec e_{\theta}
\label{forceirrCondpolar}
\end{eqnarray}

Plugging these results
into Eq. \ref{eigenleftforcecond}
yields that the energy $E(k)$ and the conditioned potential $U^{C[k]}_*(r) $
satisfy
\begin{eqnarray}
 E(k) +  \frac{  [ F_{\theta}(r)+ 2 k  A^{[{\cal O}]}_{\theta}(r)]^2}{ 4D  } 
+ && \frac{  \left( \frac{ d U^{C[k]}_*(r)  }{dr} \right)^2 }{ 4D  } 
- \frac{1}{2r} \frac{d \left[ r \frac{ d U^{C[k]}_*(r)  }{dr}\right]}{dr} 
 \nonumber \\  
 =  \frac{  [ F_{\theta}(r)]^2}{ 4D  } 
+ && \frac{ \left[ U'(r) \right]^2 }{ 4D  }  
- \frac{1}{2r} \frac{d \left[ r U'(r)\right]}{dr} 
 + \frac{k}{D} V^{[{\cal O}]}(r) 
\label{eigenleftforcecondscalarpotirrpolar}
\end{eqnarray}

The link with the previous subsection can be understood via the identification of the radial part of the conditioned force of Eq. \ref{forcerevCondpolar}
using Eqs \ref{forcePolarderi}
and \ref{changetowardshatrlPolar} with Eq. \ref{eigenrightleftringpolar}
\begin{eqnarray}
- \frac{ d U^{C[k]}_*(r)  }{dr} 
&& =   F_r(r)+ 2 k  A^{[{\cal O}]}_r(r) +2 D  \frac{ d  \ln l_k (r)  }{dr} 
\nonumber \\
&& = -  U'(r)  + 2 k  A^{[{\cal O}]}_r(r) +2 D  \frac{ d   }{dr} \left[ {\mathring \Phi}_k(r)
+ \ln {\mathring \psi}_k (r ) \right] 
\nonumber \\
&& = 2 D  \frac{ d  \ln {\mathring \psi}_k (r ) }{dr} 
\label{forcerevCondpolaridentification}
\end{eqnarray}
where the last simplification comes from the explicit expression of Eq. \ref{vectorpotppolar} for $ {\mathring \Phi}_k'(r)  $. This very simple change of functions between the conditioned potential $U^{C[k]}_*(r)  $
and $\ln {\mathring \psi}_k (r ) $ transforms indeed
Eqs \ref{eigenrightleftringpolarexplicit} and \ref{eigenleftforcecondscalarpotirrpolar}
into each other.


\subsubsection{ Variance of the trajectory observable via the second-order perturbation theory of $E(k)$ in the parameter $k$   }

The variance of the trajectory observable of Eq. \ref{cumulant2}
is governed the second-order correction $ E^{[2]} $
of Eq. \ref{energy2fromlfactorizedcond}.
The first-order correction $\vec  F^{C_1}(\vec x) $ in $k$ of the conditioned force $\vec F^{C[k]} (\vec x) $ of Eq. \ref{forceDoobDiffpolar} with Eq. \ref{forcerevCondpolar} 
only involves the first-order conditioned potential $U^{C_1}_*(r) $
\begin{eqnarray}
  F^{C_1}_r(r)
&&  = - \frac{ d U^{C_1}_*(r)  }{dr}
  \nonumber \\
  F^{C_1}_{\theta}(r)
&&  =  2   A^{[{\cal O}]}_{\theta}(r)   
\label{forceDoobDiffpolar1}
\end{eqnarray}
and satisfies Eq. \ref{eigenleftseriesl1F1} 
\begin{eqnarray}
 V^{[{\cal O}]} ( r)  +O_*  
 &&
   =    \frac{ D \vec \nabla . \vec  F^{C_1}(\vec x)}{2}  +   \frac{ \vec F(\vec x) . \vec  F^{C_1}(\vec x)}{2}
   \nonumber \\
&&   = - \frac{D}{2 r} \frac{d \left[ r \frac{ d U^{C[k]}_*(r)  }{dr}\right]}{dr}
   +  \frac{1}{2}   U'(r) \frac{ d U^{C_1}_*(r)  }{dr} 
   +    F_{\theta}(r)    A^{[{\cal O}]}_{\theta}(r)  
 \label{eigenleftseriesl1F1polar}
\end{eqnarray}
that corresponds indeed to the first-order in $k$ of Eq. \ref{eigenleftforcecondscalarpotirrpolar}.

The second order correction of Eq. \ref{energy2fromlfactorizedcond}
then reads in terms of the first-order conditioned potential $U^{C_1}_*(r) $
\begin{eqnarray}
 E^{[2]} &&  =  -  \int d^d \vec x P_*(\vec x)   \frac{ [\vec  F^{C_1}(\vec x) ]^2}{4 D}
 = - \int_0^{+\infty} dr 2 \pi r P_*(r) \frac{ [  F^{C_1}_r(r) ]^2 + [F^{C_1}_{\theta}(r)]^2}{4 D}
 \nonumber \\
 && = - \int_0^{+\infty} dr 2 \pi r P_*(r) \frac{ \left[  \frac{ d U^{C_1}_*(r)  }{dr} \right]^2 + 4[A^{[{\cal O}]}_{\theta}(r)]^2}{4 D}
  \label{energy2fromlfactorizedcondPolar}
\end{eqnarray}


\subsection{ Large deviations  in the double limit of large time $T \to + \infty$
and small diffusion coefficient $D \to 0$  }

For $ D \to 0$, the steady state still given by Eq. \ref{Polarsteady} 
can be used to evaluate the steady value $O_*$ via Eq. \ref{additiveSteadyrhoPolar}.

\subsubsection{ Equation for the rescaled energy $e(k)$ and the conditioned steady potential $u^{C[k]}_*(r) $ 
for $D \to 0$  }

The rescaled energy $e(k)$ and the conditioned steady potential $u^{C[k]}_*(r) $ for $D \to 0$
satisfy Eq. \ref{eigenleftforcecondsimply} that corresponds to the leading order $\frac{1}{D}$ of Eq. \ref{eigenleftforcecondscalarpotirrpolar}
\begin{eqnarray}
e(k) +  \frac{  [ F_{\theta}(r)+ 2 k  A^{[{\cal O}]}_{\theta}(r)]^2}{ 4  } 
+  \frac{ \left( \frac{ d u^{C[k]}_*(r)  }{dr} \right)^2 }{ 4  } 
=  \frac{  [ F_{\theta}(r)]^2}{ 4  } 
+  \frac{ \left( \frac{ d U(r)  }{dr} \right)^2 }{ 4  }   
+ k V^{[{\cal O}]}(r) 
\label{eigenleftforcecondscalarpotirrDpolar}
\end{eqnarray}
that determines how
the square of the derivative $\frac{ d u^{C[k]}_*(r)  }{dr} $ depends on $r$
\begin{eqnarray}
  \left( \frac{ d u^{C[k]}_*(r)  }{dr} \right)^2 
&& =  \left( \frac{ d U(r)  }{dr} \right)^2   
+  [ F_{\theta}(r)]^2 -  [ F_{\theta}(r)+ 2 k  A^{[{\cal O}]}_{\theta}(r)]^2
+ 4 k V^{[{\cal O}]}(r) - 4e(k)
\nonumber \\
&& =  \left( \frac{ d U(r)  }{dr} \right)^2   
- 4 k  A^{[{\cal O}]}_{\theta}(r)[ F_{\theta}(r) +  k  A^{[{\cal O}]}_{\theta}(r)]
+ 4 k V^{[{\cal O}]}(r) - 4e(k)
\label{eigenleftforcecondscalarpotirrDpolar2}
\end{eqnarray}


\subsubsection{ Simplifications for the variance of the trajectory observable   }

For $D \to 0$, Eq. \ref{eigenleftseriesl1F1polar} simplifies into
\begin{eqnarray}
 V^{[{\cal O}]} ( r)   +O_* 
   =   \frac{1}{2}   U'(r) \frac{ d u^{C_1}_*(r)  }{dr} 
   +    F_{\theta}(r)      A^{[{\cal O}]}_{\theta}(r)  
 \label{eigenleftseriesl1F1polarD}
\end{eqnarray}
that yields 
\begin{eqnarray}
 \frac{ d u^{C_1}_*(r)  }{dr} 
 = \frac{2}{U'(r) } \left[ O_* + V^{[{\cal O}]} ( r)     -     F_{\theta}(r)      A^{[{\cal O}]}_{\theta}(r)  \right]
 \label{eigenleftseriesl1F1polarDgrad}
\end{eqnarray}
that can be plugged into Eq. \ref{energy2fromlfactorizedcondPolar}
to obtain the rescaled second order correction 
\begin{eqnarray}
 e^{[2]}  && = - \int_0^{+\infty} dr 2 \pi r P_*(r) \frac{ \left[  \frac{ d u^{C_1}_*(r)  }{dr} \right]^2 + 4[A^{[{\cal O}]}_{\theta}(r)]^2}{4 }
 \nonumber \\
 && = - \int_0^{+\infty} dr 2 \pi r P_*(r) 
\left(  [A^{[{\cal O}]}_{\theta}(r)]^2
 +   \frac{ \left[ O_* + V^{[{\cal O}]} ( r)  - F_{\theta}(r)  A^{[{\cal O}]}_{\theta}(r)  \right]^2 }{[U'(r)]^2}  \right)
    \label{energy2fromlfactorizedcondPolarD}
\end{eqnarray}

In summary, the rescaled second order correction $ e^{[2]} $
can be evaluated via an integration over the steady state $P_*(r) $
for any trajectory observable parametrized by 
the two functions $[V^{[{\cal O}]} ( r);A^{[{\cal O}]}_{\theta}(r) ] $.


 \section{ Conclusions }
 
 \label{seC_conclusions}

For diffusion processes in dimension $d>1$, we have first recalled how the statistics of trajectory observables over the time-window $[0,T]$ can be studied via the appropriate Feynman-Kac deformations of the Fokker-Planck generator, that can be interpreted as euclidean non-hermitian electromagnetic quantum Hamiltonians.
We have then compared the four regimes corresponding to the time $T$ either finite or large and to the diffusion coefficient $D$ either finite or small :

 (1) For finite $T$ and finite $D$, one needs to consider the full time-dependent quantum problem that involves the full spectrum of the Hamiltonian.
 
  (2) For large time $T \to + \infty$ and finite $D$, one only needs to consider the ground-state properties of the quantum Hamiltonian to obtain the generating function of rescaled cumulants and to construct the corresponding canonical conditioned processes.
  
   (3) For finite $T$ and $D \to 0$, one only needs to consider the dominant classical trajectory and its action satisfying the Hamilton-Jacobi equation, as in the semi-classical WKB approximation of quantum mechanics. 
   
   (4) In the double limit $T \to + \infty$ and $D \to 0$, the simplifications in the large deviations in $\frac{T}{D}$ of trajectory observables can be analyzed via the two orders of limits, i.e. either from the limit $D \to 0$ of the ground-state properties of the quantum Hamiltonians of (2), or from the limit of long classical trajectories $T \to +\infty$ in the semi-classical WKB approximation of (3). 

The comparison of these different limits $ T \to + \infty$ and/or $D \to 0$
are summarized for the propagator $P_T(\vec x \vert \vec x_0)$ in Table \ref{tablePropagator},
for the generating function $Z^{[k]}(\vec x \vert \vec x_0) $ in Table \ref{tableGenerating},
and for the joint distribution ${\cal P}_T({\cal O} ; \vec x \vert \vec x_0 )$ in Table \ref{tableTrajObs}.
Finally, this general framework has been illustrated in dimension $d=2$ with rotational invariance.

As a final remark, let us stress that in the present paper we have focused only on the simplest scenario,
where large deviations of trajectory observables 
in the double limit of large time $T \to + \infty$ and small diffusion coefficient 
$D \to 0$ involve the scaling variable $\frac{T}{D}$ in front of rate functions,
and can be analyzed in the two orders of limits to obtain the same results.
This scenario has been checked previously with detailed explicit calculations 
for the simplest geometry of the one-dimensional ring \cite{Engel_Seifert,derrida-ring,bertin-conditioned}.
However in dimension $d>1$, there are also many other interesting possibilities that may occur, in particular :

(i) the two orders of limits can give different results in some cases \cite{Bertini};

(ii) the scaling variable $\frac{T}{D}$ can be replaced by $T$ in some cases \cite{renaud};

(iii) the deterministic dynamics for $D=0$ can have several coexisting attractors 
that induce singularities in the steady potentials for small $D$ \cite{graham1986},
or can be chaotic with a fractal strange attractor 
leading to multifractal steady potentials for small $D$ \cite{graham1991}.
Note that for chaotic deterministic systems, 
the large deviations properties of trajectory observables are already very interesting without noise $D=0$
and can be analyzed 
via the same idea of the deformed generators, either in continuous time \cite{tailleur_thesis,tailleur_Lyapunov,laffargue}
 or in discrete time for chaotic non-invertible maps \cite{naftali,spain,c_chaoticmap}.


\begin{table}[!h]
\setcellgapes{3pt}
\begin{tabular}{|p{3cm}|p{7cm}|p{7cm}|}
\hline
Propagator $P_T(\vec x \vert \vec x_0)$
& Finite time $T$   
& Large time $T \to + \infty$ 
\\
\hline  
Finite noise $D$  

& Fokker-Planck dynamics of Eqs \ref{fokkerplanck} and \ref{fokkerplanckj} $ \ \ \ \ \ \ \ \ \ \ \ $ 
$$\partial_T P_T(\vec x \vert \vec x_0)   
 =  \vec \nabla . \left(   D  \vec \nabla -   \vec F(\vec x)\right) P_T(\vec x \vert \vec x_0)  \ \ \ \ \ \ \ \ $$
 
 Path-integral of Eq. \ref{pathintegral} 
 $$P_T(\vec x \vert \vec x_0) 
= \int_{\vec x(t=0)=\vec x_0}^{\vec x(t=T)=\vec x} {\cal D}   \vec x(.)   e^{ - {\cal S}(\vec x (0 \leq t \leq T) ) }  $$
involving the classical action ${\cal S}(\vec x (0 \leq t \leq T) ) $

& Convergence towards the steady state $P_*(\vec x) $ 
$$P_T(\vec x \vert \vec x_0) \oppropto_{ T \to +\infty}P_*(\vec x)$$ 
The steady current $\vec J_*(\vec x) \equiv P_*(\vec x)  \vec F(\vec x ) -D \vec \nabla   P_*(\vec x) $ 
should be divergenceless (Eq. \ref{fokkerplanckst})

$$ 0 = \vec \nabla . \vec J_*(\vec x)= \vec \nabla . \left(  P_*(\vec x)  \vec F(\vec x ) -D \vec \nabla   P_*(\vec x) \right) $$
 \\
\hline 
Small noise $D \to 0$  
& Propagator of Eq. \ref{pathintegralD} 
$$ P_T(\vec x \vert \vec x_0) \oppropto_{ D \to 0}  
 e^{ -    \frac{ {\hat S}_T(\vec x  \vert \vec x_0) }{ D} } $$
 dominated by the rescaled action ${\hat S}_T(\vec x  \vert \vec x_0) $
of the classical trajectory
 satisfying the Hamilton-Jacobi Eqs \ref{HamiltonJacobi} or \ref{HamiltonJacobibis}
 $$\partial_T {\hat S}_T(\vec x  \vert \vec x_0) 
 = - \left[ \vec F( \vec x)  + \vec \nabla {\hat S}_T(\vec x  \vert \vec x_0)\right] .  \vec \nabla {\hat S}_T(\vec x  \vert \vec x_0)$$
 
& Steady state of the form of Eq. \ref{rhosteadyD}
$$
P_T(\vec x \vert \vec x_0) 
\oppropto_{\substack{T \to +\infty\\ D \to 0}}  e^{ \displaystyle - \frac{ u_* (\vec x)}{D}  }
$$
with the steady Hamilton-Jacobi Eq. \ref{divforceirrevD1} for $ u_* (\vec x) $ 
$$
0  =    \left[ \vec F(\vec x)+  \vec \nabla u_* (\vec x) \right] .   \vec \nabla u_* (\vec x) 
$$
  \\
\hline
\end{tabular}
\caption{ Simplifications for the propagator $P_T(\vec x \vert \vec x_0)$ for large time $T \to + \infty$ and/or small diffusion coefficient $D \to 0$
} 
\label{tablePropagator}
\end{table}


\newpage

\begin{table}[!h]
\setcellgapes{3pt}
\begin{tabular}{|p{3.5cm}|p{7cm}|p{7cm}|}
\hline
Generating function of Eq. \ref{gene}
\begin{eqnarray}
&& Z^{[k]}(\vec x \vert \vec x_0) \equiv
 \nonumber \\
&& \int d {\cal O}    {\cal P}_T({\cal O} ; \vec x \vert \vec x_0 ) \ e^{ \frac{k}{D}{\cal O}}
\nonumber
\end{eqnarray}
& Finite time $T$   
& Large time $T \to + \infty$ 
\\
\hline  
Finite noise $D$  

& Euclidean Schr\"odinger Eq. \ref{EuclideanZ} with quantum non-hermitian electromagnetic Hamiltonian $H_k$
$$\partial_T Z^{[k]}(\vec x \vert \vec x_0)   
 =   -  H_k Z^{[k]}(\vec x \vert \vec x_0)   $$
 
 Path-integral of Eq. \ref{gene} 
 $$Z^{[k]}(\vec x \vert \vec x_0) 
= \int_{\vec x(t=0)=\vec x_0}^{\vec x(t=T)=\vec x} {\cal D}   \vec x(.)   e^{ - {\cal S}^{[k]}(\vec x (0 \leq t \leq T) ) }  $$
with deformed action ${\cal S}^{[k]}(\vec x (0 \leq t \leq T) ) $

& Convergence for large $T$ (Eq. \ref{genelargeT})
$$Z^{[k]}(\vec x \vert \vec x_0) \opsimeq_{T \to +\infty}
 e^{- T E(k) } r_k(\vec x) l_k(\vec x_0)$$ 
with the positive right eigenvector $r_k(.)$ and
left eigenvector $l_k(.)$ associated to 
the ground state energy $E(k)$ of $H_k $
(Eqs \ref{eigenright} and \ref{eigenleft})
\begin{eqnarray}
 E(k)  r_k( \vec x) && = H_k  r_k( \vec x) 
 \nonumber \\
 E(k)  l_k( \vec x) && = H_k^{\dagger}  l_k( \vec x)
\nonumber
\end{eqnarray}
 \\
\hline 
Small noise $D \to 0$  
& Generating function of Eq. \ref{geneD}
$$ Z^{[k]}(\vec x \vert \vec x_0) \oppropto_{ D \to 0}  
 e^{ -    \frac{ {\hat S}_T^{[k]}(\vec x  \vert \vec x_0) }{ D} } $$
 dominated by the rescaled action ${\hat S}^{[k]}_T(\vec x  \vert \vec x_0) $
of the classical trajectory $x^{Class}(0 \leq t \leq T) $
 satisfying the forward and backward Hamilton-Jacobi Eqs \ref{HamiltonJacobik} and \ref{HamiltonJacobiBackwardk}
 or Eqs \ref{HamiltonJacobikbis} and \ref{HamiltonJacobiBackkwardbis}
  governed by the rescaled classical electromagnetic Hamiltonian $ {\hat {\cal H}}_k ({\vec x},  {\vec p} ) $
\begin{eqnarray}
 \partial_T {\hat S}^{[k]}_T(\vec x  \vert \vec x_0) 
&& = - {\hat {\cal H}}_k ({\vec x},  {\vec p} = \vec \nabla {\hat S}^{[k]}_T(\vec x  \vert \vec x_0))
\nonumber \\
 \partial_T {\hat S}^{[k]}_T(\vec y  \vert \vec x) 
&& = - {\hat {\cal H}}_k ({\vec x},  {\vec p} = - \vec \nabla {\hat S}^{[k]}_T(\vec y  \vert \vec x))
\nonumber
\end{eqnarray}
 
& Convergence towards Eq. \ref{genelargeTD} with $E(k) \simeq \frac{e(k)}{D}$
$$
Z^{[k]}_T(\vec x \vert \vec x_0) 
\opsimeq_{\substack{T \to +\infty\\ D \to 0}}
 e^{ \displaystyle - \frac{T}{D}  e(k) - \frac{u_k(\vec x)}{D} - \frac{{\tilde u}_k(\vec x_0)}{D}} 
$$
with the forward and backward energy Hamilton-Jacobi Equations for $u_k(\vec x) $ and ${\tilde u}_k(\vec x) $
(Eqs \ref{eigenrightlogD} and \ref{eigenleftlogD} or Eqs \ref{HamiltonJacobiWkbis} and \ref{HamiltonJacobiWkbackwardbis})
\begin{eqnarray}
- e(k) && =  {\hat {\cal H}}^{[k]} \left({\vec x},  {\vec p} = \vec \nabla u_k(\vec x) \right)
 \nonumber \\
 - e(k) && =  {\hat {\cal H}}^{[k]} \left({\vec x},  {\vec p} = - \vec \nabla {\tilde u}_k ( \vec x) \right)
 \nonumber 
\end{eqnarray}
  \\
\hline
\end{tabular}
\caption{ Simplifications for the generating function $Z^{[k]}(\vec x \vert \vec x_0)$ of the trajectory observable ${\cal O}$
for large time $T \to + \infty$ and/or small diffusion coefficient $D \to 0$.
The translation for the joint distribution ${\cal P}_T({\cal O} ; \vec x \vert \vec x_0 )$ 
is given in the next table \ref{tableTrajObs}.
} 
\label{tableGenerating}
\end{table}


\newpage

\begin{table}[!h]
\setcellgapes{3pt}
\begin{tabular}{|p{3cm}|p{7cm}|p{7cm}|}
\hline
Joint distribution ${\cal P}_T({\cal O} ; \vec x \vert \vec x_0 )$ of the trajectory observable ${\cal O}$ and position $\vec x$
& Finite time $T$   
& Large time $T \to + \infty$ 
\\
\hline  
Finite noise $D$  

& Constrained path-integral of Eq. \ref{pathintegralPO}
\begin{eqnarray}
&&{\cal P}_T({\cal O} ; \vec x \vert \vec x_0 )
= \int_{\vec x(t=0)=\vec x_0}^{\vec x(t=T)=\vec x} {\cal D}   \vec x(.)   e^{ - {\cal S}(\vec x (0 \leq t \leq T) ) }
\nonumber \\
&& \delta \left( \int_0^T dt \left[   - V^{[{\cal O}]} (\vec x(t) ) 
+  \vec A^{[{\cal O}]}( \vec x(t)) . \dot {\vec x}(t)  \right] - {\cal O}\right)
\nonumber
\end{eqnarray}
imposing the value ${\cal O} $ for the extensive trajectory observable 
parametrized by the two functions $[V^{[{\cal O}]}(.); \vec A^{[{\cal O}]}(.)] $.

& Large deviations of Eq. \ref{ratefunctionCalO} for the intensive trajectory observable $O = \frac{{\cal O}}{T} $ 
$${\cal P}_T({\cal O} = T O  ) \oppropto_{ T \to + \infty} e^{- T I (O) }$$ 
The rate function $I(O) \geq 0 $ 
is related to the ground-state energy $E(k)$ via the Legendre transform of Eq. \ref{legendrereci}
\begin{eqnarray}
  I(O)  && = E(k) +  \frac{k}{D}  O
 \nonumber \\
 0   && = E'(k) +  \frac{1}{D}  O
\nonumber
\end{eqnarray}
 \\
\hline 
Small noise $D \to 0$  
& Large deviations of Eq. \ref{ratefunctionD} for the joint distribution of ${\cal O} $ and $x$
$$
 {\cal P}_T({\cal O} ; \vec x \vert \vec x_0 ) \oppropto_{ D \to 0} e^{ \displaystyle 
 - \frac{{\hat I}_T({\cal O} ;\vec x  \vert \vec x_0)}{D}  }
$$
The rate function ${\hat I}_T({\cal O} ;\vec x  \vert \vec x_0) \geq 0 $ 
is related to the rescaled classical action ${\hat S}^{[k]}_T(\vec x  \vert \vec x_0) $
 via the Legendre transform of Eq. \ref{legendrereciD}
\begin{eqnarray}
{\hat I}_T({\cal O} ;\vec x  \vert \vec x_0)&& = k {\cal O}  + {\hat S}_T^{[k]}(\vec x  \vert \vec x_0)
 \nonumber \\
 0   && = {\cal O}  + \partial_k {\hat S}_T^{[k]}(\vec x  \vert \vec x_0)
\nonumber
\end{eqnarray}

& Large deviations for the intensive value $O = \frac{{\cal O}}{T} $ of Eq. \ref{ratefunctionCalOrecaled}
$$
 {\cal P}_T({\cal O} = T O  )  
\opsimeq_{\substack{T \to +\infty\\ D \to 0}}
 e^{ \displaystyle - \frac{T}{D}  i(O)} 
$$
The rescaled rate function $i(O) \geq 0 $ 
is related to rescaled eigenvalue $e(k)$ via the Legendre transform of Eq. \ref{legendrerecirescaped}
or Eq. \ref{legendrereciDpuisT}
   \begin{eqnarray}
 i(O)  && = e(k) +  k  O
 \nonumber \\
 0   && = e'(k) +   O
 \nonumber
\end{eqnarray}
  \\
\hline
\end{tabular}
\caption{ Large deviations for the joint distribution ${\cal P}_T({\cal O} ; \vec x \vert \vec x_0 )$ of the trajectory observable ${\cal O}$ and position $\vec x$ for large time $T \to + \infty$ and/or small diffusion coefficient $D \to 0$, as translated from the previous Table \ref{tableGenerating} concerning the corresponding generating function $ Z^{[k]}(\vec x \vert \vec x_0)$.
} 
\label{tableTrajObs}
\end{table}


\newpage

 \appendix
 

\section{ Simplifications of the large deviations at level 2.5 
in the double limit $T \to + \infty$ and $D \to 0$   }

In this Appendix, we recall the large deviations at level 2.5 for large time $T \to + \infty$
for finite $D$ in order to analyze the simplifications when the diffusion coefficient 
becomes small $D \to 0$, and to make the link with the statistics of trajectory observables
discussed in the main text.

\label{app_2.5}

\subsection{ Reminder on the large deviations at level 2.5 
for large time $T \to + \infty$ and finite diffusion coefficient $D$   }

Whenever the Fokker-Planck dynamics of Eq. \ref{fokkerplanck}
converges towards some steady state $P_*(\vec x)$ with its steady current $\vec J_*(\vec x) $
of Eq. \ref{jsteady}, it is interesting to analyze how the 
empirical observables defined as time-averages over the large time-window $[0,T]$ 
can fluctuate around their steady values.

\subsubsection{ Empirical density $ P^E(\vec x)  $ and empirical current $\vec J^E(\vec x) $ with their constitutive constraints  }

For a diffusive trajectory $\vec x(0 \leq t \leq T)$ over the large time-window $[0,T]$, 
the empirical density
\begin{eqnarray}
 P^E(\vec x)  \equiv \frac{1}{T} \int_0^T dt \  \delta^{(d)} ( \vec x(t)- \vec x)  
\label{rhodiff}
\end{eqnarray}
is normalized to unity
\begin{eqnarray}
\int d^d \vec x \  P^E(\vec x)  = 1
\label{rho1ptnormadiff}
\end{eqnarray}
while the empirical current 
\begin{eqnarray} 
\vec J^E(\vec x) \equiv   \frac{1}{T} \int_0^T dt \ \frac{d \vec x(t)}{dt}   \delta^{(d)}( \vec x(t)- \vec x)  
\label{diffj}
\end{eqnarray}
should be divergence-free 
\begin{eqnarray}
 \vec \nabla . \vec J^E(\vec x) =0
\label{divergencenulle}
\end{eqnarray}
in order to be consistent with stationarity.


\subsubsection{ Large deviations at Level 2.5 for 
the joint distribution of the empirical density $ P^E(\vec x)  $ and the empirical current $\vec J^E(\vec x) $ }

The joint probability distribution 
of the empirical density $ P^E(\vec x)  $ and the empirical current $\vec J^E(\vec x) $
satisfies the large deviation form for large time $T \to + \infty$
\cite{wynants_thesis,maes_diffusion,chetrite_formal,engel,chetrite_HDR,c_lyapunov,c_inference,c_susyboundarydriven,c_diffReg,c_inverse}
\begin{eqnarray}
 P_T^{[2.5]}[ P^E(.), \vec J^E(.)]   \opsimeq_{T \to +\infty}  \delta \left(\int d^d \vec x P^E(\vec x) -1  \right)
\left[ \prod_{\vec x }  \delta \left(  \vec \nabla . \vec J^E(\vec x) \right) \right] 
e^{- \displaystyle T I_{2.5} \left[ P^E(.);\vec J^E(.) \right]    }
\label{ld2.5diff}
\end{eqnarray}
where the constitutive constraints have been discussed in Eqs \ref{rho1ptnormadiff} and \ref{divergencenulle},
while the explicit rate function 
\begin{eqnarray}
  I_{2.5} \left[ P^E(.);\vec J^E(.) \right]   &&  =
  \int \frac{d^d \vec x}{ 4 D P^E(\vec x) } \left[ \vec J^E(\vec x) - P^E(\vec x) \vec F(\vec x)+D \vec \nabla P^E(\vec x) \right]^2
  \label{rate2.5diff}
\end{eqnarray}
vanishes only when the empirical observables $[ P^E(.);\vec J^E(.) ] $ satisfying the constraints
coincide with their steady values $[ P_*(.);\vec J_*(.) ] $.


\subsubsection{ Large deviations at Level 2.5 for 
the joint distribution of the empirical density $ P^E(\vec x)  $ and the empirical force $\vec F^E(\vec x) $ }

It is often convenient to replace the empirical current $\vec J^E(\vec x)  $ by the empirical force $\vec F^E(\vec x) $ via
the formula analog to Eq. \ref{jsteady}
\begin{eqnarray} 
\vec J^E(\vec x)  = P^E(\vec x)  \vec F^E(\vec x ) -D \vec \nabla   P^E(\vec x) 
\label{empiricalforce}
\end{eqnarray}
in particular in the field of inference of the model from data on a long trajectory \cite{c_inference}.
Then 
Eq. \ref{ld2.5diff} translates for the 
joint distribution of the empirical density $ P^E(\vec x)  $ and the empirical force $\vec F^E(\vec x) $
into
\begin{eqnarray}
 P_T^{[2.5]}[ P^E(.), \vec F^E(.)]   \opsimeq_{T \to +\infty}  \delta \left(\int d^d \vec x P^E(\vec x) -1  \right)
\left[ \prod_{\vec x }  \delta \left(  \vec \nabla . \left[ P^E(\vec x)  \vec F^E(\vec x ) -D \vec \nabla   P^E(\vec x) \right] \right) \right] 
e^{- \displaystyle T I_{2.5} \left[ P^E(.);\vec F^E(.) \right]    }
\label{ld2.5infer}
\end{eqnarray}
where the rate function obtained from Eq. \ref{rate2.5diff}
reduces to the gaussian form with respect to the empirical force $\vec F^E(\vec x) $
\begin{eqnarray}
  I_{2.5} \left[ P^E(.);\vec F^E(.) \right]     =
  \int \frac{d^d \vec x}{ 4 D  } P^E(\vec x) \left[ \vec F^E(\vec x) - \vec F(\vec x) \right]^2
  \label{rate2.5diffinfer}
\end{eqnarray}
while the stationarity constraint 
that means that the empirical density $P^E(.) $
is the steady state associated to the Fokker-Planck generator involving the empirical force $\vec F^E(.) $
can also be rewritten as
\begin{eqnarray}
0 && = D  \vec \nabla . \left( \vec F^E(\vec x ) -D \vec \nabla \ln  P^E(\vec x) \right)
+  \left( \vec F^E(\vec x ) -D \vec \nabla \ln  P^E(\vec x) \right) . D \vec \nabla \ln P^E(\vec x)
\nonumber \\
&& = D  \vec \nabla .  \vec F^{Eirr}(\vec x )
+   \vec F^{Eirr}(\vec x ) . \vec F^{Erev}(\vec x )
\label{statioinfer}
\end{eqnarray}
in terms of the reversible and irreversible empirical forces
\begin{eqnarray}
 \vec F^{Erev}(\vec x ) && \equiv D \vec \nabla \ln  P^E(\vec x) 
\nonumber \\
 \vec F^{Eirr}(\vec x ) && \equiv \vec F^E(\vec x ) -D \vec \nabla \ln  P^E(\vec x) 
\label{empirevirr}
\end{eqnarray}


\subsubsection{ Link with the large deviations of trajectory observables described in the main text }

Any trajectory observable of Eq. \ref{additive}
can be rewritten in terms of the empirical density $ P^E(\vec x)  $ of Eq. \ref{rhodiff}
and of the empirical current $\vec J^E(\vec x) $ of Eq. \ref{diffj}
\begin{eqnarray}
{\cal O} [\vec x(0 \leq t \leq T) ] &&  = \int_0^T dt \left[   - V^{[{\cal O}]} (\vec x(t) ) 
+  \vec A^{[{\cal O}]}( \vec x(t)) . \dot {\vec x}(t)  \right]
\nonumber \\
&& = T \int d^d \vec x  \left[   - V^{[{\cal O}]} (\vec x ) P^E(\vec x)
+  \vec A^{[{\cal O}]}( \vec x) . \vec J^E(\vec x)  \right] 
\label{additiveEmpi}
\end{eqnarray}
or in terms of the empirical density $ P^E(\vec x)  $ of
and of the empirical force $\vec F^E(\vec x) $ using Eq. \ref{empiricalforce}
\begin{eqnarray}
{\cal O} [\vec x(0 \leq t \leq T) ]  && = T \int d^d \vec x P^E(\vec x)  \left(   - V^{[{\cal O}]} (\vec x ) 
+  \vec A^{[{\cal O}]}( \vec x) . \left[  \vec F^E(\vec x ) -D \vec \nabla \ln  P^E(\vec x) \right]  \right)
\nonumber \\
&&
\equiv T O^E[ P^E(.), \vec F^E(.)] 
\label{additiveEmpiforce}
\end{eqnarray}

As a consequence, the generating function of the trajectory observable of Eq. \ref{gene}
can be also computed for large $T$ from the joint distribution of Eq. \ref{ld2.5infer}
\begin{eqnarray}
 Z^{[k]}_T
&& \equiv  \langle e^{ \displaystyle \frac{k}{D}{\cal O} [\vec x(0 \leq t \leq T) ]} \rangle
 = \int {\cal D}P^E(.)  {\cal D} \vec F^E(.)  P_T^{[2.5]}[ P^E(.), \vec F^E(.)] e^{ \displaystyle \frac{k}{D} T O^E[ P^E(.), \vec F^E(.)] }
 \nonumber \\
 &&
\opsimeq_{T \to +\infty}  \int {\cal D}P^E(.)  {\cal D} \vec F^E(.)  
\delta \left(\int d^d \vec x P^E(\vec x) -1  \right)
\left[ \prod_{\vec x }  \delta \left(  \vec \nabla . \left[ P^E(\vec x)  \vec F^E(\vec x ) -D \vec \nabla   P^E(\vec x) \right] \right) \right] 
 \nonumber \\
 && e^{- \displaystyle T \left( I_{2.5} \left[ P^E(.);\vec F^E(.) \right]  - \frac{k}{D}  O^E[ P^E(.), \vec F^E(.)]\right)  }
 \opsimeq_{T \to +\infty}  e^{- \displaystyle T E(k) }
 \label{geneEmpi}
\end{eqnarray}
so that the energy $E(k)$ of Eq. \ref{genelargeT} also corresponds to the optimization
of the function in the exponential over the empirical observables $[ P^E(.);\vec F^E(.) ] $
satisfying the constitutive constraints.
The solution of this optimization is the 
conditioned steady state $P^{C[k]}_* (\vec x) $ and the conditioned force $\vec  F^{C[k]}(\vec x) $
discussed in subsection \ref{subsec_canocond} of the main text.


\subsection{ Simplifications of the large deviations at level 2.5  when the diffusion coefficient is small $D \to 0$   }

\subsubsection{ Large deviations for 
the joint distribution of the empirical potential $ u^E(\vec x)  $ and the empirical force $\vec f^E(\vec x) $ }

When the diffusion coefficient is small $D \to 0$, the steady state $p_*(\vec x)$
can be written as Eq. \ref{rhosteadyD} in terms of the steady potential $u_*(\vec x)$,
so it is natural to parametrize the empirical density $ p^E(\vec x)  $ similarly
\begin{eqnarray}
 p^E(\vec x)   =
 \frac{ e^{ - \frac{ u^E(\vec x) }{  D }  } }{ \int d^d \vec y \ e^{ - \frac{ u^E(\vec y) }{  D }  }} 
\label{rhoempiricalD}
\end{eqnarray}
in order to analyze how the empirical potential $ u^E(\vec x)   $ can fluctuate around the steady potential $u_*(\vec x) $.
So the reversible and irreversible empirical forces of Eq. \ref{empirevirr} become
\begin{eqnarray}
 \vec f^{Erev}(\vec x ) && \equiv - \vec \nabla u^E(\vec x) 
\nonumber \\
 \vec f^{Eirr}(\vec x ) && \equiv \vec f^E(\vec x ) +\vec \nabla u^E(\vec x) 
\label{empirevirrD}
\end{eqnarray}
while the stationarity constraint of Eq. \ref{statioinfer}
reduces in the limit $D \to 0$ to the orthogonality constraint
\begin{eqnarray}
0 && =    \vec f^{Eirr}(\vec x ) . \vec f^{Erev}(\vec x )
= -  \left[ \vec f^E(\vec x ) +\vec \nabla u^E(\vec x) \right] . \vec \nabla u^E(\vec x)
\label{statioinferD}
\end{eqnarray}

Putting everything together, one obtains that Eq. \ref{ld2.5infer}
translates for the 
joint distribution of the empirical potential $ u^E(\vec x)  $ and the empirical force $\vec f^E(\vec x) $ into
\begin{eqnarray}
 p_T^{[2.5]}[ u^E(.), \vec f^E(.)]   \opsimeq_{\substack{T \to +\infty \\ D \to 0 }}  
\left[ \prod_{\vec x }  \delta \left( \left[ \vec f^E(\vec x ) +\vec \nabla u^E(\vec x) \right] . \vec \nabla u^E(\vec x)  \right) \right] 
e^{- \displaystyle \frac{T}{D} i_{2.5} \left[ u^E(.);\vec f^E(.) \right]    }
\label{ld2.5inferD}
\end{eqnarray}
where the rate function obtained from Eq. \ref{rate2.5diffinfer}
\begin{eqnarray}
  i_{2.5} \left[ u^E(.);\vec f^E(.) \right]     =
  \int \frac{d^d \vec x}{ 4   } \left( \frac{ e^{ - \frac{ u^E(\vec x) }{  D }  } }{ \int d^d \vec y \ e^{ - \frac{ u^E(\vec y) }{  D }  }} \right)  \left[ \vec f^E(\vec x) - \vec F(\vec x) \right]^2
  \label{rate2.5diffinferD}
\end{eqnarray}
vanishes only when the empirical potential $u_*(\vec x) $ and the empirical force $\vec f^E(\vec x) $ 
satisfying the constraints coincide
with the steady potential $u_*(\vec x) $ and the true force $\vec F(\vec x)$.


\subsubsection{ Link with the large deviations of trajectory observables described in the main text }

Using Eq. \ref{rhoempiricalD}, the trajectory observable of Eq. \ref{additiveEmpiforce}
becomes
\begin{eqnarray}
{\cal O} [\vec x(0 \leq t \leq T) ]  && = T \int d^d \vec x 
\left( \frac{ e^{ - \frac{ u^E(\vec x) }{  D }  } }{ \int d^d \vec y \ e^{ - \frac{ u^E(\vec y) }{  D }  }} \right)
  \left(   - V^{[{\cal O}]} (\vec x ) 
+  \vec A^{[{\cal O}]}( \vec x) . \left[  \vec f^E(\vec x ) +\vec \nabla u^E(\vec x)  \right]  \right)
\nonumber \\
&& \equiv T o^E[ u^E(.), \vec f^E(.)] 
\label{additiveEmpiforceD}
\end{eqnarray}
Then the generating function of Eq. \ref{geneEmpi}
becomes using Eq. \ref{ld2.5infer}
\begin{eqnarray}
 Z^{[k]}_T
&& \equiv  \langle e^{ \displaystyle \frac{k}{D}{\cal O} [\vec x(0 \leq t \leq T) ]} \rangle
 = \int {\cal D}u^E(.)  {\cal D} \vec f^E(.)  p_T^{[2.5]}[ u^E(.), \vec f^E(.)] 
 e^{ \displaystyle \frac{k}{D} T o^E[ u^E(.), \vec f^E(.)] }
 \nonumber \\
 &&
\opsimeq_{T \to +\infty}  \int {\cal D}u^E(.)  {\cal D} \vec f^E(.)  
\left[ \prod_{\vec x }  \delta \left( \left[ \vec f^E(\vec x ) +\vec \nabla u^E(\vec x) \right] . \vec \nabla u^E(\vec x)  \right) \right] 
  e^{- \displaystyle  \frac{T}{D} \left( i_{2.5} \left[ u^E(.);\vec f^E(.) \right]  - k  o^E[ u^E(.), \vec f^E(.)]\right)  }
\nonumber \\
 && \opsimeq_{T \to +\infty}  e^{- \displaystyle \frac{T}{D} e(k) }
 \label{geneEmpiD}
\end{eqnarray}
so that the rescaled energy $e(k)$ also corresponds to the optimization
of the function in the exponential over the empirical observables $[ u^E(.);\vec f^E(.) ] $
satisfying the stationarity constraint.
The solution of this optimization is the 
conditioned steady state $p^{C[k]}_* (\vec x) $ and the conditioned force $\vec  f^{C[k]}(\vec x) $
discussed in subsection \ref{subsec_canocondD} of the main text.
 

\section{ Gauge transformations for the euclidean non-hermitian electromagnetic quantum problem }

\label{app_gauge}

In this Appendix, we describe the effects of gauge transformations 
for the euclidean non-hermitian electromagnetic quantum problem
that appear in the main text (see \cite{us_gyrator} for more detailed discussions).


\subsection{ Gauge transformation for the quantum problem 
associated to the generating function $ Z^{[k]}_T(\vec x \vert \vec x_0)  $ }

The effect of the gauge transformation of the vector potential $\vec A^{[k]} (\vec x ) $ of Eq. \ref{vectorpotp}
\begin{eqnarray}
  \vec A^{[k]} (\vec x )   = - \vec \nabla {\mathring \Phi}_k( r) + \vec {\mathring A}^{[k]}(\vec x )
 \label{vectorpotgaugeirr}
\end{eqnarray}
on the path-integral of Eq. \ref{gene} 
can be analyzed from the integral over time of 
the term involving the vector potential in the Lagrangian of Eq. \ref{lagrangiank} 
\begin{eqnarray}
  \int_0^T dt \dot {\vec x} (t).   \vec A^{[k]}( \vec x(t)) 
  = {\mathring \Phi}(\vec x(0)) -  {\mathring \Phi}(\vec x(t)) 
 +   \int_0^T dt \dot {\vec x} (t).   \vec {\mathring A}^{[k]}( \vec x(t)) 
\label{pathintegrallast}
\end{eqnarray}
So the appropriate change of variable for the generating function of Eq. \ref{gene} reads
\begin{eqnarray}
 Z^{[k]}_T(\vec x \vert \vec x_0) 
 = e^{ \displaystyle {\mathring \Phi}_k( \vec x_0) - {\mathring \Phi}_k( \vec x)}  {\mathring Z}^{[k]}_T(\vec x \vert \vec x_0) 
\label{changetowardshat}
\end{eqnarray}
where the new function $ {\mathring Z}^{[k]}_T(\vec x \vert \vec x_0) $ corresponds to the path-integral
\begin{eqnarray}
{\mathring Z}^{[k]}_T(\vec x \vert \vec x_0) 
 =  \int_{\vec x(t=0)=\vec x_0}^{\vec x(t=T)=\vec x} {\cal D}   \vec x(.)  
 e^{ - \displaystyle \int_{0}^T d t
{\mathring{\cal L}}_k (\vec x(t), \dot {\vec x}(t) )
 }
\label{pathintegralring}
\end{eqnarray}
governed by the new Lagrangian
\begin{eqnarray}
{\mathring{\cal L}}_k (\vec x(t), \dot {\vec x}(t) )
  \equiv 
 \frac{\dot {\vec x}^2  (t) }{4D}  
 - \dot {\vec x } (t) . \vec {\mathring A}^{[k]}( \vec x(t))
 + V^{[k]}(\vec x(t))
 \label{lagrangianring}
\end{eqnarray}
that involves the same scalar potential $V^{[k]}( \vec x) $ 
but the new vector potential $\vec {\mathring A}^{[k]}( \vec x)  $.

Equivalently, the effect of a gauge transformation can be analyzed
from the point of view of the quantum Hamiltonian as follows.
Via the change of function of Eq. \ref{changetowardshat}, one obtains that 
the euclidean Schr\"odinger Eq. \ref{EuclideanZ} for $Z_T^{[k]}(\vec x \vert \vec x_0)   $
translates for $ {\mathring Z}^{[k]}_T(\vec x \vert \vec x_0)$ into
\begin{eqnarray}
 \partial_T {\mathring Z}^{[k]}_T(\vec x \vert \vec x_0)
 =  -  {\mathring H}_k {\mathring Z}^{[k]}_T(\vec x \vert \vec x_0)
\label{EuclideanZring}
\end{eqnarray}
where the new quantum Hamiltonian ${\mathring H}_k $
\begin{eqnarray}
 {\mathring H}_k   && =     -  D  \left( \vec \nabla -   \vec {\mathring A}^{[k]}(\vec x) \right)^2    +  V^{[k]}(\vec x)
 \label{FPhamiltonianring}
\end{eqnarray}
involves the same scalar potential $V^{[k]}( \vec x) $ 
but the new vector potential $\vec {\mathring A}^{[k]}( \vec x)  $, as it should for consistency with the 
Lagrangian of Eq. \ref{lagrangianring}.


\subsection{ Consequences for the eigenvalue problem governing
 $Z_T^{[k]}(\vec x \vert \vec x_0)   $  for large time $T \to + \infty$ and finite $D$  }

For the positive right and left eigenvectors of Eqs \ref{eigenright} and \ref{eigenleft}
governing the large time behavior of the generating function $Z_T^{[k]}(\vec x \vert \vec x_0) $ in Eq. \ref{genelargeT},
the change of functions of Eq. \ref{changetowardshat}
corresponds to
\begin{eqnarray}
 r_k(\vec x)  && = e^{ - {\mathring \Phi}_k(\vec x)}  {\mathring r}_k (\vec x ) 
 \nonumber \\
  l_k(\vec x_0)  && = e^{  {\mathring \Phi}_k(\vec x_0)}  {\mathring l}_k (\vec x_0 ) 
\label{changetowardshatrl}
\end{eqnarray}
where ${\mathring r}^{[k]} (. ) $ and ${\mathring l}^{[k]} (. ) $ are the positive right and left eigenvectors associated 
to the Hamiltonian ${\mathring H}_k  $ for the same energy $E(k)$
\begin{eqnarray}
 E(k)  {\mathring r}_k (\vec x ) && =  {\mathring H}_k  {\mathring r}^{[k]} (\vec x )
 \nonumber \\
E(k)  {\mathring l}_k( \vec x) && =  {\mathring H}_k^{\dagger}  {\mathring l}_k( \vec x)
\label{eigenrightleftring}
\end{eqnarray}
with the normalization translated from Eq. \ref{eigennorma}
\begin{eqnarray}
1 = \int d^d \vec x \   l_k( \vec x)  r_k( \vec x) =\int d^d \vec x \ {\mathring l}_k (\vec x )  {\mathring r}_k (\vec x ) 
  \label{eigennormaring}
\end{eqnarray}

In summary, the ground state energy $E(k)$ can be analyzed for the 
gauge-transformed quantum Hamiltonian ${\mathring H}_k$ of Eq. \ref{FPhamiltonianring} :
in practice, it is thus useful to choose the simplest new vector potential $\vec {\mathring A}^{[k]}( \vec x)  $
in Eq. \ref{vectorpotgaugeirr}, as discussed in the exemple of section \ref{sec_2DPolar}.


\end{document}